\ifx\mnmacrosloaded\undefined 
%
%
%
%

\catcode `\@=11 

\def\@version{1.4}
\def\@verdate{22nd Feb 1994}

%
%
%
%


\newif\ifprod@font

\ifx\@typeface\undefined
  \def\@typeface{Comp. Modern}\prod@fontfalse
\else
  \prod@fonttrue 
\fi

\def\newfam{\alloc@8\fam\chardef\sixt@@n} 

\ifprod@font
\font\fiverm=mtr10 at 5pt
\font\fivebf=mtbx10 at 5pt
\font\fiveit=mtti10 at 5pt
\font\fivesl=mtsl10 at 5pt
\font\fivett=mttt10 at 5pt     \hyphenchar\fivett=-1
\font\fivecsc=mtcsc10 at 5pt
\font\fivesf=mtss10 at 5pt
\font\fivei=mtmi10 at 5pt      \skewchar\fivei='177
\font\fivemib=mtmib10 at 5pt   \skewchar\fivemib='177
\font\fivesy=mtsy10 at 5pt     \skewchar\fivesy='60
\font\fivesyb=mtbsy10 at 5pt   \skewchar\fivesyb='60

\font\sixrm=mtr10 at 6pt
\font\sixbf=mtbx10 at 6pt
\font\sixit=mtti10 at 6pt
\font\sixsl=mtsl10 at 6pt
\font\sixtt=mttt10 at 6pt      \hyphenchar\sixtt=-1
\font\sixcsc=mtcsc10 at 6pt
\font\sixsf=mtss10 at 6pt
\font\sixi=mtmi10 at 6pt       \skewchar\sixi='177
\font\sixmib=mtmib10 at 6pt    \skewchar\sixmib='177
\font\sixsy=mtsy10 at 6pt      \skewchar\sixsy='60
\font\sixsyb=mtbsy10 at 6pt    \skewchar\sixsyb='60

\font\sevenrm=mtr10 at 7pt
\font\sevenbf=mtbx10 at 7pt
\font\sevenit=mtti10 at 7pt
\font\sevensl=mtsl10 at 7pt
\font\seventt=mttt10 at 7pt     \hyphenchar\seventt=-1
\font\sevencsc=mtcsc10 at 7pt
\font\sevensf=mtss10 at 7pt
\font\seveni=mtmi10 at 7pt      \skewchar\seveni='177
\font\sevenmib=mtmib10 at 7pt   \skewchar\sevenmib='177
\font\sevensy=mtsy10 at 7pt     \skewchar\sevensy='60
\font\sevensyb=mtbsy10 at 7pt   \skewchar\sevensyb='60

\font\eightrm=mtr10 at 8pt
\font\eightbf=mtbx10 at 8pt
\font\eightit=mtti10 at 8pt
\font\eighti=mtmi10 at 8pt      \skewchar\eighti='177
\font\eightmib=mtmib10 at 8pt   \skewchar\eightmib='177
\font\eightsy=mtsy10 at 8pt     \skewchar\eightsy='60
\font\eightsyb=mtbsy10 at 8pt   \skewchar\eightsyb='60
\font\eightsl=mtsl10 at 8pt
\font\eighttt=mttt10 at 8pt     \hyphenchar\eighttt=-1
\font\eightcsc=mtcsc10 at 8pt
\font\eightsf=mtss10 at 8pt

\font\ninerm=mtr10 at 9pt
\font\ninebf=mtbx10 at 9pt
\font\nineit=mtti10 at 9pt
\font\ninei=mtmi10 at 9pt      \skewchar\ninei='177
\font\ninemib=mtmib10 at 9pt   \skewchar\ninemib='177
\font\ninesy=mtsy10 at 9pt     \skewchar\ninesy='60
\font\ninesyb=mtbsy10 at 9pt   \skewchar\ninesyb='60
\font\ninesl=mtsl10 at 9pt
\font\ninett=mttt10 at 9pt     \hyphenchar\ninett=-1
\font\ninecsc=mtcsc10 at 9pt
\font\ninesf=mtss10 at 9pt

\font\tenrm=mtr10
\font\tenbf=mtbx10
\font\tenit=mtti10
\font\teni=mtmi10		\skewchar\teni='177
\font\tenmib=mtmib10	\skewchar\tenmib='177
\font\tensy=mtsy10		\skewchar\tensy='60
\font\tensyb=mtbsy10	\skewchar\tensyb='60
\font\tenex=cmex10
\font\tensl=mtsl10
\font\tentt=mttt10		\hyphenchar\tentt=-1
\font\tencsc=mtcsc10
\font\tensf=mtss10

\font\elevenrm=mtr10 at 11pt
\font\elevenbf=mtbx10 at 11pt
\font\elevenit=mtti10 at 11pt
\font\eleveni=mtmi10 at 11pt      \skewchar\eleveni='177
\font\elevenmib=mtmib10 at 11pt   \skewchar\elevenmib='177
\font\elevensy=mtsy10 at 11pt     \skewchar\elevensy='60
\font\elevensyb=mtbsy10 at 11pt   \skewchar\elevensyb='60
\font\elevensl=mtsl10 at 11pt
\font\eleventt=mttt10 at 11pt     \hyphenchar\eleventt=-1
\font\elevencsc=mtcsc10 at 11pt
\font\elevensf=mtss10 at 11pt

\font\twelverm=mtr10 at 12pt
\font\twelvebf=mtbx10 at 12pt
\font\twelveit=mtti10 at 12pt
\font\twelvesl=mtsl10 at 12pt
\font\twelvett=mttt10 at 12pt     \hyphenchar\twelvett=-1
\font\twelvecsc=mtcsc10 at 12pt
\font\twelvesf=mtss10 at 12pt
\font\twelvei=mtmi10 at 12pt      \skewchar\twelvei='177
\font\twelvemib=mtmib10 at 12pt   \skewchar\twelvemib='177
\font\twelvesy=mtsy10 at 12pt     \skewchar\twelvesy='60
\font\twelvesyb=mtbsy10 at 12pt   \skewchar\twelvesyb='60

\font\fourteenrm=mtr10 at 14pt
\font\fourteenbf=mtbx10 at 14pt
\font\fourteenit=mtti10 at 14pt
\font\fourteeni=mtmi10 at 14pt      \skewchar\fourteeni='177
\font\fourteenmib=mtmib10 at 14pt   \skewchar\fourteenmib='177
\font\fourteensy=mtsy10 at 14pt     \skewchar\fourteensy='60
\font\fourteensyb=mtbsy10 at 14pt   \skewchar\fourteensyb='60
\font\fourteensl=mtsl10 at 14pt
\font\fourteentt=mttt10 at 14pt     \hyphenchar\fourteentt=-1
\font\fourteencsc=mtcsc10 at 14pt
\font\fourteensf=mtss10 at 14pt

\font\seventeenrm=mtr10 at 17pt
\font\seventeenbf=mtbx10 at 17pt
\font\seventeenit=mtti10 at 17pt
\font\seventeeni=mtmi10 at 17pt      \skewchar\seventeeni='177
\font\seventeenmib=mtmib10 at 17pt   \skewchar\seventeenmib='177
\font\seventeensy=mtsy10 at 17pt     \skewchar\seventeensy='60
\font\seventeensyb=mtbsy10 at 17pt   \skewchar\seventeensyb='60
\font\seventeensl=mtsl10 at 17pt
\font\seventeentt=mttt10 at 17pt     \hyphenchar\seventeentt=-1
\font\seventeencsc=mtcsc10 at 17pt
\font\seventeensf=mtss10 at 17pt


\newfam\xmfam
\newfam\ymfam

\font\fivexm=mtxm10 at 5pt
\font\sixxm=mtxm10 at 6pt
\font\sevenxm=mtxm10 at 7pt
\font\eightxm=mtxm10 at 8pt
\font\ninexm=mtxm10 at 9pt
\font\tenxm=mtxm10
\font\elevenxm=mtxm10 at 11pt
\font\twelvexm=mtxm10 at 12pt
\font\fourteenxm=mtxm10 at 14pt
\font\seventeenxm=mtxm10 at 17pt

\font\fiveym=mtym10 at 5pt
\font\sixym=mtym10 at 6pt
\font\sevenym=mtym10 at 7pt
\font\eightym=mtym10 at 8pt
\font\nineym=mtym10 at 9pt
\font\tenym=mtym10
\font\elevenym=mtym10 at 11pt
\font\twelveym=mtym10 at 12pt
\font\fourteenym=mtym10 at 14pt
\font\seventeenym=mtym10 at 17pt
\else
\font\fiverm=cmr5
\font\fivei=cmmi5             \skewchar\fivei='177
\font\fivemib=cmmib10 at 5pt  \skewchar\fivemib='177
\font\fivesy=cmsy5            \skewchar\fivesy='60
\font\fivesyb=cmbsy10 at 5pt  \skewchar\fivesyb='60
\font\fivebf=cmbx5

\font\sixrm=cmr6
\font\sixi=cmmi6             \skewchar\sixi='177
\font\sixmib=cmmib10 at 6pt  \skewchar\sixmib='177
\font\sixsy=cmsy6            \skewchar\sixsy='60
\font\sixsyb=cmbsy10 at 6pt  \skewchar\sixsyb='60
\font\sixbf=cmbx6

\font\sevenrm=cmr7
\font\seveni=cmmi7             \skewchar\seveni='177
\font\sevenmib=cmmib10 at 7pt  \skewchar\sevenmib='177
\font\sevensy=cmsy7            \skewchar\sevensy='60
\font\sevensyb=cmbsy10 at 7pt  \skewchar\sevensyb='60
\font\sevenbf=cmbx7

\font\eightrm=cmr8
\font\eightbf=cmbx8
\font\eightit=cmti8
\font\eighti=cmmi8			\skewchar\eighti='177
\font\eightmib=cmmib10 at 8pt	\skewchar\eightmib='177
\font\eightsy=cmsy8			\skewchar\eightsy='60
\font\eightsyb=cmbsy10 at 8pt	\skewchar\eightsyb='60
\font\eightsl=cmsl8
\font\eighttt=cmtt8			\hyphenchar\eighttt=-1
\font\eightcsc=cmcsc10 at 8pt
\font\eightsf=cmss8

\font\ninerm=cmr9
\font\ninebf=cmbx9
\font\nineit=cmti9
\font\ninei=cmmi9			\skewchar\ninei='177
\font\ninemib=cmmib10 at 9pt	\skewchar\ninemib='177
\font\ninesy=cmsy9			\skewchar\ninesy='60
\font\ninesyb=cmbsy10 at 9pt	\skewchar\ninesyb='60
\font\ninesl=cmsl9
\font\ninett=cmtt9			\hyphenchar\ninett=-1
\font\ninecsc=cmcsc10 at 9pt
\font\ninesf=cmss9

\font\tenrm=cmr10
\font\tenbf=cmbx10
\font\tenit=cmti10
\font\teni=cmmi10		\skewchar\teni='177
\font\tenmib=cmmib10	\skewchar\tenmib='177
\font\tensy=cmsy10		\skewchar\tensy='60
\font\tensyb=cmbsy10	\skewchar\tensyb='60
\font\tenex=cmex10
\font\tensl=cmsl10
\font\tentt=cmtt10		\hyphenchar\tentt=-1
\font\tencsc=cmcsc10
\font\tensf=cmss10

\font\elevenrm=cmr10 scaled \magstephalf
\font\elevenbf=cmbx10 scaled \magstephalf
\font\elevenit=cmti10 scaled \magstephalf
\font\eleveni=cmmi10 scaled \magstephalf	\skewchar\eleveni='177
\font\elevenmib=cmmib10 scaled \magstephalf	\skewchar\elevenmib='177
\font\elevensy=cmsy10 scaled \magstephalf	\skewchar\elevensy='60
\font\elevensyb=cmbsy10 scaled \magstephalf	\skewchar\elevensyb='60
\font\elevensl=cmsl10 scaled \magstephalf
\font\eleventt=cmtt10 scaled \magstephalf	\hyphenchar\eleventt=-1
\font\elevencsc=cmcsc10 scaled \magstephalf
\font\elevensf=cmss10 scaled \magstephalf

\font\twelverm=cmr10 scaled \magstep1
\font\twelvebf=cmbx10 scaled \magstep1
\font\twelvei=cmmi10 scaled \magstep1      \skewchar\twelvei='177
\font\twelvemib=cmmib10 scaled \magstep1   \skewchar\twelvemib='177
\font\twelvesy=cmsy10 scaled \magstep1     \skewchar\twelvesy='60
\font\twelvesyb=cmbsy10 scaled \magstep1   \skewchar\twelvesyb='60

\font\fourteenrm=cmr10 scaled \magstep2
\font\fourteenbf=cmbx10 scaled \magstep2
\font\fourteenit=cmti10 scaled \magstep2
\font\fourteeni=cmmi10 scaled \magstep2		\skewchar\fourteeni='177
\font\fourteenmib=cmmib10 scaled \magstep2	\skewchar\fourteenmib='177
\font\fourteensy=cmsy10 scaled \magstep2	\skewchar\fourteensy='60
\font\fourteensyb=cmbsy10 scaled \magstep2	\skewchar\fourteensyb='60
\font\fourteensl=cmsl10 scaled \magstep2
\font\fourteentt=cmtt10 scaled \magstep2	\hyphenchar\fourteentt=-1
\font\fourteencsc=cmcsc10 scaled \magstep2
\font\fourteensf=cmss10 scaled \magstep2

\font\seventeenrm=cmr10 scaled \magstep3
\font\seventeenbf=cmbx10 scaled \magstep3
\font\seventeenit=cmti10 scaled \magstep3
\font\seventeeni=cmmi10 scaled \magstep3	\skewchar\seventeeni='177
\font\seventeenmib=cmmib10 scaled \magstep3	\skewchar\seventeenmib='177
\font\seventeensy=cmsy10 scaled \magstep3	\skewchar\seventeensy='60
\font\seventeensyb=cmbsy10 scaled \magstep3	\skewchar\seventeensyb='60
\font\seventeensl=cmsl10 scaled \magstep3
\font\seventeentt=cmtt10 scaled \magstep3	\hyphenchar\seventeentt=-1
\font\seventeencsc=cmcsc10 scaled \magstep3
\font\seventeensf=cmss10 scaled \magstep3
\fi

\def\hexnumber#1{\ifcase#1 0\or1\or2\or3\or4\or5\or6\or7\or8\or9\or
  A\or B\or C\or D\or E\or F\fi}

\ifprod@font
  \edef\@xm{\hexnumber\xmfam}
  \edef\@ym{\hexnumber\ymfam}
\fi

\def\makestrut{%
  \setbox\strutbox=\hbox{%
    \vrule height.7\baselineskip depth.3\baselineskip width \z@}%
}

\def\baselinestretch{1}
\newskip\tmp@bls

\def\b@ls#1{
  \tmp@bls=#1\relax
  \baselineskip=#1\relax\makestrut
  \normalbaselineskip=\baselinestretch\tmp@bls
  \normalbaselines
}

\def\nostb@ls#1{
  \normalbaselineskip=#1\relax
  \normalbaselines
  \makestrut
}

%

\newfam\mibfam 
\newfam\sybfam 
\newfam\scfam  
\newfam\sffam  

\def\mit{\fam\@ne}

\def\cal{\fam\tw@}

\def\em{\ifdim\fontdimen1\font>\z@ \rm\else\it\fi}

\textfont3=\tenex
\scriptfont3=\tenex
\scriptscriptfont3=\tenex

\setbox0=\hbox{\tenex B} \p@renwd=\wd0 

\def\eightpoint{
  \def\rm{\fam0\eightrm}%
  \textfont0=\eightrm \scriptfont0=\sixrm \scriptscriptfont0=\fiverm%
  \textfont1=\eighti  \scriptfont1=\sixi  \scriptscriptfont1=\fivei%
  \textfont2=\eightsy \scriptfont2=\sixsy \scriptscriptfont2=\fivesy%
  \textfont\itfam=\eightit\def\it{\fam\itfam\eightit}%
  \ifprod@font
    \scriptfont\itfam=\sixit
      \scriptscriptfont\itfam=\fiveit
  \else
    \scriptfont\itfam=\eightit
      \scriptscriptfont\itfam=\eightit
  \fi
  \textfont\bffam=\eightbf%
    \scriptfont\bffam=\sixbf%
      \scriptscriptfont\bffam=\fivebf%
  \def\bf{\fam\bffam\eightbf}%
  \textfont\slfam=\eightsl\def\sl{\fam\slfam\eightsl}%
  \ifprod@font
    \scriptfont\slfam=\sixsl
      \scriptscriptfont\slfam=\fivesl
  \else
    \scriptfont\slfam=\eightsl
      \scriptscriptfont\slfam=\eightsl
  \fi
  \textfont\ttfam=\eighttt\def\tt{\fam\ttfam\eighttt}%
  \ifprod@font
    \scriptfont\ttfam=\sixtt
      \scriptscriptfont\ttfam=\fivett
  \else
    \scriptfont\ttfam=\eighttt
      \scriptscriptfont\ttfam=\eighttt
  \fi
  \textfont\scfam=\eightcsc\def\sc{\fam\scfam\eightcsc}%
  \ifprod@font
    \scriptfont\scfam=\sixcsc
      \scriptscriptfont\scfam=\fivecsc
  \else
    \scriptfont\scfam=\eightcsc
      \scriptscriptfont\scfam=\eightcsc
  \fi
  \textfont\sffam=\eightsf\def\sf{\fam\sffam\eightsf}%
  \ifprod@font
    \scriptfont\sffam=\sixsf
      \scriptscriptfont\sffam=\fivesf
  \else
    \scriptfont\sffam=\eightsf
      \scriptscriptfont\sffam=\eightsf
  \fi
  \textfont\mibfam=\eightmib
    \scriptfont\mibfam=\sixmib
      \scriptscriptfont\mibfam=\fivemib
  \textfont\sybfam=\eightsyb
    \scriptfont\sybfam=\sixsyb
      \scriptscriptfont\sybfam=\fivesyb
  \ifprod@font
    \textfont\xmfam=\eightxm
      \scriptfont\xmfam=\sixxm
        \scriptscriptfont\xmfam=\fivexm
    \textfont\ymfam=\eightym
      \scriptfont\ymfam=\sixym
        \scriptscriptfont\ymfam=\fiveym
  \fi
  \def\oldstyle{\fam\@ne\eighti}%
  \def\boldstyle{\fam\mibfam\eightmib}%
  \b@ls{10pt}\rm%
}

\def\ninepoint{
  \def\rm{\fam0\ninerm}%
  \textfont0=\ninerm \scriptfont0=\sixrm \scriptscriptfont0=\fiverm%
  \textfont1=\ninei  \scriptfont1=\sixi  \scriptscriptfont1=\fivei%
  \textfont2=\ninesy \scriptfont2=\sixsy \scriptscriptfont2=\fivesy%
  \textfont\itfam=\nineit\def\it{\fam\itfam\nineit}%
  \ifprod@font
    \scriptfont\itfam=\sixit
      \scriptscriptfont\itfam=\fiveit
  \else
    \scriptfont\itfam=\nineit
      \scriptscriptfont\itfam=\nineit
  \fi
  \textfont\bffam=\ninebf%
    \scriptfont\bffam=\sixbf%
      \scriptscriptfont\bffam=\fivebf%
  \def\bf{\fam\bffam\ninebf}%
  \textfont\slfam=\ninesl\def\sl{\fam\slfam\ninesl}%
  \ifprod@font
    \scriptfont\slfam=\sixsl
      \scriptscriptfont\slfam=\fivesl
  \else
    \scriptfont\slfam=\ninesl
      \scriptscriptfont\slfam=\ninesl
  \fi
  \textfont\ttfam=\ninett\def\tt{\fam\ttfam\ninett}%
  \ifprod@font
    \scriptfont\ttfam=\sixtt
      \scriptscriptfont\ttfam=\fivett
  \else
    \scriptfont\ttfam=\ninett
      \scriptscriptfont\ttfam=\ninett
  \fi
  \textfont\scfam=\ninecsc\def\sc{\fam\scfam\ninecsc}%
  \ifprod@font
    \scriptfont\scfam=\sixcsc
      \scriptscriptfont\scfam=\fivecsc
  \else
    \scriptfont\scfam=\ninecsc
      \scriptscriptfont\scfam=\ninecsc
  \fi
  \textfont\sffam=\ninesf\def\sf{\fam\sffam\ninesf}%
  \ifprod@font
    \scriptfont\sffam=\sixsf
      \scriptscriptfont\sffam=\fivesf
  \else
    \scriptfont\sffam=\ninesf
      \scriptscriptfont\sffam=\ninesf
  \fi
  \textfont\mibfam=\ninemib
    \scriptfont\mibfam=\sixmib
      \scriptscriptfont\mibfam=\fivemib
  \textfont\sybfam=\ninesyb
    \scriptfont\sybfam=\sixsyb
      \scriptscriptfont\sybfam=\fivesyb
  \ifprod@font
    \textfont\xmfam=\ninexm
      \scriptfont\xmfam=\sixxm
        \scriptscriptfont\xmfam=\fivexm
    \textfont\ymfam=\nineym
      \scriptfont\ymfam=\sixym
        \scriptscriptfont\ymfam=\fiveym
  \fi
  \def\oldstyle{\fam\@ne\ninei}%
  \def\boldstyle{\fam\mibfam\ninemib}%
  \b@ls{\TextLeading plus \Feathering}\rm%
}

\def\tenpoint{
  \def\rm{\fam0\tenrm}%
  \textfont0=\tenrm \scriptfont0=\sevenrm \scriptscriptfont0=\fiverm%
  \textfont1=\teni  \scriptfont1=\seveni  \scriptscriptfont1=\fivei%
  \textfont2=\tensy \scriptfont2=\sevensy \scriptscriptfont2=\fivesy%
  \textfont\itfam=\tenit\def\it{\fam\itfam\tenit}%
  \ifprod@font
    \scriptfont\itfam=\sevenit
      \scriptscriptfont\itfam=\fiveit
  \else
    \scriptfont\itfam=\tenit
      \scriptscriptfont\itfam=\tenit
  \fi
  \textfont\bffam=\tenbf%
    \scriptfont\bffam=\sevenbf%
      \scriptscriptfont\bffam=\fivebf%
  \def\bf{\fam\bffam\tenbf}%
  \textfont\slfam=\tensl\def\sl{\fam\slfam\tensl}%
  \ifprod@font
    \scriptfont\slfam=\sevensl
      \scriptscriptfont\slfam=\fivesl
  \else
    \scriptfont\slfam=\tensl
      \scriptscriptfont\slfam=\tensl
  \fi
  \textfont\ttfam=\tentt\def\tt{\fam\ttfam\tentt}%
  \ifprod@font
    \scriptfont\ttfam=\seventt
      \scriptscriptfont\ttfam=\fivett
  \else
    \scriptfont\ttfam=\tentt
      \scriptscriptfont\ttfam=\tentt
  \fi
  \textfont\scfam=\tencsc\def\sc{\fam\scfam\tencsc}%
  \ifprod@font
    \scriptfont\scfam=\sevencsc
      \scriptscriptfont\scfam=\fivecsc
  \else
    \scriptfont\scfam=\tencsc
      \scriptscriptfont\scfam=\tencsc
  \fi
  \textfont\sffam=\tensf\def\sf{\fam\sffam\tensf}%
  \ifprod@font
    \scriptfont\sffam=\sevensf
      \scriptscriptfont\sffam=\fivesf
  \else
    \scriptfont\sffam=\tensf
      \scriptscriptfont\sffam=\tensf
  \fi
  \textfont\mibfam=\tenmib
    \scriptfont\mibfam=\sevenmib
      \scriptscriptfont\mibfam=\fivemib
  \textfont\sybfam=\tensyb
    \scriptfont\sybfam=\sevensyb
      \scriptscriptfont\sybfam=\fivesyb
  \ifprod@font
    \textfont\xmfam=\tenxm
      \scriptfont\xmfam=\sevenxm
        \scriptscriptfont\xmfam=\fivexm
    \textfont\ymfam=\tenym
      \scriptfont\ymfam=\sevenym
        \scriptscriptfont\ymfam=\fiveym
  \fi
  \def\oldstyle{\fam\@ne\teni}%
  \def\boldstyle{\fam\mibfam\tenmib}%
  \b@ls{11pt}\rm%
}

\def\elevenpoint{
  \def\rm{\fam0\elevenrm}%
  \textfont0=\elevenrm \scriptfont0=\eightrm \scriptscriptfont0=\sixrm%
  \textfont1=\eleveni  \scriptfont1=\eighti  \scriptscriptfont1=\sixi%
  \textfont2=\elevensy \scriptfont2=\eightsy \scriptscriptfont2=\sixsy%
  \textfont\itfam=\elevenit\def\it{\fam\itfam\elevenit}%
  \ifprod@font
    \scriptfont\itfam=\eightit
      \scriptscriptfont\itfam=\sixit
  \else
    \scriptfont\itfam=\elevenit
      \scriptscriptfont\itfam=\elevenit
  \fi
  \textfont\bffam=\elevenbf%
    \scriptfont\bffam=\eightbf%
      \scriptscriptfont\bffam=\sixbf%
  \def\bf{\fam\bffam\elevenbf}%
  \textfont\slfam=\elevensl\def\sl{\fam\slfam\elevensl}%
  \ifprod@font
    \scriptfont\slfam=\eightsl
      \scriptscriptfont\slfam=\sixsl
  \else
    \scriptfont\slfam=\elevensl
      \scriptscriptfont\slfam=\elevensl
  \fi
  \textfont\ttfam=\eleventt\def\tt{\fam\ttfam\eleventt}%
  \ifprod@font
    \scriptfont\ttfam=\eighttt
      \scriptscriptfont\ttfam=\sixtt
  \else
    \scriptfont\ttfam=\eleventt
      \scriptscriptfont\ttfam=\eleventt
  \fi
  \textfont\scfam=\elevencsc\def\sc{\fam\scfam\elevencsc}%
  \ifprod@font
    \scriptfont\scfam=\eightcsc
      \scriptscriptfont\scfam=\sixcsc
  \else
    \scriptfont\scfam=\elevencsc
      \scriptscriptfont\scfam=\elevencsc
  \fi
  \textfont\sffam=\elevensf\def\sf{\fam\sffam\elevensf}%
  \ifprod@font
    \scriptfont\sffam=\eightsf
      \scriptscriptfont\sffam=\sixsf
  \else
    \scriptfont\sffam=\elevensf
      \scriptscriptfont\sffam=\elevensf
  \fi
  \textfont\mibfam=\elevenmib
    \scriptfont\mibfam=\eightmib
      \scriptscriptfont\mibfam=\sixmib
  \textfont\sybfam=\elevensyb
    \scriptfont\sybfam=\eightsyb
      \scriptscriptfont\sybfam=\sixsyb
  \ifprod@font
    \textfont\xmfam=\elevenxm
      \scriptfont\xmfam=\eightxm
       \scriptscriptfont\xmfam=\sixxm
    \textfont\ymfam=\elevenym
      \scriptfont\ymfam=\eightym
        \scriptscriptfont\ymfam=\sixym
   \fi
  \def\oldstyle{\fam\@ne\eleveni}%
  \def\boldstyle{\fam\mibfam\elevenmib}%
  \b@ls{13pt}\rm%
}

\def\fourteenpoint{
  \def\rm{\fam0\fourteenrm}%
  \textfont0\fourteenrm  \scriptfont0\tenrm  \scriptscriptfont0\sevenrm%
  \textfont1\fourteeni   \scriptfont1\teni   \scriptscriptfont1\seveni%
  \textfont2\fourteensy  \scriptfont2\tensy  \scriptscriptfont2\sevensy%
  \textfont\itfam=\fourteenit\def\it{\fam\itfam\fourteenit}%
  \ifprod@font
    \scriptfont\itfam=\tenit
      \scriptscriptfont\itfam=\sevenit
  \else
    \scriptfont\itfam=\fourteenit
      \scriptscriptfont\itfam=\fourteenit
  \fi
  \textfont\bffam=\fourteenbf%
    \scriptfont\bffam=\tenbf%
      \scriptscriptfont\bffam=\sevenbf%
  \def\bf{\fam\bffam\fourteenbf}%
  \textfont\slfam=\fourteensl\def\sl{\fam\slfam\fourteensl}%
  \ifprod@font
    \scriptfont\slfam=\tensl
      \scriptscriptfont\slfam=\sevensl
  \else
    \scriptfont\slfam=\fourteensl
      \scriptscriptfont\slfam=\fourteensl
  \fi
  \textfont\ttfam=\fourteentt\def\tt{\fam\ttfam\fourteentt}%
  \ifprod@font
    \scriptfont\ttfam=\tentt
      \scriptscriptfont\ttfam=\seventt
  \else
    \scriptfont\ttfam=\fourteentt
      \scriptscriptfont\ttfam=\fourteentt
  \fi
  \textfont\scfam=\fourteencsc\def\sc{\fam\scfam\fourteencsc}%
  \ifprod@font
    \scriptfont\scfam=\tencsc
      \scriptscriptfont\scfam=\sevencsc
  \else
    \scriptfont\scfam=\fourteencsc
      \scriptscriptfont\scfam=\fourteencsc
  \fi
  \textfont\sffam=\fourteensf\def\sf{\fam\sffam\fourteensf}%
  \ifprod@font
    \scriptfont\sffam=\tensf
      \scriptscriptfont\sffam=\sevensf
  \else
    \scriptfont\sffam=\fourteensf
      \scriptscriptfont\sffam=\fourteensf
  \fi
  \textfont\mibfam=\fourteenmib
    \scriptfont\mibfam=\tenmib
      \scriptscriptfont\mibfam=\sevenmib
  \textfont\sybfam=\fourteensyb
    \scriptfont\sybfam=\tensyb
      \scriptscriptfont\sybfam=\sevensyb
  \ifprod@font
    \textfont\xmfam=\fourteenxm
      \scriptfont\xmfam=\tenxm
        \scriptscriptfont\xmfam=\sevenxm
   \textfont\ymfam=\fourteenym
      \scriptfont\ymfam=\tenym
        \scriptscriptfont\ymfam=\sevenym
  \fi
  \def\oldstyle{\fam\@ne\fourteeni}%
  \def\boldstyle{\fam\mibfam\fourteenmib}%
  \b@ls{17pt}\rm%
}

\def\seventeenpoint{
  \def\rm{\fam0\seventeenrm}%
  \textfont0\seventeenrm  \scriptfont0\twelverm  \scriptscriptfont0\tenrm%
  \textfont1\seventeeni   \scriptfont1\twelvei   \scriptscriptfont1\teni%
  \textfont2\seventeensy  \scriptfont2\twelvesy  \scriptscriptfont2\tensy%
  \textfont\itfam=\seventeenit\def\it{\fam\itfam\seventeenit}%
  \ifprod@font
    \scriptfont\itfam=\twelveit
      \scriptscriptfont\itfam=\tenit
  \else
    \scriptfont\itfam=\seventeenit
      \scriptscriptfont\itfam=\seventeenit
  \fi
  \textfont\bffam=\seventeenbf%
    \scriptfont\bffam=\twelvebf%
      \scriptscriptfont\bffam=\tenbf%
  \def\bf{\fam\bffam\seventeenbf}%
  \textfont\slfam=\seventeensl\def\sl{\fam\slfam\seventeensl}%
  \ifprod@font
    \scriptfont\slfam=\twelvesl
      \scriptscriptfont\slfam=\tensl
  \else
    \scriptfont\slfam=\seventeensl
      \scriptscriptfont\slfam=\seventeensl
  \fi
  \textfont\ttfam=\seventeentt\def\tt{\fam\ttfam\seventeentt}%
  \ifprod@font
    \scriptfont\ttfam=\twelvett
      \scriptscriptfont\ttfam=\tentt
  \else
    \scriptfont\ttfam=\seventeentt
      \scriptscriptfont\ttfam=\seventeentt
  \fi
  \textfont\scfam=\seventeencsc\def\sc{\fam\scfam\seventeencsc}%
  \ifprod@font
    \scriptfont\scfam=\twelvecsc
      \scriptscriptfont\scfam=\tencsc
  \else
    \scriptfont\scfam=\seventeencsc
      \scriptscriptfont\scfam=\seventeencsc
  \fi
  \textfont\sffam=\seventeensf\def\sf{\fam\sffam\seventeensf}%
  \ifprod@font
    \scriptfont\sffam=\twelvesf
      \scriptscriptfont\sffam=\tensf
  \else
    \scriptfont\sffam=\seventeensf
      \scriptscriptfont\sffam=\seventeensf
  \fi
  \textfont\mibfam=\seventeenmib
    \scriptfont\mibfam=\twelvemib
      \scriptscriptfont\mibfam=\tenmib
  \textfont\sybfam=\seventeensyb
    \scriptfont\sybfam=\twelvesyb
      \scriptscriptfont\sybfam=\tensyb
  \ifprod@font
    \textfont\xmfam=\seventeenxm
      \scriptfont\xmfam=\twelvexm
        \scriptscriptfont\xmfam=\tenxm
    \textfont\ymfam=\seventeenym
      \scriptfont\ymfam=\twelveym
        \scriptscriptfont\ymfam=\tenym
  \fi
  \def\oldstyle{\fam\@ne\seventeeni}%
  \def\boldstyle{\fam\mibfam\seventeenmib}%
  \b@ls{20pt}\rm%
}

\lineskip=1pt      \normallineskip=\lineskip
\lineskiplimit=\z@ \normallineskiplimit=\lineskiplimit



\def\la{\mathrel{\mathchoice {\vcenter{\offinterlineskip\halign{\hfil
$\displaystyle##$\hfil\cr<\cr\sim\cr}}}
{\vcenter{\offinterlineskip\halign{\hfil$\textstyle##$\hfil\cr
<\cr\sim\cr}}}
{\vcenter{\offinterlineskip\halign{\hfil$\scriptstyle##$\hfil\cr
<\cr\sim\cr}}}
{\vcenter{\offinterlineskip\halign{\hfil$\scriptscriptstyle##$\hfil\cr
<\cr\sim\cr}}}}}

\def\ga{\mathrel{\mathchoice {\vcenter{\offinterlineskip\halign{\hfil
$\displaystyle##$\hfil\cr>\cr\sim\cr}}}
{\vcenter{\offinterlineskip\halign{\hfil$\textstyle##$\hfil\cr
>\cr\sim\cr}}}
{\vcenter{\offinterlineskip\halign{\hfil$\scriptstyle##$\hfil\cr
>\cr\sim\cr}}}
{\vcenter{\offinterlineskip\halign{\hfil$\scriptscriptstyle##$\hfil\cr
>\cr\sim\cr}}}}}

\def\getsto{\mathrel{\mathchoice {\vcenter{\offinterlineskip
\halign{\hfil
$\displaystyle##$\hfil\cr\gets\cr\to\cr}}}
{\vcenter{\offinterlineskip\halign{\hfil$\textstyle##$\hfil\cr\gets
\cr\to\cr}}}
{\vcenter{\offinterlineskip\halign{\hfil$\scriptstyle##$\hfil\cr\gets
\cr\to\cr}}}
{\vcenter{\offinterlineskip\halign{\hfil$\scriptscriptstyle##$\hfil\cr
\gets\cr\to\cr}}}}}

\def\lid{\mathrel{\mathchoice {\vcenter{\offinterlineskip\halign{\hfil
$\displaystyle##$\hfil\cr<\cr\noalign{\vskip1.2pt}=\cr}}}
{\vcenter{\offinterlineskip\halign{\hfil$\textstyle##$\hfil\cr<\cr
\noalign{\vskip1.2pt}=\cr}}}
{\vcenter{\offinterlineskip\halign{\hfil$\scriptstyle##$\hfil\cr<\cr
\noalign{\vskip1pt}=\cr}}}
{\vcenter{\offinterlineskip\halign{\hfil$\scriptscriptstyle##$\hfil\cr
<\cr
\noalign{\vskip0.9pt}=\cr}}}}}

\def\gid{\mathrel{\mathchoice {\vcenter{\offinterlineskip\halign{\hfil
$\displaystyle##$\hfil\cr>\cr\noalign{\vskip1.2pt}=\cr}}}
{\vcenter{\offinterlineskip\halign{\hfil$\textstyle##$\hfil\cr>\cr
\noalign{\vskip1.2pt}=\cr}}}
{\vcenter{\offinterlineskip\halign{\hfil$\scriptstyle##$\hfil\cr>\cr
\noalign{\vskip1pt}=\cr}}}
{\vcenter{\offinterlineskip\halign{\hfil$\scriptscriptstyle##$\hfil\cr
>\cr
\noalign{\vskip0.9pt}=\cr}}}}}

\def\grole{\mathrel{\mathchoice {\vcenter{\offinterlineskip\halign{\hfil
$\displaystyle##$\hfil\cr>\cr\noalign{\vskip-1.5pt}<\cr}}}
{\vcenter{\offinterlineskip\halign{\hfil$\textstyle##$\hfil\cr
>\cr\noalign{\vskip-1.5pt}<\cr}}}
{\vcenter{\offinterlineskip\halign{\hfil$\scriptstyle##$\hfil\cr
>\cr\noalign{\vskip-1pt}<\cr}}}
{\vcenter{\offinterlineskip\halign{\hfil$\scriptscriptstyle##$\hfil\cr
>\cr\noalign{\vskip-0.5pt}<\cr}}}}}

\def\leogr{\mathrel{\mathchoice {\vcenter{\offinterlineskip\halign{\hfil
$\displaystyle##$\hfil\cr<\cr\noalign{\vskip-1.5pt}>\cr}}}
{\vcenter{\offinterlineskip\halign{\hfil$\textstyle##$\hfil\cr
<\cr\noalign{\vskip-1.5pt}>\cr}}}
{\vcenter{\offinterlineskip\halign{\hfil$\scriptstyle##$\hfil\cr
<\cr\noalign{\vskip-1pt}>\cr}}}
{\vcenter{\offinterlineskip\halign{\hfil$\scriptscriptstyle##$\hfil\cr
<\cr\noalign{\vskip-0.5pt}>\cr}}}}}

\def\loa{\mathrel{\mathchoice {\vcenter{\offinterlineskip\halign{\hfil
$\displaystyle##$\hfil\cr<\cr\approx\cr}}}
{\vcenter{\offinterlineskip\halign{\hfil$\textstyle##$\hfil\cr
<\cr\approx\cr}}}
{\vcenter{\offinterlineskip\halign{\hfil$\scriptstyle##$\hfil\cr
<\cr\approx\cr}}}
{\vcenter{\offinterlineskip\halign{\hfil$\scriptscriptstyle##$\hfil\cr
<\cr\approx\cr}}}}}

\def\goa{\mathrel{\mathchoice {\vcenter{\offinterlineskip\halign{\hfil
$\displaystyle##$\hfil\cr>\cr\approx\cr}}}
{\vcenter{\offinterlineskip\halign{\hfil$\textstyle##$\hfil\cr
>\cr\approx\cr}}}
{\vcenter{\offinterlineskip\halign{\hfil$\scriptstyle##$\hfil\cr
>\cr\approx\cr}}}
{\vcenter{\offinterlineskip\halign{\hfil$\scriptscriptstyle##$\hfil\cr
>\cr\approx\cr}}}}}

\def\diameter{{\ifmmode\mathchoice
{\ooalign{\hfil\hbox{$\displaystyle/$}\hfil\crcr
{\hbox{$\displaystyle\mathchar"20D$}}}}
{\ooalign{\hfil\hbox{$\textstyle/$}\hfil\crcr
{\hbox{$\textstyle\mathchar"20D$}}}}
{\ooalign{\hfil\hbox{$\scriptstyle/$}\hfil\crcr
{\hbox{$\scriptstyle\mathchar"20D$}}}}
{\ooalign{\hfil\hbox{$\scriptscriptstyle/$}\hfil\crcr
{\hbox{$\scriptscriptstyle\mathchar"20D$}}}}
\else{\ooalign{\hfil/\hfil\crcr\mathhexbox20D}}%
\fi}}

\def\sq{\ifmmode\squareforqed\else{\unskip\nobreak\hfil
\penalty50\hskip1em\null\nobreak\hfil\squareforqed
\parfillskip=0pt\finalhyphendemerits=0\endgraf}\fi}
\def\squareforqed{\hbox{\rlap{$\sqcap$}$\sqcup$}}


\def\bbbc{{\mathchoice {\setbox0=\hbox{$\displaystyle\rm C$}\hbox{\hbox
to0pt{\kern0.4\wd0\vrule height0.9\ht0\hss}\box0}}
{\setbox0=\hbox{$\textstyle\rm C$}\hbox{\hbox
to0pt{\kern0.4\wd0\vrule height0.9\ht0\hss}\box0}}
{\setbox0=\hbox{$\scriptstyle\rm C$}\hbox{\hbox
to0pt{\kern0.4\wd0\vrule height0.9\ht0\hss}\box0}}
{\setbox0=\hbox{$\scriptscriptstyle\rm C$}\hbox{\hbox
to0pt{\kern0.4\wd0\vrule height0.9\ht0\hss}\box0}}}}
\def\bbbq{{\mathchoice {\setbox0=\hbox{$\displaystyle\rm
Q$}\hbox{\raise
0.15\ht0\hbox to0pt{\kern0.4\wd0\vrule height0.8\ht0\hss}\box0}}
{\setbox0=\hbox{$\textstyle\rm Q$}\hbox{\raise
0.15\ht0\hbox to0pt{\kern0.4\wd0\vrule height0.8\ht0\hss}\box0}}
{\setbox0=\hbox{$\scriptstyle\rm Q$}\hbox{\raise
0.15\ht0\hbox to0pt{\kern0.4\wd0\vrule height0.7\ht0\hss}\box0}}
{\setbox0=\hbox{$\scriptscriptstyle\rm Q$}\hbox{\raise
0.15\ht0\hbox to0pt{\kern0.4\wd0\vrule height0.7\ht0\hss}\box0}}}}
\def\bbbt{{\mathchoice {\setbox0=\hbox{$\displaystyle\rm
T$}\hbox{\hbox to0pt{\kern0.3\wd0\vrule height0.9\ht0\hss}\box0}}
{\setbox0=\hbox{$\textstyle\rm T$}\hbox{\hbox
to0pt{\kern0.3\wd0\vrule height0.9\ht0\hss}\box0}}
{\setbox0=\hbox{$\scriptstyle\rm T$}\hbox{\hbox
to0pt{\kern0.3\wd0\vrule height0.9\ht0\hss}\box0}}
{\setbox0=\hbox{$\scriptscriptstyle\rm T$}\hbox{\hbox
to0pt{\kern0.3\wd0\vrule height0.9\ht0\hss}\box0}}}}
\def\bbbs{{\mathchoice
{\setbox0=\hbox{$\displaystyle     \rm S$}\hbox{\raise0.5\ht0\hbox
to0pt{\kern0.35\wd0\vrule height0.45\ht0\hss}\hbox
to0pt{\kern0.55\wd0\vrule height0.5\ht0\hss}\box0}}
{\setbox0=\hbox{$\textstyle        \rm S$}\hbox{\raise0.5\ht0\hbox
to0pt{\kern0.35\wd0\vrule height0.45\ht0\hss}\hbox
to0pt{\kern0.55\wd0\vrule height0.5\ht0\hss}\box0}}
{\setbox0=\hbox{$\scriptstyle      \rm S$}\hbox{\raise0.5\ht0\hbox
to0pt{\kern0.35\wd0\vrule height0.45\ht0\hss}\raise0.05\ht0\hbox
to0pt{\kern0.5\wd0\vrule height0.45\ht0\hss}\box0}}
{\setbox0=\hbox{$\scriptscriptstyle\rm S$}\hbox{\raise0.5\ht0\hbox
to0pt{\kern0.4\wd0\vrule height0.45\ht0\hss}\raise0.05\ht0\hbox
to0pt{\kern0.55\wd0\vrule height0.45\ht0\hss}\box0}}}}
\def\bbbz{{\mathchoice {\hbox{$\sf\textstyle Z\kern-0.4em Z$}}
{\hbox{$\sf\textstyle Z\kern-0.4em Z$}}
{\hbox{$\sf\scriptstyle Z\kern-0.3em Z$}}
{\hbox{$\sf\scriptscriptstyle Z\kern-0.2em Z$}}}}


\ifprod@font
  \mathchardef\la="3\@xm2E
  \mathchardef\getsto="3\@xm1C
  \mathchardef\lid="3\@xm35
  \mathchardef\grole="3\@xm3F
  \mathchardef\loa="3\@xm2F
  \mathchardef\ga="3\@xm26
  \mathchardef\gid="3\@xm3D
  \mathchardef\leogr="3\@xm37
  \mathchardef\goa="3\@xm27
  \mathchardef\sq="0\@xm03
%
%
\def\diameter{{%
  \ifmmode
    \mathchoice
    {\ooalign{\hfil\hbox{$\displaystyle/$}\hfil\crcr
    {\lower.2ex\hbox{$\displaystyle\mathchar"20D$}}}}%
    {\ooalign{\hfil\hbox{$\textstyle/$}\hfil\crcr
    {\lower.2ex\hbox{$\textstyle\mathchar"20D$}}}}%
    {\ooalign{\hfil\hbox{$\scriptstyle/$}\hfil\crcr
    {\lower.1ex\hbox{$\scriptstyle\mathchar"20D$}}}}%
    {\ooalign{\hfil\hbox{$\scriptscriptstyle/$}\hfil\crcr
    {\lower.1ex\hbox{$\scriptscriptstyle\mathchar"20D$}}}}%
  \else
    {\ooalign{\hfil/\hfil\crcr\lower.2ex\hbox{\mathhexbox20D}}}%
  \fi
}}
%
%

\def\bbbc{{\Bbb{C}}}
\def\bbbq{{\Bbb{Q}}}
\def\bbbt{{\Bbb{T}}}
\def\bbbs{{\Bbb{S}}}
\def\bbbz{{\Bbb{Z}}}
\fi


\ifprod@font
\mathchardef\boxdot="2\@xm00
\mathchardef\boxplus="2\@xm01
\mathchardef\boxtimes="2\@xm02
\mathchardef\square="0\@xm03
\mathchardef\blacksquare="0\@xm04
\mathchardef\centerdot="2\@xm05
\mathchardef\lozenge="0\@xm06
\mathchardef\blacklozenge="0\@xm07
\mathchardef\circlearrowright="3\@xm08
\mathchardef\circlearrowleft="3\@xm09
\mathchardef\rightleftharpoons="3\@xm0A
\mathchardef\leftrightharpoons="3\@xm0B
\mathchardef\boxminus="2\@xm0C
\mathchardef\Vdash="3\@xm0D
\mathchardef\Vvdash="3\@xm0E
\mathchardef\vDash="3\@xm0F
\mathchardef\twoheadrightarrow="3\@xm10
\mathchardef\twoheadleftarrow="3\@xm11
\mathchardef\leftleftarrows="3\@xm12
\mathchardef\rightrightarrows="3\@xm13
\mathchardef\upuparrows="3\@xm14
\mathchardef\downdownarrows="3\@xm15
\mathchardef\upharpoonright="3\@xm16

\mathchardef\downharpoonright="3\@xm17
\mathchardef\upharpoonleft="3\@xm18
\mathchardef\downharpoonleft="3\@xm19
\mathchardef\rightarrowtail="3\@xm1A
\mathchardef\leftarrowtail="3\@xm1B
\mathchardef\leftrightarrows="3\@xm1C
\mathchardef\rightleftarrows="3\@xm1D
\mathchardef\Lsh="3\@xm1E
\mathchardef\Rsh="3\@xm1F
\mathchardef\rightsquigarrow="3\@xm20
\mathchardef\leftrightsquigarrow="3\@xm21
\mathchardef\looparrowleft="3\@xm22
\mathchardef\looparrowright="3\@xm23
\mathchardef\circeq="3\@xm24
\mathchardef\succsim="3\@xm25
\mathchardef\gtrsim="3\@xm26
\mathchardef\gtrapprox="3\@xm27
\mathchardef\multimap="3\@xm28
\mathchardef\therefore="3\@xm29
\mathchardef\because="3\@xm2A
\mathchardef\doteqdot="3\@xm2B

\mathchardef\triangleq="3\@xm2C
\mathchardef\precsim="3\@xm2D
\mathchardef\lesssim="3\@xm2E
\mathchardef\lessapprox="3\@xm2F
\mathchardef\eqslantless="3\@xm30
\mathchardef\eqslantgtr="3\@xm31
\mathchardef\curlyeqprec="3\@xm32
\mathchardef\curlyeqsucc="3\@xm33
\mathchardef\preccurlyeq="3\@xm34
\mathchardef\leqq="3\@xm35
\mathchardef\leqslant="3\@xm36
\mathchardef\lessgtr="3\@xm37
\mathchardef\backprime="0\@xm38
\mathchardef\risingdotseq="3\@xm3A
\mathchardef\fallingdotseq="3\@xm3B
\mathchardef\succcurlyeq="3\@xm3C
\mathchardef\geqq="3\@xm3D
\mathchardef\geqslant="3\@xm3E
\mathchardef\gtrless="3\@xm3F
\mathchardef\sqsubset="3\@xm40
\mathchardef\sqsupset="3\@xm41
\mathchardef\vartriangleright="3\@xm42
\mathchardef\vartriangleleft="3\@xm43
\mathchardef\trianglerighteq="3\@xm44
\mathchardef\trianglelefteq="3\@xm45
\mathchardef\bigstar="0\@xm46
\mathchardef\between="3\@xm47
\mathchardef\blacktriangledown="0\@xm48
\mathchardef\blacktriangleright="3\@xm49
\mathchardef\blacktriangleleft="3\@xm4A
\mathchardef\vartriangle="0\@xm4D
\mathchardef\blacktriangle="0\@xm4E
\mathchardef\triangledown="0\@xm4F
\mathchardef\eqcirc="3\@xm50
\mathchardef\lesseqgtr="3\@xm51
\mathchardef\gtreqless="3\@xm52
\mathchardef\lesseqqgtr="3\@xm53
\mathchardef\gtreqqless="3\@xm54
\mathchardef\Rrightarrow="3\@xm56
\mathchardef\Lleftarrow="3\@xm57
\mathchardef\veebar="2\@xm59
\mathchardef\barwedge="2\@xm5A
\mathchardef\doublebarwedge="2\@xm5B
\mathchardef\angle="0\@xm5C
\mathchardef\measuredangle="0\@xm5D
\mathchardef\sphericalangle="0\@xm5E
\mathchardef\varpropto="3\@xm5F
\mathchardef\smallsmile="3\@xm60
\mathchardef\smallfrown="3\@xm61
\mathchardef\Subset="3\@xm62
\mathchardef\Supset="3\@xm63
\mathchardef\Cup="2\@xm64

\mathchardef\Cap="2\@xm65

\mathchardef\curlywedge="2\@xm66
\mathchardef\curlyvee="2\@xm67
\mathchardef\leftthreetimes="2\@xm68
\mathchardef\rightthreetimes="2\@xm69
\mathchardef\subseteqq="3\@xm6A
\mathchardef\supseteqq="3\@xm6B
\mathchardef\bumpeq="3\@xm6C
\mathchardef\Bumpeq="3\@xm6D
\mathchardef\lll="3\@xm6E

\mathchardef\ggg="3\@xm6F

\mathchardef\circledS="0\@xm73
\mathchardef\pitchfork="3\@xm74
\mathchardef\dotplus="2\@xm75
\mathchardef\backsim="3\@xm76
\mathchardef\backsimeq="3\@xm77
\mathchardef\complement="0\@xm7B
\mathchardef\intercal="2\@xm7C
\mathchardef\circledcirc="2\@xm7D
\mathchardef\circledast="2\@xm7E
\mathchardef\circleddash="2\@xm7F
\def\ulcorner{\delimiter"4\@xm70\@xm70 }
\def\urcorner{\delimiter"5\@xm71\@xm71 }
\def\llcorner{\delimiter"4\@xm78\@xm78 }
\def\lrcorner{\delimiter"5\@xm79\@xm79 }
\def\yen{\mathhexbox\@xm55 }
\def\checkmark{\mathhexbox\@xm58 }
\def\circledR{\mathhexbox\@xm72 }
\def\maltese{\mathhexbox\@xm7A }
\mathchardef\lvertneqq="3\@ym00
\mathchardef\gvertneqq="3\@ym01
\mathchardef\nleq="3\@ym02
\mathchardef\ngeq="3\@ym03
\mathchardef\nless="3\@ym04
\mathchardef\ngtr="3\@ym05
\mathchardef\nprec="3\@ym06
\mathchardef\nsucc="3\@ym07
\mathchardef\lneqq="3\@ym08
\mathchardef\gneqq="3\@ym09
\mathchardef\nleqslant="3\@ym0A
\mathchardef\ngeqslant="3\@ym0B
\mathchardef\lneq="3\@ym0C
\mathchardef\gneq="3\@ym0D
\mathchardef\npreceq="3\@ym0E
\mathchardef\nsucceq="3\@ym0F
\mathchardef\precnsim="3\@ym10
\mathchardef\succnsim="3\@ym11
\mathchardef\lnsim="3\@ym12
\mathchardef\gnsim="3\@ym13
\mathchardef\nleqq="3\@ym14
\mathchardef\ngeqq="3\@ym15
\mathchardef\precneqq="3\@ym16
\mathchardef\succneqq="3\@ym17
\mathchardef\precnapprox="3\@ym18
\mathchardef\succnapprox="3\@ym19
\mathchardef\lnapprox="3\@ym1A
\mathchardef\gnapprox="3\@ym1B
\mathchardef\nsim="3\@ym1C
\mathchardef\ncong="3\@ym1D

\mathchardef\varsubsetneq="3\@ym20
\mathchardef\varsupsetneq="3\@ym21
\mathchardef\nsubseteqq="3\@ym22
\mathchardef\nsupseteqq="3\@ym23
\mathchardef\subsetneqq="3\@ym24
\mathchardef\supsetneqq="3\@ym25
\mathchardef\varsubsetneqq="3\@ym26
\mathchardef\varsupsetneqq="3\@ym27
\mathchardef\subsetneq="3\@ym28
\mathchardef\supsetneq="3\@ym29
\mathchardef\nsubseteq="3\@ym2A
\mathchardef\nsupseteq="3\@ym2B
\mathchardef\nparallel="3\@ym2C
\mathchardef\nmid="3\@ym2D
\mathchardef\nshortmid="3\@ym2E
\mathchardef\nshortparallel="3\@ym2F
\mathchardef\nvdash="3\@ym30
\mathchardef\nVdash="3\@ym31
\mathchardef\nvDash="3\@ym32
\mathchardef\nVDash="3\@ym33
\mathchardef\ntrianglerighteq="3\@ym34
\mathchardef\ntrianglelefteq="3\@ym35
\mathchardef\ntriangleleft="3\@ym36
\mathchardef\ntriangleright="3\@ym37
\mathchardef\nleftarrow="3\@ym38
\mathchardef\nrightarrow="3\@ym39
\mathchardef\nLeftarrow="3\@ym3A
\mathchardef\nRightarrow="3\@ym3B
\mathchardef\nLeftrightarrow="3\@ym3C
\mathchardef\nleftrightarrow="3\@ym3D
\mathchardef\divideontimes="2\@ym3E
\mathchardef\varnothing="0\@ym3F
\mathchardef\nexists="0\@ym40
\mathchardef\mho="0\@ym66
\mathchardef\eth="0\@ym67
\mathchardef\eqsim="3\@ym68
\mathchardef\beth="0\@ym69
\mathchardef\gimel="0\@ym6A
\mathchardef\daleth="0\@ym6B
\mathchardef\lessdot="3\@ym6C
\mathchardef\gtrdot="3\@ym6D
\mathchardef\ltimes="2\@ym6E
\mathchardef\rtimes="2\@ym6F
\mathchardef\shortmid="3\@ym70
\mathchardef\shortparallel="3\@ym71
\mathchardef\smallsetminus="2\@ym72
\mathchardef\thicksim="3\@ym73
\mathchardef\thickapprox="3\@ym74
\mathchardef\approxeq="3\@ym75
\mathchardef\succapprox="3\@ym76
\mathchardef\precapprox="3\@ym77
\mathchardef\curvearrowleft="3\@ym78
\mathchardef\curvearrowright="3\@ym79
\mathchardef\digamma="0\@ym7A
\mathchardef\varkappa="0\@ym7B
\mathchardef\hslash="0\@ym7D
\mathchardef\hbar="0\@ym7E
\mathchardef\backepsilon="3\@ym7F


\def\Bbb{\ifmmode\let\next\Bbb@\else
\def\next{\errmessage{Use \string\Bbb\space only in math mode}}\fi\next}
\def\Bbb@#1{{\Bbb@@{#1}}}
\def\Bbb@@#1{\fam\ymfam#1}
\fi


\def\Nulle{0} 
\def\Afe{1}   
\def\Hae{2}   
\def\Hbe{3}   
\def\Hce{4}   
\def\Hde{5}   


\newcount\LastMac       \LastMac=\Nulle

\newskip\half      \half=5.5pt plus 1.5pt minus 2.25pt
\newskip\one       \one=11pt plus 3pt minus 5.5pt
\newskip\onehalf   \onehalf=16.5pt plus 5.5pt minus 8.25pt
\newskip\two       \two=22pt plus 5.5pt minus 11pt

\def\Half{\addvspace{\half}}
\def\One{\addvspace{\one}}
\def\OneHalf{\addvspace{\onehalf}}
\def\Two{\addvspace{\two}}

\def\hdskip{\hskip 1pc\relax} 

\def\Raggedright{
  \rightskip=\z@ plus \hsize\relax
}

\def\Fullout{
  \rightskip=\z@\relax
}

\def\Hang#1#2{
  \hangindent=#1%
  \hangafter=#2\relax
}


\newif\ifsp@page
\def\pagestyle#1{\csname ps@#1\endcsname}
\def\thispagestyle#1{\global\sp@pagetrue\gdef\sp@type{#1}}

\def\ps@titlepage{%
  \def\@oddhead{\eightpoint\noindent \the\CatchLine
    \ifprod@font\else\qquad Printed\ \today\fi \hfil}%
  \let\@evenhead=\@oddhead
}

\def\ps@headings{%
  \def\@oddhead{\elevenpoint\it\noindent
    \hfill\the\RightHeader\hskip1.5em\rm\folio}%
  \def\@evenhead{\elevenpoint\noindent
    \folio\hskip1.5em\it\the\LeftHeader\hfill}%
}

\def\ps@plate{%
  \def\@oddhead{\eightpoint\noindent\plt@cap\hfil}%
  \def\@evenhead{\eightpoint\noindent\plt@cap\hfil}%
}



\def\title#1{
  \bgroup
    \vbox to 8pt{\vss}%
    \seventeenpoint
    \Raggedright
    \noindent \strut{\bf #1}\par
  \egroup
}

\def\author#1{
  \bgroup
    \ifnum\LastMac=\Afe \OneHalf\else \vskip 21pt\fi
    \fourteenpoint
    \Raggedright
    \noindent \strut #1\par
    \vskip 3pt%
  \egroup
}

\def\affiliation#1{
  \bgroup
    \vskip -4pt%
    \eightpoint
    \Raggedright
    \noindent \strut {\it #1}\par
  \egroup
  \LastMac=\Afe\relax
}

\def\acceptedline#1{
  \bgroup
    \Two
    \eightpoint
    \Raggedright
    \noindent \strut #1\par
  \egroup
}

\long\def\abstract#1{%
  \bgroup
    \vskip 20pt%
    \everypar{\Hang{11pc}{0}}%
    \noindent{\ninebf ABSTRACT}\par
    \tenpoint
    \Fullout
    \noindent #1\par
  \egroup
}

\long\def\keywords#1{
  \bgroup
    \Half
    \everypar{\Hang{11pc}{0}}%
    \tenpoint
    \Fullout
    \noindent\hbox{\bf Key words:}\ #1\par
  \egroup
}


\def\maketitle{%
  \EndOpening
  \ifsinglecol \else \MakePage\fi
}



\newif\ifAutoNumber \AutoNumberfalse
\newcount\Sec        
\newcount\SecSec
\newcount\SecSecSec

\Sec=\z@

\def\:{\let\@sptoken= } \:  
\def\:{\@xifnch} \expandafter\def\: {\futurelet\@tempc\@ifnch}

\def\@ifnextchar#1#2#3{%
  \let\@tempMACe #1%
  \def\@tempMACa{#2}%
  \def\@tempMACb{#3}%
  \futurelet \@tempMACc\@ifnch%
}

\def\@ifnch{%
\ifx \@tempMACc \@sptoken%
  \let\@tempMACd\@xifnch%
\else%
  \ifx \@tempMACc \@tempMACe%
    \let\@tempMACd\@tempMACa%
  \else%
    \let\@tempMACd\@tempMACb%
  \fi%
\fi%
\@tempMACd%
}

\def\@ifstar#1#2{\@ifnextchar *{\def\@tempMACa*{#1}\@tempMACa}{#2}}

\newskip\@tempskipb

\def\addvspace#1{%
  \ifvmode\else \endgraf\fi%
  \ifdim\lastskip=\z@%
    \vskip #1\relax%
  \else%
    \@tempskipb#1\relax\@xaddvskip%
  \fi%
}

\def\@xaddvskip{%
  \ifdim\lastskip<\@tempskipb%
    \vskip-\lastskip%
    \vskip\@tempskipb\relax%
  \else%
    \ifdim\@tempskipb<\z@%
      \ifdim\lastskip<\z@ \else%
        \advance\@tempskipb\lastskip%
        \vskip-\lastskip\vskip\@tempskipb%
      \fi%
    \fi%
  \fi%
}

\newskip\@tmpSKIP

\def\addpen#1{%
  \ifvmode
    \if@nobreak
    \else
      \ifdim\lastskip=\z@
        \penalty#1\relax
      \else
        \@tmpSKIP=\lastskip
        \vskip -\lastskip
        \penalty#1\vskip\@tmpSKIP
      \fi
    \fi
  \fi
}

\newcount\@clubpen   \@clubpen=\clubpenalty
\newif\if@nobreak    \@nobreakfalse

\def\@noafterindent{%
  \global\@nobreaktrue
  \everypar{\if@nobreak
              \global\@nobreakfalse
              \clubpenalty \@M
              {\setbox\z@\lastbox}%
              \LastMac=\Nulle\relax%
            \else
              \clubpenalty \@clubpen
              \everypar{}%
            \fi}
}

\newcount\gds@cbrk   \gds@cbrk=-300

\def\@nohdbrk{\interlinepenalty \@M\relax}

\let\@par=\par
\def\@restorepar{\def\par{\@par}}

\newif\if@endpe   \@endpefalse
 
\def\@doendpe{\@endpetrue \@nobreakfalse \LastMac=\Nulle\relax%
     \def\par{\@restorepar\everypar{}\par\@endpefalse}%
              \everypar{\setbox\z@\lastbox\everypar{}\@endpefalse}%
}

\def\section{\@ifstar{\@ssection}{\@section}}

\def\@section#1{
  \if@nobreak
    \everypar{}%
    \ifnum\LastMac=\Hae \addvspace{\half}\fi
  \else
    \addpen{\gds@cbrk}%
    \addvspace{\two}%
  \fi
  \bgroup
    \ninepoint\bf
    \Raggedright
    \ifAutoNumber
      \global\advance\Sec \@ne
      \noindent\@nohdbrk\number\Sec\hskip 1pc \uppercase{#1}\par
      \global\SecSec=\z@
    \else
      \noindent\@nohdbrk\uppercase{#1}\par
    \fi
  \egroup
  \nobreak
  \vskip\half
  \nobreak
  \@noafterindent
  \LastMac=\Hae\relax
}

\def\@ssection#1{
  \if@nobreak
    \everypar{}%
    \ifnum\LastMac=\Hae \addvspace{\half}\fi
  \else
    \addpen{\gds@cbrk}%
    \addvspace{\two}%
  \fi
  \bgroup
    \ninepoint\bf
    \Raggedright
    \noindent\@nohdbrk\uppercase{#1}\par
  \egroup
  \nobreak
  \vskip\half
  \nobreak
  \@noafterindent
  \LastMac=\Hae\relax
}

\def\subsection#1{
  \if@nobreak
    \everypar{}%
    \ifnum\LastMac=\Hae \addvspace{1pt plus 1pt minus .5pt}\fi
  \else
    \addpen{\gds@cbrk}%
    \addvspace{\onehalf}%
  \fi
  \bgroup
    \ninepoint\bf
    \Raggedright
    \ifAutoNumber
      \global\advance\SecSec \@ne
      \noindent\@nohdbrk\number\Sec.\number\SecSec \hskip 1pc\relax #1\par
      \global\SecSecSec=\z@
    \else
      \noindent\@nohdbrk #1\par
    \fi
  \egroup
  \nobreak
  \vskip\half
  \nobreak
  \@noafterindent
  \LastMac=\Hbe\relax
}

\def\subsubsection#1{
  \if@nobreak
    \everypar{}%
    \ifnum\LastMac=\Hbe \addvspace{1pt plus 1pt minus .5pt}\fi
  \else
    \addpen{\gds@cbrk}%
    \addvspace{\onehalf}%
  \fi
  \bgroup
    \ninepoint\it
    \Raggedright
    \ifAutoNumber
      \global\advance\SecSecSec \@ne
      \noindent\@nohdbrk\number\Sec.\number\SecSec.\number\SecSecSec
        \hskip 1pc\relax #1\par
    \else
      \noindent\@nohdbrk #1\par
    \fi
  \egroup
  \nobreak
  \vskip\half
  \nobreak
  \@noafterindent
  \LastMac=\Hce\relax
}

\def\paragraph#1{
  \if@nobreak
    \everypar{}%
  \else
    \addpen{\gds@cbrk}%
    \addvspace{\one}%
  \fi%
  \bgroup%
    \ninepoint\it
    \noindent #1\ \nobreak%
  \egroup
  \LastMac=\Hde\relax
  \ignorespaces
}




\def\beginlist{%
  \par\if@nobreak \else\addvspace{\half}\fi%
  \bgroup%
    \ninepoint
    \let\item=\list@item%
}

\def\list@item{%
  \par\noindent\hskip 1em\relax%
  \ignorespaces%
}

\def\endlist{\par\egroup\addvspace{\half}\@doendpe}


\def\beginrefs{%
  \par
  \bgroup
    \eightpoint
    \Raggedright
    \let\bibitem=\bib@item
}

\def\bib@item{%
  \par\parindent=1.5em\Hang{1.5em}{1}%
  \everypar={\Hang{1.5em}{1}\ignorespaces}%
  \noindent\ignorespaces
}

\def\endrefs{\par\egroup\@doendpe}


\newtoks\CatchLine

\def\@journal{Mon.\ Not.\ R.\ Astron.\ Soc.\ }  
\def\@pubyear{1994}        
\def\@pagerange{000--000}  
\def\@volume{000}          
\def\@microfiche{}         %

\def\pubyear#1{\gdef\@pubyear{#1}\@makecatchline}
\def\pagerange#1{\gdef\@pagerange{#1}\@makecatchline}
\def\volume#1{\gdef\@volume{#1}\@makecatchline}
\def\microfiche#1{\gdef\@microfiche{and Microfiche\ #1}\@makecatchline}

\def\@makecatchline{%
  \global\CatchLine{%
    {\rm \@journal {\bf \@volume},\ \@pagerange\ (\@pubyear)\ \@microfiche}}%
}

\@makecatchline 

\newtoks\LeftHeader
\def\shortauthor#1{
  \global\LeftHeader{#1}%
}

\newtoks\RightHeader
\def\shorttitle#1{
  \global\RightHeader{#1}%
}

\def\PageHead{
  \begingroup
    \ifsp@page
      \csname ps@\sp@type\endcsname
      \global\sp@pagefalse
    \fi
    \ifodd\pageno
      \let\the@head=\@oddhead
    \else
      \let\the@head=\@evenhead
    \fi
    \vbox to \z@{\vskip-22.5\p@%
      \hbox to \PageWidth{\vbox to8.5\p@{}%
        \the@head
      }%
    \vss}%
  \endgroup
  \nointerlineskip
}

\def\today{%
  \number\day\space
  \ifcase\month\or January\or February\or March\or April\or May\or June\or
    July\or August\or September\or October\or November\or December\fi
  \space\number\year%
}

\def\PageFoot{} 

\def\authorcomment#1{%
  \gdef\PageFoot{%
    \nointerlineskip%
    \vbox to 22pt{\vfil%
      \hbox to \PageWidth{\elevenpoint\noindent \hfil #1 \hfil}}%
  }%
}


\newif\ifplate@page
\newbox\plt@box

\def\beginplatepage{%
  \let\plate=\plate@head
  \let\caption=\fig@caption
  \global\setbox\plt@box=\vbox\bgroup
  \TEMPDIMEN=\PageWidth 
  \hsize=\PageWidth\relax
}

\def\endplatepage{\par\egroup\global\plate@pagetrue}
\def\plate@head#1{\gdef\plt@cap{#1}}


\def\letters{%
  \gdef\folio{\ifnum\pageno<\z@ L\romannumeral-\pageno
    \else L\number\pageno \fi}%
}


\everydisplay{\displaysetup}

\newif\ifeqno
\newif\ifleqno

\def\displaysetup#1$${%
 \displaytest#1\eqno\eqno\displaytest
}

\def\displaytest#1\eqno#2\eqno#3\displaytest{%
 \if!#3!\ldisplaytest#1\leqno\leqno\ldisplaytest
 \else\eqnotrue\leqnofalse\def\eqn{#2}\def\eq{#1}\fi
 \generaldisplay$$}

\def\ldisplaytest#1\leqno#2\leqno#3\ldisplaytest{%
 \def\eq{#1}%
 \if!#3!\eqnofalse\else\eqnotrue\leqnotrue
  \def\eqn{#2}\fi}

\def\generaldisplay{%
\ifeqno \ifleqno 
   \hbox to \hsize{\noindent
     $\displaystyle\eq$\hfil$\displaystyle\eqn$}
  \else
    \hbox to \hsize{\noindent
     $\displaystyle\eq$\hfil$\displaystyle\eqn$}
  \fi
 \else
 \hbox to \hsize{\vbox{\noindent
  $\displaystyle\eq$\hfil}}
 \fi
}


\def\@notice{%
  \par\Two%
  \noindent{\b@ls{11pt}\ninerm This paper has been produced using the
    Blackwell Scientific Publications \TeX\ macros.\par}%
}

\outer\def
\section*{REFERENCES}
\beginrefs
\bibitem Acker A., Stenholm B., 1990, A\&AS, 86, 219
\bibitem Acker A., Ochsenbein F., Stenholm B., Tylenda R., Marcout J., 
Schohn C., 1992, Strasbourg-ESO Catalog of Galactic Planetary Nebul{\ae}. 
ESO, Garching   (=A92)
\bibitem Acker A., Fresneau A., Pottasch S. R., Jasniewicz G., 1998, A\&A, 
337, 253  (=A98)
\bibitem Arenou F., Luri X., 1999, in Egret D., Heck A., eds, Harmonizing 
Cosmic Distance Scales, ASP Conference Series 167. ASP, San Francisco, p. 13 
\bibitem Benedict G. F. et al., 2003, AJ, 126, 2549
\bibitem Benedict G. F. et al., 2009, AJ, 138, 1969  (=B09)
\bibitem Bl\"ocker T., Sch\"onberner D., 1990, A\&A, 240, L11
\bibitem Bond H. E., Ciardullo R., 1999, PASP, 111, 217
\bibitem Brown A., Arenou F., van Leeuwen F., Lindegren L., Luri X., 1997, 
{\it Hipparcos} Venice '97, ESA SP-402. p. 63
\bibitem Cahn J. H., Kaler J. B., Stanghellini L., 1992, A\&AS, 94, 399  (=CKS)
\bibitem Ciardullo R., Bond H. E., Sipior M. S., Fullton L. K., Zhang C.-Y., 
Schaefer K. G., 1999, AJ, 118,488  (=C99)
\bibitem C\'orsico A. H., Althaus L. G., 2006, A\&A, 454, 863
\bibitem Daub C. T., 1982, ApJ, 260, 612
\bibitem De Marco O., Passy J.C., Frew D. J., Moe M., Jacoby G. H., 2013, 
MNRAS, 428, 2118
\bibitem Frew D. J., 2008, Ph.D. thesis, Macquarie University, Sydney  (=F08)
\bibitem Frew D. J., Madsen G. J., O'Toole S. J., Parker Q. A., 2010, PASA, 27, 
203
\bibitem Frew D. J., Parker Q., 2006, in Barlow M. J., Mendez R. H., eds., 
Planetary Nebul{\ae} in Our Galaxy and Beyond, Proc. IAU Symposium 234. 
Cambridge Univ. Press, Cambridge, p. 49 (=FP06)
\bibitem Frew D. J., Parker Q. A., 2010, PASA, 27, 129
\bibitem Frew D. J., Parker Q. A., Boji\^ci\'c I. S., 2014, MNRAS, submitted  
(=F14)
\bibitem Gathier R., Pottasch S. R., Pel J. W., 1986, A\&A, 157, 171  (=G86)
\bibitem Giammanco C. et al., 2011, A\&A, 525, 58
\bibitem Griffin, R. F., 1967, ApJ, 148,465
\bibitem Guti\'errez-Moreno A., Anguita A., Loyola P., Moreno H., 1999, 
PASP, 111, 1163
\bibitem Harris H. C., Dahn C. C., Monet D. G., Pier J. R., 1997, in 
Habing H. J., Lamers  H. J. G. L. M., eds., Planetary Nebul{\ae}, Proc. 
IAU Symposium 180. D. Reidel, Dordrecht, p. 40  (=H97)
\bibitem Harris H. C. et al., 2007, AJ, 133, 631 \ \ (=H07)
\bibitem Hewett P. C., Irwin M. J., Skillman E. D., Foltz C. B., Willis J. P., 
Warren S. J., Walton N. A., 2003, ApJ, 599, L37
\bibitem Jacob R., Sch\"onberber D., Steffen M., 2013, A\&A, 558, 78
\bibitem Kaler J. B., Lutz J. H., 1985, PASP, 97, 700
\bibitem Kingsburgh R. L., Barlow M. J., 1992, MNRAS, 257, 317
\bibitem Kohoutek L., 1977, A\&A, 59,137
\bibitem Lupton R., 1993, Statistics in Theory and Practice. Princeton Univ. 
Press, Princeton
\bibitem Lutz T. E., Kelker D. H., 1973, PASP, 85, 573 
\bibitem Maciel W. J., Pottasch S. R., 1980, A\&A, 88, 1
\bibitem Manteiga M., Arcay B., Ulla A., Aller A., Miranda L., Isasi Y.,2012, in 
A. Manchado, L. Stanghellini, D. Sch\"onberner, eds, IAU Symposium 283, p. 428
\bibitem Manteiga M., Fustes D., Dafonte C., Arcay B., Ulla A., 2014, in C. 
Morriset, G. Delgado-Inglada, S. Torres-Peimbert, eds, Asymmetrical Planetary 
Nebul{\ae} VI, p. 57
\bibitem Masson C. R., 1986,ApJL, 302, L27
\bibitem Melis C., Reid M. R., Mioduszewski A. J., Stauffer J. R., Bower G. C., 
2014, Science, 345, 1029
\bibitem Mellema G., 2004, A\&A, 416, 623
\bibitem M\'endez R. H., Groth H. G., Husfeld D., Kudritzki R.-P., Herrero A., 
1988, A\&A, 197, L25
\bibitem Napiwotzki R., 1993, Ph.D. thesis
\bibitem Napiwotzki R., 2001, A\&A, 367, 973  (=N01)
\bibitem Napiwotzki R., Sch\"onberner D., 1995, A\&A, 367, 973  (=NS95)
\bibitem O'Dell C. R., 1962, ApJ, 135, 371 (=O62)
\bibitem O'Dell C. R., 1998, AJ, 116, 1346
\bibitem Palen S., Balick B., Hajian A. R., Terzian Y., Bond H. E., 
Panagia N., 2002, AJ, 123, 2666
\bibitem Parker Q. A. et al., 2006, MNRAS, 373, 79
\bibitem Peimbert M., 1978, in Terzian Y., ed., Planetary Nebul{\ae}: Observations 
and Theory, Proc. IAU Symposium 76. D. Reidel, Dordrecht, p. 215
\bibitem Perryman et al., 2001, A\&A, 369, 339
\bibitem Phillips J. P., 2002, ApJSS, 139, 199  (=Ph02)
\bibitem Phillips J. P., 2004, MNRAS, 353, 589  (=Ph04)
\bibitem Phillips J. P., 2005, MNRAS, 357,619
\bibitem Pier J. R., Harris H. C., Dahn C. C., Monet D., 1993, in Weinberger R., 
Acker A., eds., Planetary Nebul{\ae}, Proc. IAU Symposium 155. Kluwer, 
Dordrecht, p. 175
\bibitem Pont F., 1999, in Egret D., Heck A., eds, Harmonizing Cosmic 
Distance Scales, (see above), p. 113
\bibitem Pottasch S. R., 1980, A\&A, 89, 336
\bibitem Pottasch S. R., 1996, A\&A, 307, 561  (=P96)
\bibitem Quireza C., Rocha-Pinto H. J., Maciel W. J., 2007, A\&A, 475, 217
\bibitem Rauch T., Kerber F., Pauli E. M., 2004, A\&A, 417, 647
\bibitem Reed D. S., Balick B., Hajian A. R., Klayton T. L., Giovanardi S., 
Casertano S., Panagia N., Terzian Y., 1999, AJ, 118, 2430
\bibitem Schneider S. E., Buckley D., 1996, ApJ, 459, 606
\bibitem Sch\"onberner D., Jacob R.,  Steffen M., 2005, A\&A, 441, 573 (=S05)
\bibitem Seaton M. J., 1966, MNRAS,132, 113 (=S66)
\bibitem Shaw R. A., Stanghellini L., Mutchler M., Balick B., Blades J. C., 2001, 
ApJ, 548, 727
\bibitem Shaw R. A., Stanghellini L., Villaver, E., Mutchler M., 2006, ApJS, 167, 201
\bibitem Shklovsky I. S., 1956, AJUSSR, 33, 222
\bibitem Smith H., 1976, A\&A, 53, 333
\bibitem Smith H., 2003, MNRAS, 338, 891 
\bibitem Smith H., 2006, MNRAS, 365, 469
\bibitem Smith H., Eichhorn H. K., 1996, MNRAS, 281,211  (=SE96)
\bibitem Stanghellini L., Shaw R. A., Balick B., Mutchler M., Blades J. C., Villaver E., 
2003, ApJ, 596, 997
\bibitem Stanghellini L., Shaw R. A., Mutchler M., Palen S., Balick B., Blades J. C., 
2002, ApJ, 515, 178
\bibitem Stanghellini L., Shaw R. A., Villaver E., 2008, ApJ, 689, 194  (=SSV)
\bibitem Terzian Y., 1980, QJRAS, 21, 82
\bibitem Terzian Y., 1993, in Proc. IAU Symposium 155 (see above), p. 109 (=T93)
\bibitem Terzian Y., 1997, in Proc. IAU Symposium 180 (see above), p. 29  (=T97)
\bibitem Trumpler R. J., Weaver H. F., 1953, Statistical Astronomy. Univ. 
California Press, Berkeley, p. 369
\bibitem Tweedy R. W., Kwitter K. B., 1994, ApJ, 433, L93
\bibitem Tweedy R. W., Kwitter K. B., 1996, ApJS, 107, 255
\bibitem Tweedy R. W., Martos M. A., Noriega-Crespo A., 1995, ApJ, 447, 257
\bibitem van de Steene G. C., Zijlstra A. A., 1995, A\&A, 293, 541 (=VdSZ)
\bibitem van Leeuwen F., 2007a, {\it Hipparcos}, The New Reduction of the Raw Data 
(Dordrecht: Springer)  (=VL07)
\bibitem van Leeuwen F., 2007b, A\&A, 474,653  
\bibitem Wareing C. J., Zijlstra A. A., O'Brien T. J., 2007, MNRAS, 382,1233
\bibitem Zhang C. Y., 1993, ApJ, 410, 239
\bibitem Zhang C. Y., 1995, ApJS, 98, 659  (=Z95)
\endrefs
\bye{\@notice\par\vfill\supereject\end}


\def\start@mess{%
  Monthly notices of the RAS journal style (\@typeface)\space
    v\@version,\space \@verdate.%
}

\everyjob{\Warn{\start@mess}}



\newif\if@debug \@debugfalse  

\def\Print#1{\if@debug\immediate\write16{#1}\else \fi}
\def\Warn#1{\immediate\write16{#1}}
\def\wlog#1{}

\newcount\Iteration 

\def\Single{0} \def\Double{1}                 
\def\Figure{0} \def\Table{1}                  

\def\InStack{0}  
\def\InZoneA{1}
\def\InZoneB{2}
\def\InZoneC{3}

\newcount\TEMPCOUNT 
\newdimen\TEMPDIMEN 
\newbox\TEMPBOX     
\newbox\VOIDBOX     

\newcount\LengthOfStack 
\newcount\MaxItems      
\newcount\StackPointer
\newcount\Point         
\newcount\NextFigure    
\newcount\NextTable     
\newcount\NextItem      

\newcount\StatusStack   
\newcount\NumStack      
\newcount\TypeStack     
\newcount\SpanStack     
\newcount\BoxStack      

\newcount\ItemSTATUS    
\newcount\ItemNUMBER    
\newcount\ItemTYPE      
\newcount\ItemSPAN      
\newbox\ItemBOX         
\newdimen\ItemSIZE      

\newdimen\PageHeight    
\newdimen\TextLeading   
\newdimen\Feathering    
\newcount\LinesPerPage  
\newdimen\ColumnWidth   
\newdimen\ColumnGap     
\newdimen\PageWidth     
\newdimen\BodgeHeight   
\newcount\Leading       

\newdimen\ZoneBSize  
\newdimen\TextSize   
\newbox\ZoneABOX     
\newbox\ZoneBBOX     
\newbox\ZoneCBOX     

\newif\ifFirstSingleItem
\newif\ifFirstZoneA
\newif\ifMakePageInComplete
\newif\ifMoreFigures \MoreFiguresfalse 
\newif\ifMoreTables  \MoreTablesfalse  

\newif\ifFigInZoneB 
\newif\ifFigInZoneC 
\newif\ifTabInZoneB 
\newif\ifTabInZoneC

\newif\ifZoneAFullPage

\newbox\MidBOX    
\newbox\LeftBOX
\newbox\RightBOX
\newbox\PageBOX   

\newif\ifLeftCOL  
\LeftCOLtrue

\newdimen\ZoneBAdjust

\newcount\ItemFits
\def\Yes{1}
\def\No{2}


\MaxItems=15
\NextFigure=\z@        
\NextTable=\@ne

\BodgeHeight=6pt
\TextLeading=11pt    
\Leading=11
\Feathering=\z@      
\LinesPerPage=61     
\topskip=\TextLeading
\ColumnWidth=20pc    
\ColumnGap=2pc       

\newskip\ItemSepamount  
\ItemSepamount=\TextLeading plus \TextLeading minus 4pt

\parskip=\z@ plus .1pt
\parindent=18pt
\widowpenalty=\z@
\clubpenalty=10000
\tolerance=1500
\hbadness=1500
\abovedisplayskip=6pt plus 2pt minus 2pt
\belowdisplayskip=6pt plus 2pt minus 2pt
\abovedisplayshortskip=6pt plus 2pt minus 2pt
\belowdisplayshortskip=6pt plus 2pt minus 2pt

\ninepoint 


\PageHeight=682pt

\PageWidth=2\ColumnWidth
\advance\PageWidth by \ColumnGap

\pagestyle{headings}




\newcount\DUMMY \StatusStack=\allocationnumber
\newcount\DUMMY \newcount\DUMMY \newcount\DUMMY 
\newcount\DUMMY \newcount\DUMMY \newcount\DUMMY 
\newcount\DUMMY \newcount\DUMMY \newcount\DUMMY
\newcount\DUMMY \newcount\DUMMY \newcount\DUMMY 
\newcount\DUMMY \newcount\DUMMY \newcount\DUMMY

\newcount\DUMMY \NumStack=\allocationnumber
\newcount\DUMMY \newcount\DUMMY \newcount\DUMMY 
\newcount\DUMMY \newcount\DUMMY \newcount\DUMMY 
\newcount\DUMMY \newcount\DUMMY \newcount\DUMMY 
\newcount\DUMMY \newcount\DUMMY \newcount\DUMMY 
\newcount\DUMMY \newcount\DUMMY \newcount\DUMMY

\newcount\DUMMY \TypeStack=\allocationnumber
\newcount\DUMMY \newcount\DUMMY \newcount\DUMMY 
\newcount\DUMMY \newcount\DUMMY \newcount\DUMMY 
\newcount\DUMMY \newcount\DUMMY \newcount\DUMMY 
\newcount\DUMMY \newcount\DUMMY \newcount\DUMMY 
\newcount\DUMMY \newcount\DUMMY \newcount\DUMMY

\newcount\DUMMY \SpanStack=\allocationnumber
\newcount\DUMMY \newcount\DUMMY \newcount\DUMMY 
\newcount\DUMMY \newcount\DUMMY \newcount\DUMMY 
\newcount\DUMMY \newcount\DUMMY \newcount\DUMMY 
\newcount\DUMMY \newcount\DUMMY \newcount\DUMMY 
\newcount\DUMMY \newcount\DUMMY \newcount\DUMMY

\newbox\DUMMY   \BoxStack=\allocationnumber
\newbox\DUMMY   \newbox\DUMMY \newbox\DUMMY 
\newbox\DUMMY   \newbox\DUMMY \newbox\DUMMY 
\newbox\DUMMY   \newbox\DUMMY \newbox\DUMMY 
\newbox\DUMMY   \newbox\DUMMY \newbox\DUMMY 
\newbox\DUMMY   \newbox\DUMMY \newbox\DUMMY

\def\wlog{\immediate\write\m@ne}


\def\GetItemAll#1{%
 \GetItemSTATUS{#1}
 \GetItemNUMBER{#1}
 \GetItemTYPE{#1}
 \GetItemSPAN{#1}
 \GetItemBOX{#1}
}

\def\GetItemSTATUS#1{%
 \Point=\StatusStack
 \advance\Point by #1
 \global\ItemSTATUS=\count\Point
}

\def\GetItemNUMBER#1{%
 \Point=\NumStack
 \advance\Point by #1
 \global\ItemNUMBER=\count\Point
}

\def\GetItemTYPE#1{%
 \Point=\TypeStack
 \advance\Point by #1
 \global\ItemTYPE=\count\Point
}

\def\GetItemSPAN#1{%
 \Point\SpanStack
 \advance\Point by #1
 \global\ItemSPAN=\count\Point
}

\def\GetItemBOX#1{%
 \Point=\BoxStack
 \advance\Point by #1
 \global\setbox\ItemBOX=\vbox{\copy\Point}
 \global\ItemSIZE=\ht\ItemBOX
 \global\advance\ItemSIZE by \dp\ItemBOX
 \TEMPCOUNT=\ItemSIZE
 \divide\TEMPCOUNT by \Leading
 \divide\TEMPCOUNT by 65536
 \advance\TEMPCOUNT \@ne
 \ItemSIZE=\TEMPCOUNT pt
 \global\multiply\ItemSIZE by \Leading
}


\def\JoinStack{%
 \ifnum\LengthOfStack=\MaxItems 
  \Warn{WARNING: Stack is full...some items will be lost!}
 \else
  \Point=\StatusStack
  \advance\Point by \LengthOfStack
  \global\count\Point=\ItemSTATUS
  \Point=\NumStack
  \advance\Point by \LengthOfStack
  \global\count\Point=\ItemNUMBER
  \Point=\TypeStack
  \advance\Point by \LengthOfStack
  \global\count\Point=\ItemTYPE
  \Point\SpanStack
  \advance\Point by \LengthOfStack
  \global\count\Point=\ItemSPAN
  \Point=\BoxStack
  \advance\Point by \LengthOfStack
  \global\setbox\Point=\vbox{\copy\ItemBOX}
  \global\advance\LengthOfStack \@ne
  \ifnum\ItemTYPE=\Figure 
   \global\MoreFigurestrue
  \else
   \global\MoreTablestrue
  \fi
 \fi
}


\def\LeaveStack#1{%
 {\Iteration=#1
 \loop
 \ifnum\Iteration<\LengthOfStack
  \advance\Iteration \@ne
  \GetItemSTATUS{\Iteration}
   \advance\Point by \m@ne
   \global\count\Point=\ItemSTATUS
  \GetItemNUMBER{\Iteration}
   \advance\Point by \m@ne
   \global\count\Point=\ItemNUMBER
  \GetItemTYPE{\Iteration}
   \advance\Point by \m@ne
   \global\count\Point=\ItemTYPE
  \GetItemSPAN{\Iteration}
   \advance\Point by \m@ne
   \global\count\Point=\ItemSPAN
  \GetItemBOX{\Iteration}
   \advance\Point by \m@ne
   \global\setbox\Point=\vbox{\copy\ItemBOX}
 \repeat}
 \global\advance\LengthOfStack by \m@ne
}


\newif\ifStackNotClean

\def\CleanStack{%
 \StackNotCleantrue
 {\Iteration=\z@
  \loop
   \ifStackNotClean
    \GetItemSTATUS{\Iteration}
    \ifnum\ItemSTATUS=\InStack
     \advance\Iteration \@ne
     \else
      \LeaveStack{\Iteration}
    \fi
   \ifnum\LengthOfStack<\Iteration
    \StackNotCleanfalse
   \fi
 \repeat}
}


\def\FindItem#1#2{%
 \global\StackPointer=\m@ne 
 {\Iteration=\z@
  \loop
  \ifnum\Iteration<\LengthOfStack
   \GetItemSTATUS{\Iteration}
   \ifnum\ItemSTATUS=\InStack
    \GetItemTYPE{\Iteration}
    \ifnum\ItemTYPE=#1
     \GetItemNUMBER{\Iteration}
     \ifnum\ItemNUMBER=#2
      \global\StackPointer=\Iteration
      \Iteration=\LengthOfStack 
     \fi
    \fi
   \fi
  \advance\Iteration \@ne
 \repeat}
}


\def\FindNext{%
 \global\StackPointer=\m@ne 
 {\Iteration=\z@
  \loop
  \ifnum\Iteration<\LengthOfStack
   \GetItemSTATUS{\Iteration}
   \ifnum\ItemSTATUS=\InStack
    \GetItemTYPE{\Iteration}
   \ifnum\ItemTYPE=\Figure
    \ifMoreFigures
      \global\NextItem=\Figure
      \global\StackPointer=\Iteration
      \Iteration=\LengthOfStack 
    \fi
   \fi
   \ifnum\ItemTYPE=\Table
    \ifMoreTables
      \global\NextItem=\Table
      \global\StackPointer=\Iteration
      \Iteration=\LengthOfStack 
    \fi
   \fi
  \fi
  \advance\Iteration \@ne
 \repeat}
}


\def\ChangeStatus#1#2{%
 \Point=\StatusStack
 \advance\Point by #1
 \global\count\Point=#2
}



\def\Zone{\InZoneA}

\ZoneBAdjust=\z@

\def\MakePage{
 \global\ZoneBSize=\PageHeight
 \global\TextSize=\ZoneBSize
 \global\ZoneAFullPagefalse
 \global\topskip=\TextLeading
 \MakePageInCompletetrue
 \MoreFigurestrue
 \MoreTablestrue
 \FigInZoneBfalse
 \FigInZoneCfalse
 \TabInZoneBfalse
 \TabInZoneCfalse
 \global\FirstSingleItemtrue
 \global\FirstZoneAtrue
 \global\setbox\ZoneABOX=\box\VOIDBOX
 \global\setbox\ZoneBBOX=\box\VOIDBOX
 \global\setbox\ZoneCBOX=\box\VOIDBOX
 \loop
  \ifMakePageInComplete
 \FindNext
 \ifnum\StackPointer=\m@ne
  \NextItem=\m@ne
  \MoreFiguresfalse
  \MoreTablesfalse
 \fi
 \ifnum\NextItem=\Figure
   \FindItem{\Figure}{\NextFigure}
   \ifnum\StackPointer=\m@ne \global\MoreFiguresfalse
   \else
    \GetItemSPAN{\StackPointer}
    \ifnum\ItemSPAN=\Single \def\Zone{\InZoneB}\relax
     \ifFigInZoneC \global\MoreFiguresfalse\fi
    \else
     \def\Zone{\InZoneA}
     \ifFigInZoneB \def\Zone{\InZoneC}\fi
    \fi
   \fi
   \ifMoreFigures\Print{}\FigureItems\fi
 \fi
\ifnum\NextItem=\Table
   \FindItem{\Table}{\NextTable}
   \ifnum\StackPointer=\m@ne \global\MoreTablesfalse
   \else
    \GetItemSPAN{\StackPointer}
    \ifnum\ItemSPAN=\Single\relax
     \ifTabInZoneC \global\MoreTablesfalse\fi
    \else
     \def\Zone{\InZoneA}
     \ifTabInZoneB \def\Zone{\InZoneC}\fi
    \fi
   \fi
   \ifMoreTables\Print{}\TableItems\fi
 \fi
   \MakePageInCompletefalse 
   \ifMoreFigures\MakePageInCompletetrue\fi
   \ifMoreTables\MakePageInCompletetrue\fi
 \repeat
 \ifZoneAFullPage
  \global\TextSize=\z@
  \global\ZoneBSize=\z@
  \global\vsize=\z@\relax
  \global\topskip=\z@\relax
  \vbox to \z@{\vss}
  \eject
 \else
 \global\advance\ZoneBSize by -\ZoneBAdjust
 \global\vsize=\ZoneBSize
 \global\hsize=\ColumnWidth
 \global\ZoneBAdjust=\z@
 \ifdim\TextSize<23pt
 \Warn{}
 \Warn{* Making column fall short: TextSize=\the\TextSize *}
 \vskip-\lastskip\eject\fi
 \fi
}

\def\MakeRightCol{
 \global\TextSize=\ZoneBSize
 \MakePageInCompletetrue
 \MoreFigurestrue
 \MoreTablestrue
 \global\FirstSingleItemtrue
 \global\setbox\ZoneBBOX=\box\VOIDBOX
 \def\Zone{\InZoneB}
 \loop
  \ifMakePageInComplete
 \FindNext
 \ifnum\StackPointer=\m@ne
  \NextItem=\m@ne
  \MoreFiguresfalse
  \MoreTablesfalse
 \fi
 \ifnum\NextItem=\Figure
   \FindItem{\Figure}{\NextFigure}
   \ifnum\StackPointer=\m@ne \MoreFiguresfalse
   \else
    \GetItemSPAN{\StackPointer}
    \ifnum\ItemSPAN=\Double\relax
     \MoreFiguresfalse\fi
   \fi
   \ifMoreFigures\Print{}\FigureItems\fi
 \fi
 \ifnum\NextItem=\Table
   \FindItem{\Table}{\NextTable}
   \ifnum\StackPointer=\m@ne \MoreTablesfalse
   \else
    \GetItemSPAN{\StackPointer}
    \ifnum\ItemSPAN=\Double\relax
     \MoreTablesfalse\fi
   \fi
   \ifMoreTables\Print{}\TableItems\fi
 \fi
   \MakePageInCompletefalse 
   \ifMoreFigures\MakePageInCompletetrue\fi
   \ifMoreTables\MakePageInCompletetrue\fi
 \repeat
 \ifZoneAFullPage
  \global\TextSize=\z@
  \global\ZoneBSize=\z@
  \global\vsize=\z@\relax
  \global\topskip=\z@\relax
  \vbox to \z@{\vss}
  \eject
 \else
 \global\vsize=\ZoneBSize
 \global\hsize=\ColumnWidth
 \ifdim\TextSize<23pt
 \Warn{}
 \Warn{* Making column fall short: TextSize=\the\TextSize *}
 \vskip-\lastskip\eject\fi
\fi
}

\def\FigureItems{
 \Print{Considering...}
 \ShowItem{\StackPointer}
 \GetItemBOX{\StackPointer} 
 \GetItemSPAN{\StackPointer}
  \CheckFitInZone 
  \ifnum\ItemFits=\Yes
   \ifnum\ItemSPAN=\Single
     \ChangeStatus{\StackPointer}{\InZoneB} 
     \global\FigInZoneBtrue
     \ifFirstSingleItem
      \hbox{}\vskip-\BodgeHeight
     \global\advance\ItemSIZE by \TextLeading
     \fi
     \unvbox\ItemBOX\ItemSep
     \global\FirstSingleItemfalse
     \global\advance\TextSize by -\ItemSIZE
     \global\advance\TextSize by -\TextLeading
   \else
    \ifFirstZoneA
     \global\advance\ItemSIZE by \TextLeading
     \global\FirstZoneAfalse\fi
    \global\advance\TextSize by -\ItemSIZE
    \global\advance\TextSize by -\TextLeading
    \global\advance\ZoneBSize by -\ItemSIZE
    \global\advance\ZoneBSize by -\TextLeading
    \ifFigInZoneB\relax
     \else
     \ifdim\TextSize<3\TextLeading
     \global\ZoneAFullPagetrue
     \fi
    \fi
    \ChangeStatus{\StackPointer}{\Zone}
    \ifnum\Zone=\InZoneC \global\FigInZoneCtrue\fi
  \fi
   \Print{TextSize=\the\TextSize}
   \Print{ZoneBSize=\the\ZoneBSize}
  \global\advance\NextFigure \@ne
   \Print{This figure has been placed.}
  \else
   \Print{No space available for this figure...holding over.}
   \Print{}
   \global\MoreFiguresfalse
  \fi
}

\def\TableItems{
 \Print{Considering...}
 \ShowItem{\StackPointer}
 \GetItemBOX{\StackPointer} 
 \GetItemSPAN{\StackPointer}
  \CheckFitInZone 
  \ifnum\ItemFits=\Yes
   \ifnum\ItemSPAN=\Single
    \ChangeStatus{\StackPointer}{\InZoneB}
     \global\TabInZoneBtrue
     \ifFirstSingleItem
      \hbox{}\vskip-\BodgeHeight
     \global\advance\ItemSIZE by \TextLeading
     \fi
     \unvbox\ItemBOX\ItemSep
     \global\FirstSingleItemfalse
     \global\advance\TextSize by -\ItemSIZE
     \global\advance\TextSize by -\TextLeading
   \else
    \ifFirstZoneA
    \global\advance\ItemSIZE by \TextLeading
    \global\FirstZoneAfalse\fi
    \global\advance\TextSize by -\ItemSIZE
    \global\advance\TextSize by -\TextLeading
    \global\advance\ZoneBSize by -\ItemSIZE
    \global\advance\ZoneBSize by -\TextLeading
    \ifFigInZoneB\relax
     \else
     \ifdim\TextSize<3\TextLeading
     \global\ZoneAFullPagetrue
     \fi
    \fi
    \ChangeStatus{\StackPointer}{\Zone}
    \ifnum\Zone=\InZoneC \global\TabInZoneCtrue\fi
   \fi
  \global\advance\NextTable \@ne
   \Print{This table has been placed.}
  \else
  \Print{No space available for this table...holding over.}
   \Print{}
   \global\MoreTablesfalse
  \fi
}


\def\CheckFitInZone{%
{\advance\TextSize by -\ItemSIZE
 \advance\TextSize by -\TextLeading
 \ifFirstSingleItem
  \advance\TextSize by \TextLeading
 \fi
 \ifnum\Zone=\InZoneA\relax
  \else \advance\TextSize by -\ZoneBAdjust
 \fi
 \ifdim\TextSize<3\TextLeading \global\ItemFits=\No
 \else \global\ItemFits=\Yes\fi}
}

\def\BeginOpening{%
  \thispagestyle{titlepage}%
  \global\setbox\ItemBOX=\vbox\bgroup%
    \hsize=\PageWidth%
    \hrule height \z@
    \ifsinglecol\vskip 6pt\fi 
}

\let\begintopmatter=\BeginOpening  

\def\EndOpening{%
  \One
  \egroup
  \ifsinglecol
    \box\ItemBOX%
    \vskip\TextLeading plus 2\TextLeading
    \@noafterindent
  \else
    \ItemNUMBER=\z@%
    \ItemTYPE=\Figure
    \ItemSPAN=\Double
    \ItemSTATUS=\InStack
    \JoinStack
  \fi
}


\newif\if@here  \@herefalse

\def\no@float{\global\@heretrue}
\let\nofloat=\relax 

\def\beginfigure{%
  \@ifstar{\global\@dfloattrue \@bfigure}{\global\@dfloatfalse \@bfigure}%
}

\def\@bfigure#1{%
  \par
  \if@dfloat
    \ItemSPAN=\Double
    \TEMPDIMEN=\PageWidth
  \else
    \ItemSPAN=\Single
    \TEMPDIMEN=\ColumnWidth
  \fi
  \ifsinglecol
    \TEMPDIMEN=\PageWidth
  \else
    \ItemSTATUS=\InStack
    \ItemNUMBER=#1%
    \ItemTYPE=\Figure
  \fi
  \bgroup
    \hsize=\TEMPDIMEN
    \global\setbox\ItemBOX=\vbox\bgroup
      \eightpoint\nostb@ls{10pt}%
      \let\caption=\fig@caption
      \ifsinglecol \let\nofloat=\no@float\fi
}

\def\fig@caption#1{%
  \vskip 5.5pt plus 6pt%
  \bgroup 
    \eightpoint\nostb@ls{10pt}%
    \setbox\TEMPBOX=\hbox{#1}%
    \ifdim\wd\TEMPBOX>\TEMPDIMEN
      \noindent \unhbox\TEMPBOX\par
    \else
      \hbox to \hsize{\hfil\unhbox\TEMPBOX\hfil}%
    \fi
  \egroup
}

\def\endfigure{%
  \par\egroup 
  \egroup
  \ifsinglecol
    \if@here \midinsert\global\@herefalse\else \topinsert\fi
      \unvbox\ItemBOX
    \endinsert
  \else
    \JoinStack
    \Print{Processing source for figure \the\ItemNUMBER}%
  \fi
}


\newbox\tab@cap@box
\def\tab@caption#1{\global\setbox\tab@cap@box=\hbox{#1\par}}

\newtoks\tab@txt@toks
\long\def\tab@txt#1{\global\tab@txt@toks={#1}\global\table@txttrue}

\newif\iftable@txt  \table@txtfalse
\newif\if@dfloat    \@dfloatfalse

\def\begintable{%
  \@ifstar{\global\@dfloattrue \@btable}{\global\@dfloatfalse \@btable}%
}

\def\@btable#1{%
  \par
  \if@dfloat
    \ItemSPAN=\Double
    \TEMPDIMEN=\PageWidth
  \else
    \ItemSPAN=\Single
    \TEMPDIMEN=\ColumnWidth
  \fi
  \ifsinglecol
    \TEMPDIMEN=\PageWidth
  \else
    \ItemSTATUS=\InStack
    \ItemNUMBER=#1%
    \ItemTYPE=\Table
  \fi
  \bgroup
    \eightpoint\nostb@ls{10pt}%
    \global\setbox\ItemBOX=\vbox\bgroup
      \let\caption=\tab@caption
      \let\tabletext=\tab@txt
      \ifsinglecol \let\nofloat=\no@float\fi
}

\def\endtable{%
  \par\egroup 
  \egroup
  \setbox\TEMPBOX=\hbox to \TEMPDIMEN{%
    \hss
    \vbox{%
      \hsize=\wd\ItemBOX
      \ifvoid\tab@cap@box
      \else
        \noindent\unhbox\tab@cap@box
        \vskip 5.5pt plus 6pt%
      \fi
      \box\ItemBOX
      \iftable@txt
        \vskip 10pt%
        \eightpoint\nostb@ls{10pt}%
        \noindent\the\tab@txt@toks
        \global\table@txtfalse
      \fi
    }%
    \hss
  }%
  \ifsinglecol
    \if@here \midinsert\global\@herefalse\else \topinsert\fi
      \box\TEMPBOX
    \endinsert
  \else
    \global\setbox\ItemBOX=\box\TEMPBOX
    \JoinStack
    \Print{Processing source for table \the\ItemNUMBER}%
  \fi
}

\def\UnloadZoneA{%
\FirstZoneAtrue
 \Iteration=\z@
  \loop
   \ifnum\Iteration<\LengthOfStack
    \GetItemSTATUS{\Iteration}
    \ifnum\ItemSTATUS=\InZoneA
     \GetItemBOX{\Iteration}
     \ifFirstZoneA \vbox to \BodgeHeight{\vfil}%
     \FirstZoneAfalse\fi
     \unvbox\ItemBOX\ItemSep
     \LeaveStack{\Iteration}
     \else
     \advance\Iteration \@ne
   \fi
 \repeat
}

\def\UnloadZoneC{%
\Iteration=\z@
  \loop
   \ifnum\Iteration<\LengthOfStack
    \GetItemSTATUS{\Iteration}
    \ifnum\ItemSTATUS=\InZoneC
     \GetItemBOX{\Iteration}
     \ItemSep\unvbox\ItemBOX
     \LeaveStack{\Iteration}
     \else
     \advance\Iteration \@ne
   \fi
 \repeat
}


\def\ShowItem#1{
  {\GetItemAll{#1}
  \Print{\the#1:
  {TYPE=\ifnum\ItemTYPE=\Figure Figure\else Table\fi}
  {NUMBER=\the\ItemNUMBER}
  {SPAN=\ifnum\ItemSPAN=\Single Single\else Double\fi}
  {SIZE=\the\ItemSIZE}}}
}

\def\ShowStack{%
 \Print{}
 \Print{LengthOfStack = \the\LengthOfStack}
 \ifnum\LengthOfStack=\z@ \Print{Stack is empty}\fi
 \Iteration=\z@
 \loop
 \ifnum\Iteration<\LengthOfStack
  \ShowItem{\Iteration}
  \advance\Iteration \@ne
 \repeat
}

\def\B#1#2{%
\hbox{\vrule\kern-0.4pt\vbox to #2{%
\hrule width #1\vfill\hrule}\kern-0.4pt\vrule}
}


\newif\ifsinglecol   \singlecolfalse

\def\onecolumn{%
  \global\output={\singlecoloutput}%
  \global\hsize=\PageWidth
  \global\vsize=\PageHeight
  \global\ColumnWidth=\hsize
  \global\TextLeading=12pt
  \global\Leading=12
  \global\singlecoltrue
  \global\let\onecolumn=\relax
  \global\let\footnote=\sing@footnote
  \global\let\vfootnote=\sing@vfootnote
  \ninepoint 
  \message{(Single column)}%
}

\def\singlecoloutput{%
  \shipout\vbox{\PageHead\pagebody\PageFoot}%
  \advancepageno
  \ifplate@page
    \shipout\vbox{%
      \sp@pagetrue
      \def\sp@type{plate}%
      \global\plate@pagefalse
      \PageHead\vbox to \PageHeight{\unvbox\plt@box\vfil}\PageFoot%
    }%
    \message{[plate]}%
    \advancepageno
  \fi
  \ifnum\outputpenalty>-\@MM \else\dosupereject\fi%
}

\def\ItemSep{\vskip\ItemSepamount\relax}

\def\ItemSepbreak{\par\ifdim\lastskip<\ItemSepamount
  \removelastskip\penalty-200\ItemSep\fi%
}


\let\@@endinsert=\endinsert 

\def\endinsert{\egroup 
  \if@mid \dimen@\ht\z@ \advance\dimen@\dp\z@ \advance\dimen@12\p@
    \advance\dimen@\pagetotal \advance\dimen@-\pageshrink
    \ifdim\dimen@>\pagegoal\@midfalse\p@gefalse\fi\fi
  \if@mid \ItemSep\box\z@\ItemSepbreak
  \else\insert\topins{\penalty100 
    \splittopskip\z@skip
    \splitmaxdepth\maxdimen \floatingpenalty\z@
    \ifp@ge \dimen@\dp\z@
    \vbox to\vsize{\unvbox\z@\kern-\dimen@}
    \else \box\z@\nobreak\ItemSep\fi}\fi\endgroup%
}


\def\gobbleone#1{}
\def\gobbletwo#1#2{}
\let\footnote=\gobbletwo 
\let\vfootnote=\gobbleone

\def\sing@footnote#1{\let\@sf\empty 
  \ifhmode\edef\@sf{\spacefactor\the\spacefactor}\/\fi
  \hbox{$^{\hbox{\eightpoint #1}}$}\@sf\sing@vfootnote{#1}%
}

\def\sing@vfootnote#1{\insert\footins\bgroup\eightpoint\b@ls{9pt}%
  \interlinepenalty\interfootnotelinepenalty
  \splittopskip\ht\strutbox 
  \splitmaxdepth\dp\strutbox \floatingpenalty\@MM
  \leftskip\z@skip \rightskip\z@skip \spaceskip\z@skip \xspaceskip\z@skip
  \noindent $^{\scriptstyle\hbox{#1}}$\hskip 4pt%
    \footstrut\futurelet\next\fo@t%
}

\def\footnoterule{\kern-3\p@ \hrule height \z@ \kern 3\p@}

\skip\footins=19.5pt plus 12pt minus 1pt
\count\footins=1000
\dimen\footins=\maxdimen


\def\landscape{%
  \global\TEMPDIMEN=\PageWidth
  \global\PageWidth=\PageHeight
  \global\PageHeight=\TEMPDIMEN
  \global\let\landscape=\relax
  \onecolumn
  \message{(landscape)}%
  \raggedbottom
}


\output{%
  \ifLeftCOL
    \global\setbox\LeftBOX=\vbox to \ZoneBSize{\box255\unvbox\ZoneBBOX}%
    \global\LeftCOLfalse
    \MakeRightCol
  \else
    \setbox\RightBOX=\vbox to \ZoneBSize{\box255\unvbox\ZoneBBOX}%
    \setbox\MidBOX=\hbox{\box\LeftBOX\hskip\ColumnGap\box\RightBOX}%
    \setbox\PageBOX=\vbox to \PageHeight{%
      \UnloadZoneA\box\MidBOX\UnloadZoneC}%
    \shipout\vbox{\PageHead\box\PageBOX\PageFoot}%
    \advancepageno
    \ifplate@page
      \shipout\vbox{%
        \sp@pagetrue
        \def\sp@type{plate}%
        \global\plate@pagefalse
        \PageHead\vbox to \PageHeight{\unvbox\plt@box\vfil}\PageFoot%
      }%
      \message{[plate]}%
      \advancepageno
    \fi
    \global\LeftCOLtrue
    \CleanStack
    \MakePage
  \fi
}


\Warn{\start@mess}

\def\mnmacrosloaded{} 

\catcode `\@=12 


\fi

\input miniltx

\def\Gin@driver{dvips.def}
\input graphicx.sty

\resetatcatcode

\hoffset=-2pc\relax

\begintopmatter

\title {On the distances of planetary nebul{\ae}}

\author{Haywood Smith, Jr}

\affiliation{Department of Astronomy, University of Florida, 
Gainesville, Florida U.S.A. 32611} 

\shortauthor{H. Smith jr}
\shorttitle{Distances of planetary nebulae}
\abstract{Past calibrations of statistical distance scales for planetary nebul{\ae}  
have been problematic, especially with regard to \lq short' vs. \lq long' scales. 
Reconsidering the calibration process naturally involves examining the precision and 
especially the systematic errors of various distance methods. Here we present a 
different calibration strategy, new for planetaries, that is anchored by precise 
trigonometric parallaxes for sixteen central stars published by Harris et al.~(2007) 
of USNO, with four improved by Benedict et al.~using the {\it Hubble Space Telescope}. 
We show how an internally consistent system of distances might be constructed by 
testing other methods against those and each other. In such a way systematic errors 
can be minimized. 
\par
Several of the older statistical scales have systematic errors that can account for 
the short-long dichotomy. In addition to scale-factor errors all show signs of radius 
dependence, i.e.~the distance ratio [scale/true] is some function of nebular radius. 
These systematic errors were introduced by choices of data sets for calibration, by 
the methodologies used, and by assumptions made about nebular evolution.  The 
statistical scale of Frew and collaborators (2008, 2014) is largely free of these 
errors, although there may be a radius dependence for the largest objects. One set of 
spectroscopic parallaxes was found to be consistent with the trigonometric ones while 
another set underestimates distance consistently by a factor of two, probably because 
of a calibration difference. \lq Gravity' distances seem to be overestimated for nearby 
objects but may be underestimated for distant objects, i.e.~distance-dependent. Angular 
expansion distances appear to be suitable for calibration after correction for 
astrophysical effects (e.g.~Mellema 2004). We find extinction distances to be often 
unreliable individually though sometimes approximately correct overall (total sample). 
\par
Comparison of the {\it Hipparcos} parallaxes (van Leeuwen 2007) for large planetaries 
with our \lq best estimate' distances confirms that those parallaxes are overestimated 
by a factor $2.5$, as suggested by Harris et al.'s result for PHL 932. There may be 
negative implications for {\it Gaia} parallaxes for these objects. We suggest a 
possible connection with the much smaller overestimation recently shown for the 
{\it Hipparcos} Pleiades parallaxes by Melis et al. (2014). 
}

\keywords {stars:distances -- ISM:planetary nebul{\ae} -- 
methods:statistical -- astrometry}

\maketitle

\section{1\hdskip Introduction}
\subsection{1.1\hdskip Brief history of calibration of statistical distances.}
By the mid-twentieth century, when accurate distances could not generally be obtained for 
planetary nebul{\ae} with the usual methods (e.g.~trigonometric and spectroscopic 
parallax), recourse was had to methods based on uniformity assumptions about the physical 
properties of the nebul{\ae}.  The hope with these \lq statistical' methods was that the 
individual objects' properties do not greatly deviate from the assumed universal value.  
One was the Shklovsky (1956) method, based on the assumption of identical ionized mass 
$M_i$ for all planetaries; it used recombination-line theory to obtain a relation between 
H$\beta$ surface brightness $S_\beta$  (presumably distance-independent except for 
extinction) and radius $R$ to be used with the angular diameter $\varphi$ in estimating 
distance. Shklovsky showed mathematically that the distance estimate so obtained is fairly 
insensitive to the value of $M_i$. 
\par
Later the Shklovsky method was modified and refined, as for example with postulated 
universal relations between $M_i$ (no longer assumed constant; cf.~Pottasch 1980; Maciel 
\& Pottasch 1980) and $R$ or between 5 GHz brightness temperature $T_b$ (presumably 
unaffected by extinction, unlike $S_\beta$) and $R$ (cf.  Daub 1982). Yet calibration 
remained difficult because of a dearth of accurate individual distance estimates. For a 
long time one had only a small collection of miscellaneous data, some of dubious quality 
and hardly any of high precision, to use for calibration.  In addition to the inaccuracies 
inherent in each kind there can be systematic errors differing from one kind to 
another or even from one data set to another for the same kind. 
\par
Because of the past scarcity of high-quality data the practice in calibrating statistical 
scales has usually been to follow one of two strategies: the inclusive strategy, where one 
simply includes all (or almost all) the various kinds of data in the calibration set, or 
the eclectic strategy, choosing only the \lq best' determinations for the calibration set. 
With the former strategy one hopes that the various errors, random and systematic, will 
average out. In fact the result is likely a substantially larger uncertainty than the formal 
errors lead one to expect, and there may be some residual systematic error also. The latter 
strategy is likewise potentially vulnerable to systematic error; indeed, the narrower is the 
selection the less likely that systematic errors will cancel out. Historically, then, 
different choices of data, differences in weights given to the various data, and different 
calibration methods have produced a sizeable range of calibrations for these scales, just 
as one would expect when there are systematic errors. 
\par
Broadly speaking, statistical scales have divided into \lq short' and \lq long,' the two 
groups typically differing by a factor of the order of two (cf.~e.g.~Phillips 2002, hereafter 
Ph02).  An example of the former is the scale of Cahn, Kaler, \& Stanghellini (1992, hereafter 
CKS); the latter is exemplified by Zhang's (1995, hereafter Z95) scale. This dichotomy can 
actually be traced back at least as far as O'Dell (1962; hereafter O62) for the \lq short' 
scale and Seaton (1966; hereafter S66) for the \lq long' one.  
\par
During the past few decades more and better data have become available. For example, already 
in estimating the local space density of planetaries Pottasch (1996, hereafter P96) made use 
of (among others) eight spectroscopic parallaxes of companions of central stars, six distance 
estimates from angular expansion rates, and thirty from extinction (either line or continuum) 
as a function of distance. Ciardullo et al.~(1999, hereafter C99) used the Hubble Space 
Telescope to search for more central star companions, considerably augmenting the number of 
spectroscopic parallaxes. The angular expansion method, originally applied to optical images, 
has been extended to radio images with the VLA (Terzian 1980, Masson 1986; cf.~Terzian 1997, 
hereafter T97); the optical version has been improved with the replacement of photographic 
plates by CCD's and the use of {\it HST} (e.g.~Reed et al.~1999, Palen et al.~2002). An 
astrophysical method based on fitting central star spectral line profiles to those from 
stellar atmosphere models and matching the properties to evolutionary tracks has been 
developed (M\'endez et al.~1988) yielding what are termed \lq gravity' distances (referring 
to surface gravity) that can be used for calibration.  Lastly, new techniques have been 
applied to measuring central stars' trigonometric parallaxes, finally bringing those within 
reach. Parallaxes have been obtained using the {\it Hipparcos} satellite (Acker et al.~1998, 
hereafter A98), the {\it HST} fine guidance sensors (Benedict et al.~2003), and ground-based 
CCD cameras (Pier et al.~1993; Harris et al.~1997, hereafter H97; and Guti\'errez-Moreno et 
al.~1999). 
\par
The trigonometric parallax method has the virtues that it is geometrical and thus direct 
and, in principle at least, is model-independent; at least, it involves no astrophysical 
assumptions or modelling. The parallax should be valid for the nebula provided that the 
central star is correctly identified and, if needed, the correction for the reference 
stars' parallaxes (i.e., relative to absolute) is done properly. There is also no 
need to correct for interstellar extinction. While the method's applicability is 
necessarily limited at present to nearby planetaries, it can be used to evaluate and/or 
calibrate other methods of greater reach. In the near future the range is expected to be 
greatly extended because of the {\it Gaia} observatory (Perryman et al. 2001; Manteiga et al. 
2012; Manteiga et al. 2014); however, note the remark at the beginning of Section 8.3. 
\par
Unfortunately five of the nineteen original {\it Hipparcos} central star parallaxes in A98 
were negative, while the remainder were not very precise, with a median relative parallax 
error $\lambda\equiv\sigma_\pi^\prime/\pi^\prime$ of $0.66$. (Here as usual $\pi^\prime$ is 
the measured parallax and $\sigma_\pi^\prime$ is the estimated standard error of the 
parallax; the corresponding true values are $\pi$ and $\sigma_\pi$ resp.)  The median 
$\lambda$ for the three obtained by Guti\'errez-Moreno et al.~was almost the same, $0.69$, 
while the H97 ones were much better, with median $\lambda$ of $0.34$, but on the whole still 
not highly precise.  The {\it HST} fine guidance sensors are capable of very high precision 
but until recently had yielded only one parallax measurement for a planetary. 
\par
While some of the notation we use is standard, much -- e.g.~the use of $\lambda$ for relative 
parallax error -- is not and likely is unfamiliar to the reader. At the end there is an 
Appendix with a list containing definitions and first locations in the text. 

\subsection{1.2\hdskip Accurate parallaxes and the \lq anchor' strategy for calibration.}
In the past several years the situation has improved considerably. An expanded sample 
($N=16$) of high-quality CCD parallaxes was published by the USNO group (Harris et al.~2007, 
hereafter H07). These parallaxes have a median error of $0.42$ mas and a median 
$\lambda$ of $0.17$, an improvement of a factor of two over their previous work. More 
recently four of these objects were studied using {\it HST} (Benedict et al.~2009, hereafter 
B09). The precision of those measurements is even greater, with median error $0.23$ mas and 
median $\lambda = 0.08$. The results of the two studies are in generally good agreement: The 
median parallax ratio H07/B09 is $1.17$ and the mean is $1.19\pm 0.09$, indicating that 
there might be a slight systematic difference between the two. We will discuss this question 
in the next section, arguing that there is in fact no significant systematic difference. 
\par
We believe that accurate trigonometric parallaxes can serve as a solid foundation on which to 
erect an interlocking structure of distance determinations from various methods. This idea is 
not new; of necessity that is largely what happened with stellar distances, and in O62 O'Dell 
lamented the absence of astrometric data to fill precisely this r\^ole with planetaries. For 
a long time the inclusive and eclectic strategies appeared to be the only choices. We contend 
that space observatories and CCD cameras have changed that. 
\par
In this paper we demonstrate what we term the \lq anchor' strategy for calibration. Our 
calibration is anchored by the parallaxes, which serve to check other methods which can be used 
to verify still others, and so forth. For our strategy to succeed it is essential that no 
appreciable systematic errors be present in the parallax data or be introduced by our 
methodology. We then take pains to eliminate or mitigate any systematic errors in the other data 
types by comparing those with the parallax data, either directly or, if need be, indirectly and 
applying corrections or modifying our techniques. 
\par
In order to be able to track down systematic error it is highly desirable to have large, 
reasonably homogeneous data sets. If instead one has merely a hodgepodge of meagre data from 
different sources, using different instruments and/or reduction methods, it can be difficult to 
tease out any systematic differences. As an extreme example, with only one parallax obtained from 
{\it HST} (as was the case in 2007) one could not really compare that approach with the CCD one. 
\par
For convenience the parallaxes for the USNO sample are presented in Table 1 along with 
corresponding angular diameters $\varphi$ and $R$ values. Parallax values are from H07 
except for the ones marked with asterisks, which are weighted means of the H07 and B09 
values taken from the latter source. Angular diameters are mostly optical values taken 
from the Strasbourg-ESO Catalog (Acker et al.~1992, hereafter A92). For NGC 7293 we 
have used the radio value rather than the optical one in order to leave out the faint outer 
halo; it is almost identical to the optical value 654 arcsec given in O'Dell (1998). We 
likewise have ignored the very faint extended halo found for PG 1034+001 (Rauch, Kerber, 
\& Pauli 2004) and instead used the original value from Hewett et al.~(2003). 
The values for RE 1738+665 and Ton 320 are from Tweedy and Kwitter (1996), while the value 
for Sh 2-216 is from Tweedy, Martos, \& Noriega-Crespo (1995). Our $\varphi$ values are in 
most cases fairly close to those used with the statistical scales we consider; correcting for 
obvious errors the mean ratio of those to ours is $1.00\pm 0.11$ (s.d.) for CKS and 
$0.96\pm 0.07$ (s.d.) for Z95. For the mean statistical scale of Frew (2008; hereafter F08) 
the mean is $1.07\pm 0.03$ and the median $1.04$, indicating ours are slightly smaller.   

\begintable{1}
\caption{{\bf Table 1.} Observational data and radii for the USNO trigonometric 
parallax sample of H07}
{\settabs 4 \columns
\vskip 0.1 in
\hrule
\vskip 0.1 in
\+Name&$\pi^\prime$ (mas)&$\varphi$ (arcsec)&$R$ (pc)\cr
\vskip 0.1 in
\hrule
\vskip 0.1 in
\+A 7&$1.48\pm 0.42$&760&$1.25$\cr
\+A 21&$1.85\pm 0.51$&615&$0.81$\cr
\+A 24&$1.92\pm 0.34$&355&$0.45$\cr
\+A 31&$1.61\pm 0.21*$&970&$1.46$\cr
\+A 74&$1.33\pm 0.63$&830&$1.51$\cr
\+DeHt 5&$2.90\pm 0.15*$&530&$0.44$\cr
\+HDW 4&$4.78\pm 0.40$&104&$0.053$\cr
\+NGC 6720&$1.42\pm 0.55$&76&$0.13$\cr
\+NGC 6853&$2.47\pm 0.16*$&402&$0.32$\cr
\+NGC 7293&$4.66\pm 0.27*$&660&$0.34$\cr
\+PG 1034+001&$4.75\pm 0.53$&7200&$3.68$\cr
\+PHL 932&$3.36\pm 0.62$&275&$0.20$\cr
\+PuWe 1&$2.74\pm 0.31$&1200&$1.06$\cr
\+RE 1738+665&$5.91\pm 0.42$&3600&$1.47$\cr
\+Sh 2-216&$7.76\pm 0.33$&5840&$1.83$\cr
\+Ton 320&$1.88\pm 0.33$&1800&$2.32$\cr
\vskip 0.1 in
\hrule
\vskip 0.1 in
}
\tabletext{Parallaxes with asterisks are means of H07 
and B09 taken from B09. For information about the 
values for $\varphi$ please see the text.}
\endtable

\par
We present evidence below that the H07 parallaxes themselves are with one exception free of 
systematic error such as might arise from the use of two different CCD cameras and are 
otherwise consistent with the B09 parallaxes. Of course systematic error can be introduced 
by the methodology employed when using parallaxes, e.g.~Lutz-Kelker type bias; that and 
others will be considered in Section 2. 
\par
Two important limitations of the H07 sample are evident in Table 1. First, no object is 
likely to be more distant than 1 kpc. Hence the H07 sample by itself is unsuited to 
exploring distance dependence, i.e.~ the distance ratio [scale/true] depending upon distance. 
Second, most of them are fairly large. This is to be expected, for planetaries with low surface 
brightness $S$ should cause relatively little interference with position measurements of the 
(often faint) central stars, and large $R$ goes with low $S$. Indeed, there were obviously 
problems with the {\it Hipparcos} measurements for planetaries having small angular sizes and 
high surface brightnesses, as noted in A98.  There is only one H07 object smaller in $R$ than 
$0.1$ pc, and it may not be a planetary nebula, as discussed below.  There is a preponderance 
of objects with low $T_b$ at 5 GHz as well, the sample values (shown below) almost all less 
than 1 K. For this reason the H07 sample is of limited usefulness by itself in assessing any 
systematic error depending on $R$ or $T_b$.  We show below an example of a distance dependence, 
and radius dependence (distance ratio depending on nebular radius) is common.
\par
The calibration of a statistical scale built upon a $T_b$-$R$ or $S$-$R$ relation involves 
estimation of at minimum two parameters, a zero point and (for log-log relations) a slope, 
and in principle the slope may vary with $R$. Extrapolating the slope found with the H07 sample 
to smaller $R$ risks introducing a radius dependence into the scale if the true slope differs. 
On the other hand, use of a different sample such as a set of spectroscopic parallaxes to fix 
the relation at smaller $R$ injects the problem of heterogeneity with its potential for 
systematic error, for example with a calibration for the spectroscopic parallaxes that is 
inconsistent with the trigonometric ones (an example to be provided below). 
\par
We can use the H07 sample to indirectly evaluate other methods by means of an intermediary 
distance scale. If the method to be evaluated can be assumed to have no radius or distance 
dependence (e.g.~with spectroscopic parallax) and if each sample has a substantial presence 
within a given limited range in $R$ that method's relative precision and error in zero point 
can be estimated. In this way it might be possible to establish an internally consistent 
calibration over a wider range in $R$ and estimate the variation in slope within that range, 
if any. The process can then be repeated to further extend the range. What we are outlining is 
the stepwise construction of an interlocking system of distance determinations. 

\subsection{1.3\hdskip Classification complications arising with calibration.}
Another problem, similar to the one with combining distance estimates from different 
methods, is that there is copious evidence that the objects called planetary nebul{\ae} do 
not comprise a homogeneous class. To be sure, there are some objects that have been 
misclassified as planetaries (cf.~e.g.~Acker \& Stenholm 1990), and this seems to be true 
of a few objects in the H07 sample, as noted directly below. However, apart from such cases 
there appear to be differences among the nebul{\ae} in chemical composition, kinematic 
properties, and spatial distribution (Peimbert 1978) which are interpreted as being due to 
differing progenitor masses (cf.~e.g.~the discussion in Quireza, Rocha-Pinto, \& Maciel 
2007). Unfortunately only a relatively small fraction of known planetaries have been 
classified, and among the objects we consider here fewer than half have been, so we cannot 
pursue that thread in this paper given our modest sample sizes. 
\par
Frew \& Parker (2006) identified five objects in the H07 sample -- RE 1738+665, DeHt 5, PHL 
932, HDW 4, and PG 1034+001 -- as possibly being associated with ionised ISM rather than 
being true planetary nebul{\ae}. These identifications are supported by Frew \& Parker 
(2010) and specifically for PHL 932 by Frew et al.~(2010) and for PG 1034+001 by Chu et 
al.~(2012).  A 35 was classified as an H II region in F08; it is not a member of the H07 
sample but will be considered in connection with the {\it Hipparcos} parallaxes in Section 
6.  We provisionally accept the classification of these (following F08) as \lq imposters.' 
\par
In the next section we address the bias issue for trigonometric parallaxes, both for 
the overall distance scale of the H07 sample and for a possible sample-dependent bias 
of individual parallaxes caused by the Lutz-Kelker effect. We also consider the bias in 
overall distance ratio arising with sample selection based on statistical distances and 
biases with several estimators used when comparing distance scales. In Section 3 we use 
H07 parallaxes to test our representative examples of the \lq short' and \lq long' 
statistical scales, resp.~CKS and Z95, along with the mean F08 scale. We check the C99 
spectroscopic parallaxes against H07 indirectly, using the statistical scales, with 
nebul{\ae} in the range of overlap, namely $-1 < {\rm log}\ R < 0$; then we use both the 
H07 and C99 data sets to cover a fairly wide range in $R$ in our testing of the statistical 
scales.  We examine the $T_b$-$R$ relation generally in Section 4, adding data on Magellanic 
Cloud planetaries to extend the range in $R$ even further, and briefly relate the relation 
to evolution of the star+nebula systems. The gravity, angular expansion, and interstellar 
extinction distances are tested in Section 5. In Section 6 we consider the {\it Hipparcos} 
trigonometric parallaxes, which we have not used in this paper for testing or calibration, 
comparing them to our \lq best estimates' and demonstrate their systematic error for large 
objects. We propose an explanation of the long-standing dichotomy in statistical distance 
scales in the context of calibration strategies in Section 7. Our conclusions are summarized 
and discussed in the final section, where we make a few suggestions for future work. 

\section{2\hdskip The bias problem with trigonometric parallaxes}
\subsection{2.1\hdskip Some general considerations.}
\par
When earlier trigonometric parallax data suggested the gravity distances were overestimated, 
Napiwotzki (2001, hereafter N01) pointed out that the cause might instead be Lutz-Kelker 
bias. He carried out Monte Carlo simulations using a fairly realistic model of the spatial 
distribution of planetaries and found an underestimation of approximately the right amount. 
Strictly speaking, the bias he found was of the Trumpler-Weaver type (Trumpler \& Weaver 1953), 
since he imposed a lower limit on the measured parallaxes in his synthetic samples. 
Nevertheless some similar bias might affect a distance scale comparison in the absence of a 
lower limit. In N01 the bias was evaluated numerically because the classical Lutz-Kelker 
corrections, originally devised to counter Trumpler-Weaver bias, assumed a uniform spatial 
distribution, whereas N01's model was more complicated.
\par
Trumpler-Weaver bias is an example of truncation bias, which can arise when selecting a 
sample based on a limited range of values of quantities that have errors of measurement. 
The original idea was that when a sample of parallaxes is truncated at some lower limit 
$\pi^\prime_l$ the remaining parallaxes will have an excess of positive errors as well as 
a deficiency of negative errors and hence a positive bias. The discarded parallaxes will 
include some with true parallax $\pi>\pi^\prime_l$ (negative error), while some parallaxes 
with $\pi<\pi^\prime_l$ (positive error) will be erroneously included. Sometimes the 
sample is truncated according to $\lambda$ instead of $\pi^\prime$, with an upper limit 
instead of a lower limit; again the result is a positive bias (Arenou \& Luri 1999; Pont 
1999). Obviously if one wishes to avoid truncation bias the best way is to include all 
parallaxes regardless of relative error, or sign for that matter -- in other words, no 
truncation. 
\par
Another kind of bias that can arise is transformation bias, as when one converts measured 
parallaxes with their errors into distances or magnitudes (cf. e.g.~Smith \& Eichhorn 1996, 
hereafter SE96, and Brown et al.~1997). N01 noted that the conversion of parallaxes to 
distances is problematic when comparing distance scales, for this very reason. The problem 
can be avoided by not converting the parallax; indeed, working strictly in the parallax space 
has already been suggested (e.g.~Arenou \& Luri 1999). 
\par
The beauty of trigonometric parallaxes in evaluating the calibration of distance scales is 
that we {\it can} use them without conversion, just as they are, and indeed should do so. The 
distance of an object according to a given scale, $d_S^\prime$, can be multiplied by the 
parallax for that object to get the distance ratio for that object directly. In fact, 
precisely this was done in N01. We assume that $\pi^\prime$ has something like a normal 
probability density function (pdf) around $\pi$ so that one does not have to deal with the 
transformed error pdf for distance. If, as is sometimes reasonable, we also assume that the 
distance estimate being tested has a normal error pdf centred on the correct scaled value 
then it is easy to show that the expectation of the product 
$d^\prime_S\ \pi^\prime = {\cal R}_S$ will be the product of the individual expectation values, 
viz.~the true distance ratio (the scale factor $B$), and the expectation of the variance will be 
$$\sigma_{\cal R}^2=d_S^2\sigma_\pi^2+\pi^2\sigma_S^2+\sigma_\pi^2\sigma_S^2 \ \eqno{(1)}$$
where $d_S$ is the true distance $d$ multiplied by the actual value of $B$ for the given 
distance scale, $\sigma_S$ is the standard error in $d_S$, and $\sigma_\pi$ is the standard 
error of the parallax. As a matter of fact, these properties do not require that the pdf 
of the error in the scaled distance be normal in form or even symmetric; it is sufficient 
that the means of the error pdf's equal zero. If the relative errors in parallax and 
distance are small and both pdf's are normal the pdf of the product is very nearly normal; 
for larger errors it becomes noticeably skew, as will be shown below. 
\par
On the other hand, if instead of multiplying the distance estimate by the parallax one 
computes the distance from the parallax and then divides by the comparison distance, as 
was done in A98, the expectation of the result will in general not equal the reciprocal 
of the true distance ratio. The mean of the distances computed from the measured 
parallaxes (including errors) for a given object does not equal the true distance, as was 
shown in SE96, and the mean of the reciprocals of the distance estimates being compared 
will also be biased (transformation bias). Therefore the mean of the product of the two 
will in general be biased as well.
\par
To this point, if there is no systematic error in the data there should be no bias in the 
results from our procedure. However, it is natural to weight the results according to their 
precision, especially when that precision varies widely; for example, in A98 where the 
smallest positive $\lambda$ was $0.21$ and the largest $8.45$ the distance ratios obtained 
using {\it Hipparcos} parallaxes were roughly weighted according to $\lambda$, with greater 
weight for smaller $\lambda$. In the present instance, a logical choice of weight is the 
inverse of the expected variance $\sigma_{\cal R}^2$ of ${\cal R}_S$ for each object. 
However, that choice introduces weighting bias, discussed for example in Smith (2006), 
because the weight must be calculated using measured values instead of the true ones: 
$$\sigma_{\cal R}^{\prime 2}=d_S^{\prime 2}\sigma_\pi^{\prime 2}+\pi^{\prime 2}\sigma_S^{\prime 2}+
\sigma_\pi^{\prime 2}\sigma_S^{\prime 2}\ \eqno{(1a)}$$
where $d_S^\prime$ is the estimated distance and $\sigma_S^\prime$ is the error estimated 
for $d_S^\prime$. The uncertainty $\sigma^\prime_S$ is generally assumed to be proportional 
to $d_S^\prime$, so a positive error in distance contributes to an increase in the 
estimated variance, as may be seen from Eq.~(1a), and therefore decreases the weight, while 
a negative error increases the weight. Similarly, a positive error in the parallax 
increases the estimated variance and decreases the weight, while a negative error increases 
the weight. Consequently there tends to be a negative bias in the weighted mean distance 
ratio. 
\par
This weighting bias was explored with a set of idealized Monte Carlo simulations.  Three 
different distance distributions were used: (1) uniform density in 3-d (spherical), 
(2) uniform density in 2-d (disk), and (3) uniform density in 1-d (flat). All three 
were sampled to a maximum distance $d_{max}$, which was varied to evaluate its influence on 
the amount of bias. To simulate the statistical distance estimation method the randomly 
chosen true distance $d$ of each object was multiplied by the scale factor $B$ chosen for 
that run (value $0.7$, 2, or 4) to give $d_S$; then a random error was added sampled from a 
Gaussian having standard deviation $\sigma_d=\alpha_Sd_S$ where $\alpha_S$ is a measure of 
the relative distance error for that scale (chosen as $0.35$, $0.4$, or $0.45$ 
for each run) to give a pseudo-distance; the Gaussian was truncated at $\pm 2\sigma_d$. The 
true distance was also converted into a parallax and an error added selected at random from 
a Gaussian whose standard deviation was itself randomly chosen over a variable range, from 
$0.3$ to as much as $1.3$ mas. 

\begintable{2}
\caption{{\bf Table 2.} Unweighted and weighted mean and Phillips 
$\kappa$ as estimators of distance ratio based on parallaxes, using 
synthetic data with $B=2$ and a disk distribution}
\vskip 0.1 in
{\settabs 4 \columns
\hrule
\vskip 0.1 in
\+$d_{max}$ (pc)&Unweighted&Weighted&$\kappa$\cr
\vskip 0.1 in
\hrule
\vskip 0.1 in
\+$200$&$2.005\pm 0.008$&$1.969\pm 0.008$&$1.996\pm 0.009$\cr
\+$500$&$1.982\pm 0.006$&$1.827\pm 0.007$&$1.907\pm 0.013$\cr
\+$700$&$1.995\pm 0.008$&$1.742\pm 0.009$&$1.751\pm 0.032$\cr
\+$1000$&$2.001\pm 0.012$&$1.642\pm 0.012$&$1.903\pm 0.510$\cr
\+$1500$&$2.003\pm 0.015$&$1.521\pm 0.012$&$1.554\pm 0.246$\cr
\vskip 0.1 in
\hrule
}
\endtable

\par
A fairly representative set of results of simulations for ten samples of $N=1000$ stars 
each are shown in Table 2. The spatial distribution used for these was the disk 
distribution, $B$ was 2, the spread of the parallax errors was roughly from $0.3$ to $0.5$ 
mas, and $\alpha$ was $0.4$. As expected the unweighted mean distance ratio obtained by 
multiplying the pseudo-distances by the parallaxes generally gave the correct value for $B$ 
to within the uncertainty. On the other hand, the negative weighting bias showed up quite 
clearly in the weighted mean ratios. The amount of bias depends on the limiting distance 
because of the first and third terms on the rhs in Eq.~(1a). It becomes of order -15 per 
cent when the limiting distance approaches 1 kpc. 
\par
Also included in this table are results for Phillips's (Ph02) $\kappa$ estimator for the 
distance ratio, defined as 
$$\kappa\equiv {\sum_{i=1}^N d^\prime_{2,i}\over \sum_{i=1}^N d^\prime_{1,i}}\ \eqno{(2)}$$
where $d^\prime_{1,i}$ is the distance of the $i$th object in distance scale 1 and 
$d^\prime_{2,i}$ is the distance of the same object in scale 2. Here $d^\prime_1$ was 
calculated as $1/\pi^\prime$. Not only is there a bias with $\kappa$ which is a 
transformation bias arising from the conversion of the parallaxes, but the uncertainty is 
considerably greater than with the other two estimators, probably in large part because of 
the statistical instability of the conversion from $\pi^\prime$ to $d^\prime$ that was 
remarked upon in SE96. (In brief, the rare occurrence of $\pi^\prime$ values very near 
zero can cause extraordinarily large $d^\prime$ values, either positive or negative.)  
These behaviours are typical for $\kappa$ in our experiments, and the large uncertainty 
alone renders it clearly unsuitable for estimation of a distance ratio with distances 
based on trigonometric parallaxes in the denominator. 
\par
Having said this, we must point out that $\kappa$ is actually a very good measure of the 
mean distance ratio if one uses distances in the denominator which are {\it not} derived 
from parallaxes but instead have something like a normal error distribution. Table 3 
gives a typical set of results for the same spatial distribution and $B$ as in Table 2 
but for distances $d^\prime_1$ not derived from parallaxes, ones which have relative 
errors $\alpha_1=0.40$ (roughly comparable to the typical $\lambda$ for the parallaxes 
in the previous case) and the same relative error $\alpha_2$ for the second scale. We 
have not assigned weights to the values because the relative errors are all the same. 
Obviously the average of the individual distance ratios is biased, whereas $\kappa$ is 
not. The bias of the former is formally equivalent to the bias pointed out in SE96 for 
the mean of the distances obtained from parallaxes; fig.~1 of that paper suggests that 
for $\alpha = 0.4$ there should be a positive bias of roughly 20 per cent, while the actual 
figure is 18 per cent. (The difference might largely be due to the 2$\sigma$ truncation of 
the Gaussian in our experiments.)  Indeed we should expect $\kappa$ to be asymptotically 
unbiased, since the errors in both numerator and denominator are presumably symmetric 
when there is no truncation of the sample according to $d^\prime$, as in this case, and 
therefore positive and negative errors should tend to cancel out in both places. 
Incidentally, the same would be expected for the ratio of sums of parallaxes. 
Additionally we see in these experiments that $\kappa$ seems to have a smaller 
uncertainty than does the mean distance ratio. These conclusions are supported by a 
more extensive set of numerical experiments which we will not present here. Finally, 
for a given set of objects $\kappa$ is strictly transitive, i.e.~for three distance 
scales A, B, and C the overall distance ratio A/C equals the ratio A/B times the ratio 
B/C, a nice property not shared with many other estimators.
\par
There are two important reservations concerning $\kappa$, however. First, if the sample 
does not have a smooth distance distribution like the synthetic samples just considered 
but instead has one object with a distance that is much greater than those of the rest, 
its distance ratio will dominate the result. For example, there is a planetary nebula in 
the globular cluster M 15 whose distance is at least an order of magnitude greater than 
the typical distances of planetaries in the local solar neighborhood. More generally 
$\kappa$ gives higher weight to more distant objects. As a result, if there is a 
distance-dependent systematic error in the distance ratio $\kappa$ will tend to reflect 
the value appropriate to the most distant members of the sample. Hence there needs to 
be a modification to more nearly balance the contributions to the estimator from objects 
with widely differing distances. Below we propose a weighting scheme intended to do 
precisely that. Second, $\kappa$ is only a metric for comparing distance scales on 
average, not assessing how exactly individual distances in one scale follow those in 
another. The oft-used Pearson correlation coefficient $r$ is conventionally chosen for 
the latter purpose. By its form, however, it ignores any scale factor difference that 
might be present. It, too, suffers from a sensitivity to isolated extreme values. 
\begintable{3}
\caption{{\bf Table 3.} Unweighted mean, Phillips $\kappa$, and $\zeta$ (see text) 
as estimators of distance ratio comparing two statistical distance scales, using 
synthetic data with $B=2$ and a disk distribution}
{\settabs 4 \columns
\vskip 0.1 in
\hrule
\vskip 0.1 in
\+$d_{max}$ (pc)&Unweighted&$\kappa$&$\zeta$\cr
\vskip 0.1 in
\hrule
\vskip 0.1 in
\+ 200&$2.381\pm 0.021$&$2.008\pm 0.013$&$1.900\pm 0.011$\cr
\+ 500&$2.362\pm 0.011$&$1.998\pm 0.010$&$1.892\pm 0.007$\cr
\+ 700&$2.355\pm 0.018$&$1.971\pm 0.014$&$1.913\pm 0.011$\cr
\+ 1000&$2.375\pm 0.018$&$2.006\pm 0.011$&$1.904\pm 0.012$\cr
\+ 1500&$2.336\pm 0.016$&$1.984\pm 0.008$&$1.914\pm 0.010$\cr
\vskip 0.1 in
\hrule
}
\endtable
\par
To modify $\kappa$, we introduce weights $w_i$ to be applied to the distance values 
from the two scales {$d^\prime _{1,i}$} and {$d^\prime _{2,i}$}. We choose 
$w_i = 1/(d^\prime _{1,i}+d^\prime _{2,i})$; the form of our new estimator is now 
$$\zeta = {\Sigma_{i=1}^N w_i\ d_{2,i}^\prime\over\Sigma_{i=1}^N w_i\ d_{1,i}^\prime}\ .\eqno{(3)}$$ 
If again $B$ is the true distance ratio the terms in the numerator will each 
approximate $B/(B + 1)$ while those in the denominator will approximate $1/(B + 1)$. 
Except for the effects of the distance errors each object will contribute the same 
amount of information on the distance ratio, with no one object or small set of 
distant objects dominating the result. As with $\kappa$, $\zeta$ should to some 
extent tend to asymptotically approach $B$ as the distance errors $\delta d_2$ in the 
numerator (mostly) cancel and the same for the errors $\delta d_1$ in the denominator. 
However, in Table 3 we see that there is a negative bias in $\zeta$; it is a weighting 
bias caused by $\delta d_1$ and $\delta d_2$ affecting $w_i$. When the sum 
($\delta d_1 + \delta d_2$) is negative $w_i$ is increased compared to when it is 
positive. In the particular case shown it is 5 per cent independent of distance; in general 
it depends on $B$ as well as $\alpha_1$ and $\alpha_2$. The bias affects both 
numerator and denominator, which tends to mitigate the damage. Compared to the other 
biases we have seen it is small unless one or both $\alpha$'s are large, of order $0.4$ 
or more. We also note that unlike $\kappa$ $\zeta$ is not transitive. For the 
synthetic data we have examined the precision of $\zeta$ is quite comparable to that 
of $\kappa$. In our comparisons of distance scales we will use $\zeta$ as a check on 
$\kappa$. The amount of bias is not large, ranging up to 9 per cent for $B=2$, $d_{max} = 1.5$ kpc, 
$\alpha_1 = 0$, and $\alpha_2 = 0.4$. 
\par
Individual values of log $R$ are of interest in connection with both any possible radius 
dependence of distance ratios and their relation to log $T_b$ or log $S$.  Estimates of log 
$R$ based on parallaxes are affected by transformation bias (through the log function acting 
on distance $d^\prime$) together with a double truncation bias, a combination sometimes 
mistakenly thought of as universal for absolute magnitudes $M$: Lutz-Kelker bias 
(Lutz \& Kelker 1973). As the author pointed out (Smith 2003) the effect of the combination 
is not intrinsic and universal as originally claimed but rather depends on the 
characteristics of the particular sample; this fact is clearly demonstrated by figs.~3 and 4 
of that paper. Especially in fig.~4 one sees that the mean of the error in absolute 
magnitude $\Delta M$ caused by distance error, which in fact is proportional to the error in 
log $R$ for nebul{\ae}, is a function of $\lambda$ {\bf} not necessarily given by the 
negative of the classical Lutz-Kelker corrections. Indeed it is not necessary that there be 
any Trumpler-Weaver bias for the sample in order for this Lutz-Kelker effect to act on 
individual values, even though that bias is what the corrections were originally intended to 
cancel out. Whereas the sample shown in fig.~4 is formally truncated at $\lambda=0.175$ it 
is effectively limited by the apparent magnitude cutoff at $m_l=7$; therefore there is no 
truncation bias for the parallax sample as a whole. 
\par
In evaluating the individual bias as a function of $\lambda$ in the particular case 
we started with the (assumed) distribution of true parallaxes $g(\pi)$ and added 
errors $\epsilon_\pi$ selected randomly from a normal distribution with the value of 
$\sigma_\pi$ assumed to be the same for all stars independent of $\pi$. Computing 
the bias is then straightforward. In the real world, of course, the situation 
is quite different: One often does not know $g(\pi)$ or at best only approximately; 
almost certainly $\sigma_\pi$ will have a distribution which is different for each 
value of $\pi$; and one has only an estimate of $\sigma_\pi$, namely 
$\sigma_\pi^\prime$, when forming the ratio $\lambda$. This last is probably a 
relatively minor problem; seemingly it should result in nothing more than a \lq 
smearing' of the distribution compared to the true one. On the other hand, 
variations in the distribution of $\sigma_\pi$ and thus that of $\epsilon_\pi$ 
with $\pi$ introduce a higher degree of complexity. 
\par
The inversion problem, namely going from the (normalized) observed distribution 
$\phi(\pi^\prime,\sigma_\pi^\prime)$ to the underlying distribution 
$\Phi(\pi,\sigma_\pi)$, formally seems rather difficult. Instead we will merely 
try to find a model $\Phi$ which gives a fairly decent match to the observed $\phi$, 
keeping in mind that there are probably some other such functions $\Phi$ that will 
give a fit that is as good or possibly even a little better. For our present 
purposes it is likely unnecessary to obtain an optimum fit to $\phi$, however. 
For the H07 sample we will attempt to use the selection criteria (insofar as they 
can be inferred) together with some other information to constrain the model. 

\subsection{2.2 \hdskip Bias with the H07 parallax sample.}
\par
To the best of the author's knowledge the exact criteria used in choosing the actual 
H07 sample have never been published. The nearest approaches to a specific statement 
on this subject are a comment at the end of H97 about adding nebul{\ae} that \lq are 
likely to be at a distance closer than 500 pc' and the information on the model used 
to show that the bias is small which appears near the end of H07 (see next paragraph). 
There is no explicit truncation on the basis of either $\pi^\prime$ or $\lambda$, so 
one would expect no Trumpler-Weaver bias. To be sure, there was one planetary, A 29, 
which was dropped from the USNO program list because its astrometric solution was not 
stable, but the provisional value published in H97, $2.18\pm1.30$ mas, was not 
negative or unusually small. We believe that the faintness of its central star, 
$V=18.31$, rendered it unsuitable for parallax measurement. 
\par
The model used in H07 when considering possible bias was based on an underlying spatial 
distribution of planetaries similar to that of N01, namely an exponential 
$z$-distribution with a scale height $z_0$ of 250 pc  and selection according to 
statistical distance with upper limit 550 pc for a distance scale having $\alpha_S = 0.3$ 
and $B=1$.  Their model apparently gives a fairly good fit to the marginal distributions 
of $\pi^\prime$, $\sigma_\pi^\prime$, etc.~as indicated by the first and second moments. 
\par
However, such a model has a potential problem when one evaluates overall distance ratios 
for statistical distance scales. Truncation of the sample on the basis of statistical 
distance (not parallax) might well introduce a bias in the distance ratio because of an 
excess of negative distance errors together with a deficit of positive errors. This bias 
could arise to some extent even if the statistical scale used for selection is different 
from the one being tested, if both are based on essentially the same approach and use 
the same or related data (such as H$\beta$ or photored flux and 5 GHz flux density).  We 
emphasise that this problem has nothing to do with any bias of the trigonometric 
parallaxes themselves. 
\par
To assess this effect we have looked into the selection of the H07 sample as related in 
the USNO group's papers (Pier {\it et al.}~1993; H97; and H07). Briefly, half of the H07 
sample objects appear to have been selected because they were on the list of nearby 
planetary nebul{\ae} compiled by Terzian (1993, hereafter T93) with distances smaller than 
300 pc largely based on combinations of estimates from five variants of the Shklovsky 
method. We refer to this group as the T93 subsample; it is at risk of the selection bias 
described above.  The remaining objects in the H07 sample were selected for other reasons 
-- association with nearby white dwarfs such as Ton 320 (Tweedy \& Kwitter 1994), perhaps 
large $\varphi$ as with PG 1034+001 (Hewett et al.~2003), or possibly because it is a 
well-known planetary with a suitably faint central star (NGC 6720). The latter are not 
at risk of this bias.  However, five of these -- DeHt5, HDW 4, PG 1034+001, PHL 932, and 
RE 1738+665 -- have been classified as H II regions (see Sect.~1.1) and therefore cannot 
be used to test statistical scales (but can be used for others). Those five are our 
imposters (following F08); the three others are out non-T93 subsample and have 
non-statistical distance estimates ranging up to 700 or 800 pc.  

\beginfigure{1}
\includegraphics[height=6cm,width=7.5cm]{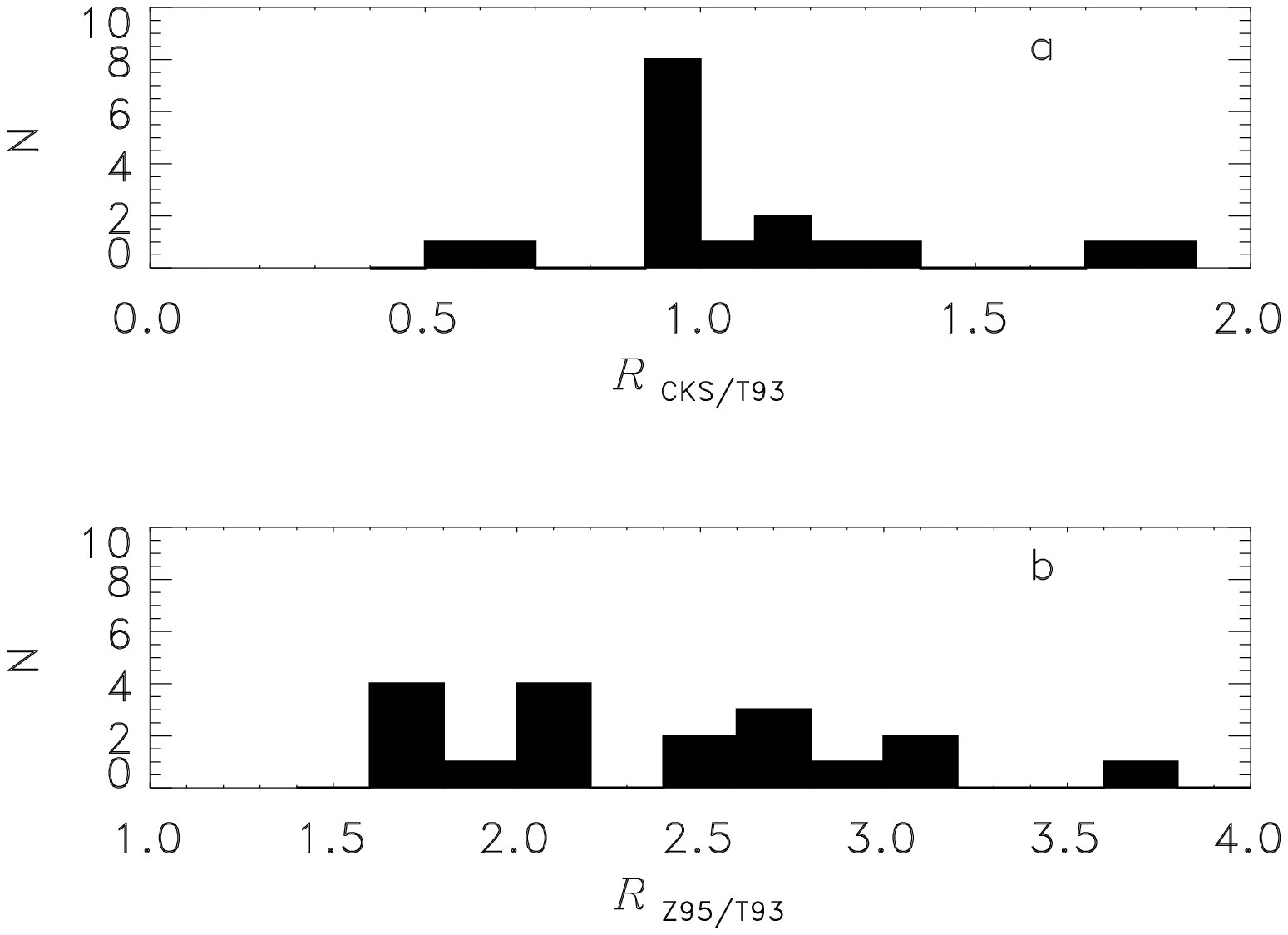}
\caption{{\bf Figure 1.} (a) Distribution of distance ratios 
${\cal R}_{\rm CKS/T93}$ for nebul{\ae} common to both those lists, with 
the mean value being $1.1$; (b) distribution of distance ratios 
${\cal R}_{\rm Z95/T93}$ for nebul{\ae} in common to both those lists, with 
the mean value being approximately $2.5$. Both values leave out the object 
LoTr 5, which has anomalously high values for both scales, respectively $15.74$ and 
$21.73$, and has been omitted from both figures. (A referee pointed out that $\varphi$ 
for this object is in error.)}
\endfigure

\par
The key issue here is the degree of correlation between whatever statistical distance scale 
is being studied and the T93 distances that were used (in part) for sample selection. 
Representative examples of the \lq short' and \lq long' statistical scales are CKS and Z95 
respectively.  They are also fairly  comprehensive, having two of the largest overlaps 
($\sim$50 per cent) with the H07 sample of the various scales. Fig.~1(a) shows the ratios 
of the CKS distance estimates to the T93 ones (for the entire overlap, not just the H07 
sample).  Essentially half the ratios lie between $0.9$ and $1.1$, indicating a very close 
connection between the two scales. By contrast, the ratios of the Z95 distances to the T93 
ones, shown in Fig.~1(b), are widely scattered around the mean value. Obviously there is 
little correlation between the Z95 and T93 distance estimates, so there should be no 
significant bias in that comparison.  
\par
To roughly estimate the effect of this bias with the CKS scale we have generated synthetic 
samples of 4000 nebul{\ae} resembling the H07 model, with an underlying disk distribution 
having scale height $z_0=250$ pc and selected according to a pseudo-distance limit with 
several different choices of $B$, the distance scale factor, and $\alpha_S$. Two values of 
this limiting pseudo-distance were chosen, 300 pc and 500 pc; anticipating our later 
results we used $\alpha_S=0.3,0.35,$ and $0.4$ and $B$ equal to $0.7$, $1.0$, and $2.0$.  
The results were found to be essentially the same for both limiting distances, namely an 
underestimation of $B$ by a factor $0.70$. The amount of bias depends significantly on 
$\alpha$: if $\alpha_S$ is $0.30$ the underestimation factor is $0.79$, whereas if it is 
$0.40$ this factor is $0.59$. The greater the relative spread in pseudo-distance around the 
true scaled value the greater the underestimation, as would be expected. 
\par
Following the reasoning laid out in the preceding paragraphs, we have generated multiple 
realizations of two synthetic subsamples which together model the H07 sample (with imposters 
removed). One subsample was selected from a model distribution very similar to that of H07 
(with $z_0=250$ pc) according to pseudo-distance with $\alpha_S = 0.35$, the limiting $d_S$ 
chosen as 300 pc, and $B=0.8$ as above, essentially imitating the T93 sample. The other 
subsample was chosen from the same underlying spatial distribution purely according to 
limiting true distance; the value 800 pc was chosen based on the parallax distribution for 
the non-T93 objects. The values of $\sigma_\pi$ were chosen at random between $0.3$ mas 
and $0.8$ mas for both synthetic samples. Figs.~2(a-c) show the parallax distribution, 
a plot of $\pi^\prime$ vs.~$\lambda$, and the distribution in galactic latitude $b$ 
for the T93 subsample together with corresponding plots for one realization of the model. 
The same plots are shown for the non-T93 objects and the second synthetic subsample in 
Figs.~3(a-c). In Figs.~2(b) and 3(b) the sharply defined lower envelope for the synthetic 
subsamples is caused by the sharp lower limit on $\sigma_\pi^\prime$, which is $0.3$ mas. 
Altogether the model seems to give a fairly good fit to the H07 points, but of course the 
latter are quite sparse, especially for the non-T93 subsample. Figs.~2(d) and 3(d) show 
the error in log $R$, $\Delta\ {\rm log}\ R$, resulting from the parallax errors as a 
function of $\lambda$ for the respective synthetic subsamples, with the filled circles 
being the means for bins of width $0.10$ in $\lambda$. (As noted above 
$\Delta\ {\rm log}\ R$ is proportional to $\Delta M$ that is commonly shown in such plots.)  
There is only a very slight underestimation of ${\rm log}\ R$ for small $\lambda$, whereas 
there is a modest overestimation at the largest values of $\lambda$. The nebul{\ae} in the 
H07 sample most likely to be affected by the latter are A 74 ($\lambda=0.47$) from the T93 
subsample and NGC 6720 ($\lambda=0.39$) from the non-T93 subsample.  Based on several 
realizations of the synthetic subsamples we estimate that these two objects need 
corrections in log $R$ of $-0.10$ and $-0.06$ respectively. These amounts are actually less 
than the respective errors in log $R$. 
\par
Similar plots for the imposters are not shown because they cannot have trustworthy 
statistical distances. While we have not used them for tests of statistical distances we 
have sometimes employed them in tests of other distance methods when it seemed appropriate. 
Simulations for them suggest the error $\Delta\ {\log}\ R$ is negligible in all cases. 
\par
Later we will briefly consider the effect of observational selection respecting $S$, which 
implies selection according to $R$. While the latter could in principle affect the form of 
$g(\pi)$ and hence the value of $\Delta$ log $R$ we do not consider NGC 6720 in particular 
to have been selected for the H07 sample in such a way. Hence we do not see a need to modify 
$g(\pi)$. 

\subsection{2.3 \hdskip Estimation of typical relative errors in statistical distances 
using trigonometric parallaxes.}
\par
We have so far mainly used assumed values of $\alpha_S$ for the statistical distance 
estimates. However, this quantity can at least in principle be estimated from the sample 
variance of the distance ratios on the assumption that the relative distance error 
$\alpha$ is the same for all nebul{\ae}. We rewrite the expression for the variance in 
terms of $\alpha_S$ in Eq.~(1), now for the entire sample, as 
$$\sigma_S^2={1\over N}\sum_{i=1}^N ({\cal R}_i-\bar{\cal R}_S)^2$$
$$\ \ \ \ \  = {1\over N} \sum_{i=1}^N(d_i^2\sigma_{\pi,i}^2+\pi_i^2\alpha_S^2
d_i^2+\sigma_{\pi,i}^2\alpha_S^2d_i^2)\ \eqno{(4)}$$
where $N$ is the sample size and $\bar{\cal R}_S$ is the unweighted mean estimate for the 
overall distance ratio for the given scale. We solve Eq.~(4) to obtain the estimate 
$\alpha_S^\prime$ (after substituting the observed values) from 
$$\alpha_S^\prime=\sqrt{{\sum_{i=1}^N[({\cal R}_i-\bar{\cal R}_S)^2-d_i^{\prime 2}\sigma_\pi^{\prime 2}]
\over\sum_{i=1}^N d_i^{\prime 2}(\pi^{\prime 2}+\sigma_{\pi,i}^{\prime 2})}}
\ .\eqno{(5)}$$
\par
Our idealized Monte Carlo experiments indicate that this approach can sometimes lead to an 
underestimation of $\alpha_S$ that depends on $d_{max}$. The reason is that the term 
$d_i^{\prime 2}\sigma_\pi^{\prime 2}=(d_i+\epsilon_{d,i})^2\sigma_\pi^{\prime 2}$ has on 
average an extra contribution due to the square of the error $\epsilon_{d,i}$ which 
increases with $d_{max}$ and is subtracted from the numerator but added to the denominator. 
There is also an effect due to the use of $\pi^\prime$ in place of $\pi$. As Table 4 
demonstrates, this underestimation can be appreciable. When the actual values are used in 
place of the observed ones the true $\alpha$ is approximately recovered. 
\par
The statistics in Table 4 are for relatively large samples. With a small sample such as 
those considered below Eq.~(5) may not have a real solution. Numerical experiments for 
samples with $N=10$, which is comparable to the typical sample sizes used here, indicate 
that the fraction of samples that do not yield a real solution increases with 
$d_{max}$ and with $\sigma_\pi$ and decreases with increasing $\alpha_S$. For example, for 
$\alpha_S=0.36$ and $d_{max}=700$ pc with typical error sizes only 11 of 100 samples failed, 
whereas for $\alpha=0.18$ that number was 36. We infer that such failure is suggestive of 
small $\alpha_S$. 
\begintable{4}
\caption{{\bf Table 4.} Estimation of $\alpha_S$ for ten synthetic samples 
($N=1000$) with different values of $d_{max}$; $\alpha_{true}=0.36$.} 
{\settabs 3 \columns
\vskip 0.1 in
\hrule
\vskip 0.1 in
\+$d_{max}$ (pc)&$\alpha_{S}^\prime$&$\alpha_{S}^\prime/\alpha_{S}$\cr
\vskip 0.1 in
\hrule
\vskip 0.1 in
\+200&$0.333\pm 0.003$&$0.925\pm 0.008$\cr
\+500&$0.326\pm 0.002$&$0.906\pm 0.006$\cr
\+700&$0.319\pm 0.003$&$0.886\pm 0.008$\cr
\+1000&$0.312\pm 0.009$&$0.867\pm 0.025$\cr
\+1500&$0.274\pm 0.006$&$0.761\pm 0.017$\cr
\vskip 0.1 in
\hrule
}
\endtable
\par
Eq.~(1) can be expressed in terms of ${\cal R}_S$, $\alpha$, and $\lambda$ as 
$$\sigma_{\cal R}^2={\cal R}_S^2(\alpha^2+\lambda^2+\alpha^2\lambda^2)
\ .\eqno{(6)}$$
Obviously the third term in parentheses is not very important so long as both $\alpha$ 
and $\lambda$ are much less than unity, so in that case we are usually justified in 
ignoring it. If $\alpha$ is indeed around $0.4$ then $\lambda$ is not very important for 
$\sigma_{\cal R}$ if it is smaller than that, as with most of the H07 parallaxes, but it 
is dominant for the majority of {\it Hipparcos} parallaxes recalling that the median 
$\lambda$ for those is $0.66$). 

\beginfigure{2}
\includegraphics[height=7cm,width=7.5cm]{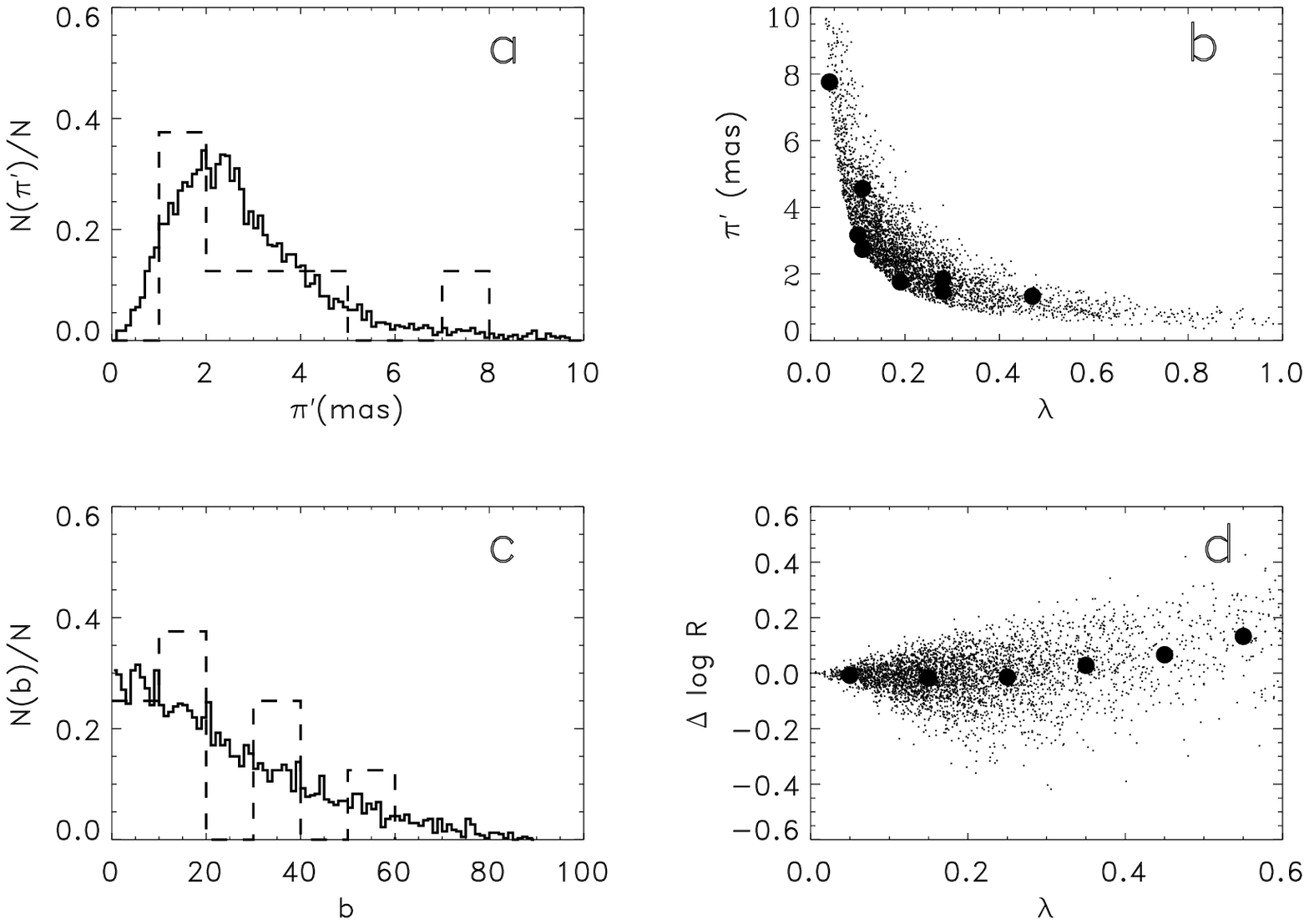}
\caption{{\bf Figure 2.} Representative properties of the T93 synthetic 
sample: (a) distribution of $\pi^\prime$; {\it dashed histogram} is for H07 
sample; (b) plot of $\pi^\prime$ vs.~$\lambda$ with the 
{\it filled circles} representing the H07 objects; (c) the distribution in 
galactic latitude $b$, H07 sample as in (a); and (d) 
the plot of $\Delta$ log $R$ {\it vs.} $\lambda$, with the {\it filled circles} 
now being the means for bins of width $0.1$ in $\lambda$. For more details 
about both the synthetic samples see the text.}
\endfigure
\beginfigure{3}
\includegraphics[height=7cm,width=7.5cm]{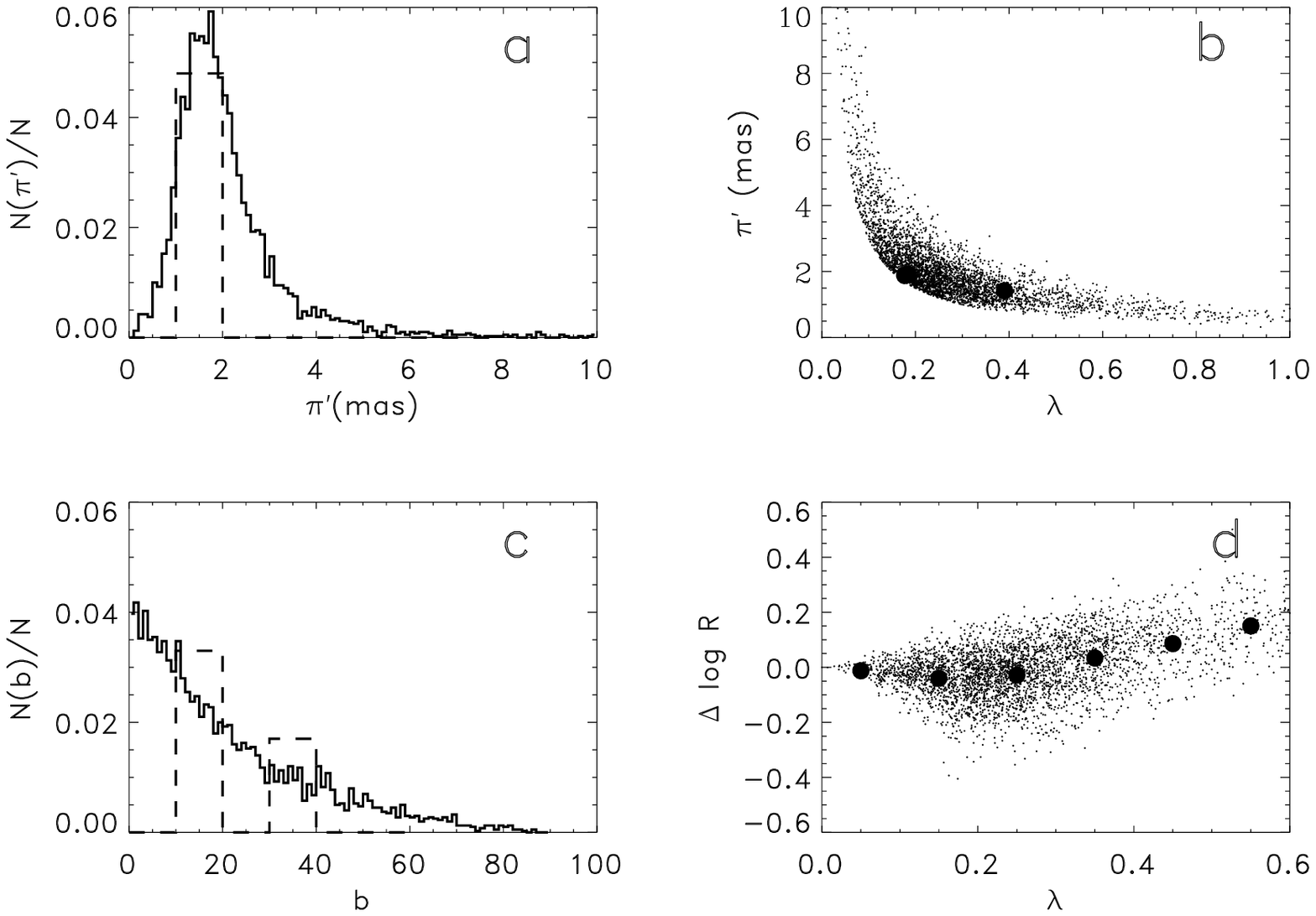}
\caption{{\bf Figure 3.} Same as in Fig. 2 but for the non-T93 subsample.}
\vskip 0.1 in
\endfigure

\par
We now consider the estimated characteristics of synthetic samples which model the H07 
subsamples referred to above. The CKS and Z95 scales are treated separately because 
we believe the sample selection distances are correlated for the former but not the latter. 
Five synthetic samples with $N=4000$ were used for each case to evaluate the bias. If 
the entire T93 subsample had been selected using the CKS distances with an assumed 
$B=0.8$ and $\alpha_S = 0.35$ the unweighted mean distance ratio would be $0.575\pm 0.001$ 
and the weighted mean $0.425\pm 0.001$; the weighting bias is $-0.15$. The reduction factor 
for the unweighted mean arising from the correlation in distance, $0.72$, is close to the 
value of $0.70$ found with our idealized calculations.  For the same subsample but with a 
set of distances different from that used for selection (yet still having the same 
underlying spatial distribution) the corresponding values would be $0.804\pm 0.002$ (no 
appreciable bias, as expected) and $0.511\pm 0.004$ (weighting bias $-0.29$). The non-T93 
subsample gives values $0.798\pm 0.001$ (practically no bias) and $0.584\pm 0.002$ 
(weighting bias $-0.21$) respectively. 
\par
In the case of Z95 we use instead the value $B = 1.5$ while keeping $\alpha_S$ at $0.35$. 
The unweighted mean for the T93 sample, once again with uncorrelated statistical distances, 
is $1.507\pm 0.003$ and the weighted mean $1.103\pm 0.005$ (weighting bias $-0.40$). The 
corresponding values for the non-T93 sample are $1.495\pm 0.001$ and $1.094\pm 0.003$ 
respectively (weighting bias the same). As will become evident later, these values ought to 
be fairly close for the two cases. 
\par
The estimation of $\alpha_S$ using Eq.~(5) is complicated; not only is there underestimation 
caused by the errors in distance and parallax and the numerical difficulty, but when the 
distances are correlated there can be overestimation instead of underestimation. With $B=0.8$ 
and $\alpha_S=0.35$ (appropriate to CKS) the T93 subsample yields 
$\alpha_S^\prime = 0.364\pm 0.001$ with correlation and $0.311\pm 0.003$ without. The 
magnitude of these effects depends on $\alpha$: For $\alpha_S=0.3$ with correlation we have 
$0.295\pm 0.002$ (apparently the effect of the errors is virtually cancelled out by the 
effect of the distance correlation) and without correlation we get $0.244\pm 0.002$, while 
when $\alpha_S = 0.4$ the respective values are $0.457\pm 0.006$ (evidently the correlation 
has the upper hand) and $0.321\pm 0.002$. The non-T93 subsample with $\alpha_S = 0.35$ yields 
$\alpha_S^\prime = 0.283\pm 0.001$ or $0.81$ times the true value; the latter ratio only 
changes slightly when one increases or decreases the true $\alpha$ by $0.05$, and it stays 
the same when one goes to $B=1.5$ at $\alpha_S = 0.35$ (the Z95 case). The T93 subsample with 
$B=1.5$ (for the tested distances, not those used for selection) gives a value 
$0.276\pm 0.003$, essentially the same as for the non-T93 subsample. 
\subsection{2.4 \hdskip Homogeneity of the H07 parallax sample}
\par
The original H07 data were obtained using two different CCD cameras, a TI800 and a Tek2048. 
Six objects were observed with the former, thirteen with the latter, and three objects with 
both. The ratio of the sums of parallaxes Tek2048/TI800 for the three nebul{\ae} in common 
is $1.08\pm 0.10$.  The error estimate has been obtained using the jackknife method 
(cf.~Lupton 1993 p.~46). However, the ratio for NGC 6853 by itself is $1.45$, and the 
difference between the two is 2$\sigma$ according to H07. On the other hand, its TI800 value, 
$2.63\pm 0.43$ mas, is not far from the value obtained with the {\it HST} FGS, while the 
Tek2048 value is $3.38\sigma$ away. The other two planetaries studied using both CCD cameras, 
Sh 2-216 and PuWe 1, have a combined ratio of $0.99\pm 0.12$. 
\par
All four of the H07 nebul{\ae} studied in B09 have parallaxes obtained with the Tek2048. 
Excluding NGC 6853, the ratio of the sums of parallaxes for the B09 sample in the sense 
Tek2048/FGS is $1.08\pm 0.09$. Therefore we conclude that there is no evidence for an 
appreciable systematic difference among the H07 parallaxes or between the H07 and B09 
parallaxes. The Tek2048 parallax for NGC 6853 appears to be anomalous. In H07 the authors 
suggested the cause was contamination by light from a faint companion which depended on 
seeing. Despite the problem we nonetheless incorporate it into the weighted mean, following 
B09. 
\section{3 \hdskip Comparison of statistical distance scales with parallaxes}

\subsection{3.1 \hdskip The \lq short' CKS scale.}
\par
Table 5 presents the distance values from CKS and the individual distance ratios 
${\cal R}_{CKS}$ together with their uncertainties $\sigma_{\cal R}$ according to Eq.~(1a). 
The imposter PHL 932 is included in the table but omitted from our analysis.  For estimating 
$\sigma_{\cal R}$ the value $\alpha_{CKS}=0.35$ was used, obtained after correcting for the 
bias mentioned in Section 2.3 using the combined data from the synthetic subsamples as 
described below. 

\begintable{5}
\caption{{\bf Table 5.} CKS distances and distance ratios ${\cal R}_{CKS}$ 
relative to H07 with their uncertainties}
{\settabs  3 \columns
\vskip 0.1 in
\hrule
\vskip 0.1 in
\+Name&$d_{CKS}$ (pc)&${\cal R}_{CKS}\pm\sigma_{\cal R}$\cr
\vskip 0.1 in
\hrule
\vskip 0.1 in
\+A 7&216&$0.32\pm 0.15$\cr
\+A 21&243&$0.45\pm 0.20$\cr
\+A 24&525&$1.01\pm 0.40$\cr
\+A 31&233&$0.38\pm 0.14$\cr
\+NGC 6720&872&$1.24\pm 0.67$\cr
\+NGC 6853&262&$0.65\pm 0.31$\cr
\+NGC 7293&157&$0.73\pm 0.26$\cr
\+PHL 932&819&$2.75\pm 1.10$\cr
\+PuWe 1&141&$0.39\pm 0.14$\cr
\vskip 0.1 in
\hrule
}
\endtable

\par
Using Eq.~(5) we find that $\alpha_{CKS}^\prime=0.330$. To estimate $\alpha_{CKS}$ we 
combine the values for the synthetic subsamples. We consider half the T93 objects to have 
the pseudo-distances that were used for selection and the other half to have independently 
generated ones.  Their values are combined quadratically with the value for the non-T93 
objects.  We then find that for $\alpha_S = 0.35$ we get $\alpha_S^\prime = 0.331$, while 
for $\alpha_S = 0.3$ we have $\alpha_S^\prime = 0.263$ and for $\alpha_S=0.4$ the value 
$0.375$. Our adopted value is then $\alpha_{CKS}=0.35$. 
\par
The straight mean distance ratio is $0.65\pm 0.12$. The median, a more robust statistic, is 
$0.55$.  The weighted mean, with weights assigned that are inversely proportional to 
$\sigma_{\cal R}^2$, is $0.45\pm 0.07$. The difference between the two means, $-0.20$, is 
consistent with weighting bias; our combined estimate for that from the synthetic subsamples 
is $-0.22$. The weighted mean for the actual sample is fairly insensitive to the exact value 
of $\alpha_{CKS}$ used in calculating $\sigma_R$: even for $\alpha_{CKS}=0.45$, which is 
almost 30 per cent larger, the result is essentially the same, $0.45\pm 0.09$. As H07 noted, 
many but not all nebul{\ae} have ${\cal R}_{CKS} < 1$.  In fact, all five of those belong to 
the T93 subsample, for which we expect underestimation (though not in every case). 
\par
Bias due to sample selection by statistical distance as discussed in Section 2.2 can be 
corrected by using data from the synthetic samples. The appropriate correction factor is 
$1.11$ for an $\alpha_S$ of $0.35$. Then the revised mean is $0.72\pm 0.13$ and the weighted 
mean $0.50\pm 0.08$.  Whereas the unweighted mean is barely significantly different from 
unity the weighted mean corrected for weighting bias, $0.72\pm 0.08$, definitely is. This 
conclusion must still hold when one takes into account any reasonable estimate of the error 
in the bias correction. 
\par
That we should find an underestimation overall with the CKS scale in all three estimators 
after correcting for the negative selection bias and weighting bias might seem puzzling in 
light of the fact that A98 found the CKS distances {\it over}estimated when compared to 
those from the {\it Hipparcos} parallaxes. Their conclusion was based on a flawed estimate 
of the distance ratio (Section 2.1); however, C99 also suggested that the CKS distances 
might be overestimated.  There have been suggestions of a relation between the distance 
ratio [CKS/reference] and nebular radius by Van de Steene \& Zijlstra (1995, hereafter VdSZ), 
N01, and C99. Thus it is not just a matter of a scale factor as implied by the short-long 
dichotomy; we demonstrate the radius dependence below. 
\par
Table 6 below presents data for the C99 sample compared to CKS. Only those nebul{\ae} for 
which the association is \lq probable' are considered. Also included is NGC 246 because it 
is mentioned in C99, using data from Bond \& Ciardullo (1999) and the same method as in C99. 
We have used the modified distance for NGC 7008 from F08 table 6.2 instead of the much 
smaller one from C99, based on Frew's corrected extinction. The sample consists of nebul{\ae} 
that on the whole are smaller than those typical of the H07 sample, as is obvious comparing 
Table 6 with Table 1. The unweighted mean distance ratio from Table 6 is $1.05\pm 0.11$.  As 
shown in Table 3 the mean distance ratio has a positive bias, but this value does appear 
significantly different from the H07 one. (The amount of overestimation found in C99 was in 
part the result of their including possible companions and partly the lower distance values 
for NGC 246 and NGC 7008.  The mean for our sample with their original distance values is 
$1.14\pm 0.17$.)  
A radius dependence might well account for the difference between the H07 and C99 results. 

\begintable{6}
\caption{{\bf Table 6.} Spectroscopic distances and their logarithmic 
uncertainties, angular diameters, and radii from C99 together with CKS 
distances and distance ratios}
\vskip 0.1 in
{\settabs  8 \columns
\hrule
\vskip 0.1 in
\+Name&&$d$ (pc)&$\sigma_{{\rm log}\ d}$&$\varphi$ ({\tt "})&$R$ (pc)&
$d_{CKS}$&${\cal R}_{CKS}\pm \sigma_{\cal R}$\cr
\vskip 0.1 in
\hrule
\vskip 0.1 in
\+ A33&&1160&$0.062$&$270.0$&$0.76$&$751$&$0.65\pm 0.25$\cr
\+K 1-14&&3000&$0.066$&$47.0$&$0.34$&$3378$&$1.12\pm 0.43$\cr
\+K 1-22&&1330&$0.066$&$180.0$&$0.58$&$988$&$0.74\pm 0.28$\cr
\+K 1-27&&470&$0.106$&$46.0$&$0.05$&--&--\cr
\+Mz 2&&2160&$0.096$&$23.0$&$0.12$&$2341$&$1.08\pm 0.45$\cr
\+NGC 246&&580$^a$&$0.10$&$245.0$&$0.34$&$470$&$0.81\pm 0.36$\cr
\+NGC 1535&&2310&$0.074$&$21.0$&$0.12$&$2283$&$0.98\pm 0.39$\cr
\+NGC 3132&&770&$0.142$&$45.0$&$0.08$&$1251$&$1.63\pm 0.81$\cr
\+NGC 7008&&690$^b$&$0.14$&$86.0$&$0.14$&$860$&$1.43\pm 0.47$\cr
\+Sp 3&&2380&$0.106$&$35.5$&$0.20$&$1877$&$0.79\pm 0.34$\cr
\vskip 0.1 in
\hrule
\vskip 0.1 in
}
\tabletext{Values for $\varphi$ are from the Strasbourg-ESO Catalog except for K 1-27, 
for which the geometric mean of the dimensions from Kohoutek (1977) was used. 
$^a$ Based on ($V - I$)$_0$ from Bond \& Ciardullo (1999); $^b$ Based on revised 
extinction in F08.}
\endtable

\par
Because of the difference in typical $R$ between H07 and C99 the two samples are 
complementary,  especially important since the only small H07 object may not be a planetary. 
(Even with the C99 sample there is only one very small object.)  Therefore we take the 
sample in Table 6 as a potential calibration set. There are no objects common to H07 and C99, 
but fortunately there is sufficient overlap in $R$ to allow indirect comparison of the two 
(Section 1.2), with CKS as intermediary. 

\subsection{3.2 \hdskip The C99 spectroscopic parallaxes and CKS $R$-dependence.}
\par
The C99 \lq spectroscopic' parallaxes (actually photometric) used a ($V - I$) - $M_V$ 
relation calibrated using a combination of data: USNO trigonometric parallaxes for faint 
stars using the CCD technique and for brighter stars using photography together with some 
parallaxes of nearby stars from other observatories (see C99 for more information). 
Although we have no concrete reason to expect bias in this calibration, we do not know the 
selection criteria for the calibration samples or the analysis procedure(s) used, in 
particular the corrections for bias (if any).  Therefore even though USNO parallaxes were 
used in part for calibration we cannot simply assume that this calibration is entirely 
consistent with the H07 parallaxes.  
\par
Comparing the angular diameters with the ones in CKS as we did for the H07 sample, we find 
the mean ratio CKS/A92 to be $0.98\pm 0.05$ (s.d.) and the median $1.00$. For Z95 the mean 
Z95/A92 with the C99 sample is $0.97\pm 0.05$ and the median $1.00$.  Only two objects have 
ratios that differ noticeably from unity, both about 10 per cent smaller: NGC 246 and NGC 
1535. With the F08 scale (examined in a later section) the mean ratio is $1.21\pm 0.25$ 
(s.d.), and all but one are greater than unity, indicating that for this sample the F08 
ones are systematically higher. The errors $\sigma_{{\rm log}\ d}$ have been taken from 
table~7 in C99, as $0.2\sigma_m$. The median value is $0.086$, which corresponds to a 
median relative error of about $0.20$, comparable to the median $\lambda$ for the original 
H07 parallaxes. 
\par
C99 based their likelihood of association in part on comparison of the spectroscopic 
distance with statistical distances, especially CKS and to a lesser extent Z95. 
While understandable, this procedure complicates use of the C99 distances to evaluate 
the statistical scales, as it favors distances that conform to those. Later we will 
see a possible effect of this selection. 
\par
Because the unweighted mean distance ratio is biased we need a different estimator 
to look for systematic differences in scale. The $\kappa$ estimator is a possible 
choice; however, the errors in the spectroscopic distances presumably have a 
lognormal distribution rather than normal, the latter being what we assumed earlier 
when we evaluated $\kappa$ as an estimator. Consequently one would expect a bias in 
the estimate. An alternative is Phillips's $\Gamma$ (Phillips 2004, hereafter Ph04), 
which is defined as 
$$\Gamma\equiv {1\over N}\ \sum_{i=1}^N {\rm log}\ (d^\prime_{2,i}/d^\prime_{1,i})\ .\eqno{(7)}$$
If both sets of distances have lognormal error distributions (with mean zero, of 
course) then we expect no bias for that estimator. Unfortunately the CKS distances 
do not (as far as we know, and we have explicitly assumed otherwise) have lognormal 
errors, so there will be a bias with $\Gamma$ as well. Therefore the question is 
which is superior. 
\par
We can model the C99 sample as we did the H07 sample, with the Monte Carlo 
approach. The best fit to the cumulative distribution function (cdf) is with a 
linear, not disk, distribution, as may be seen in Fig.~4. 

\beginfigure{4}
\includegraphics[height=6cm,width=7.5cm]{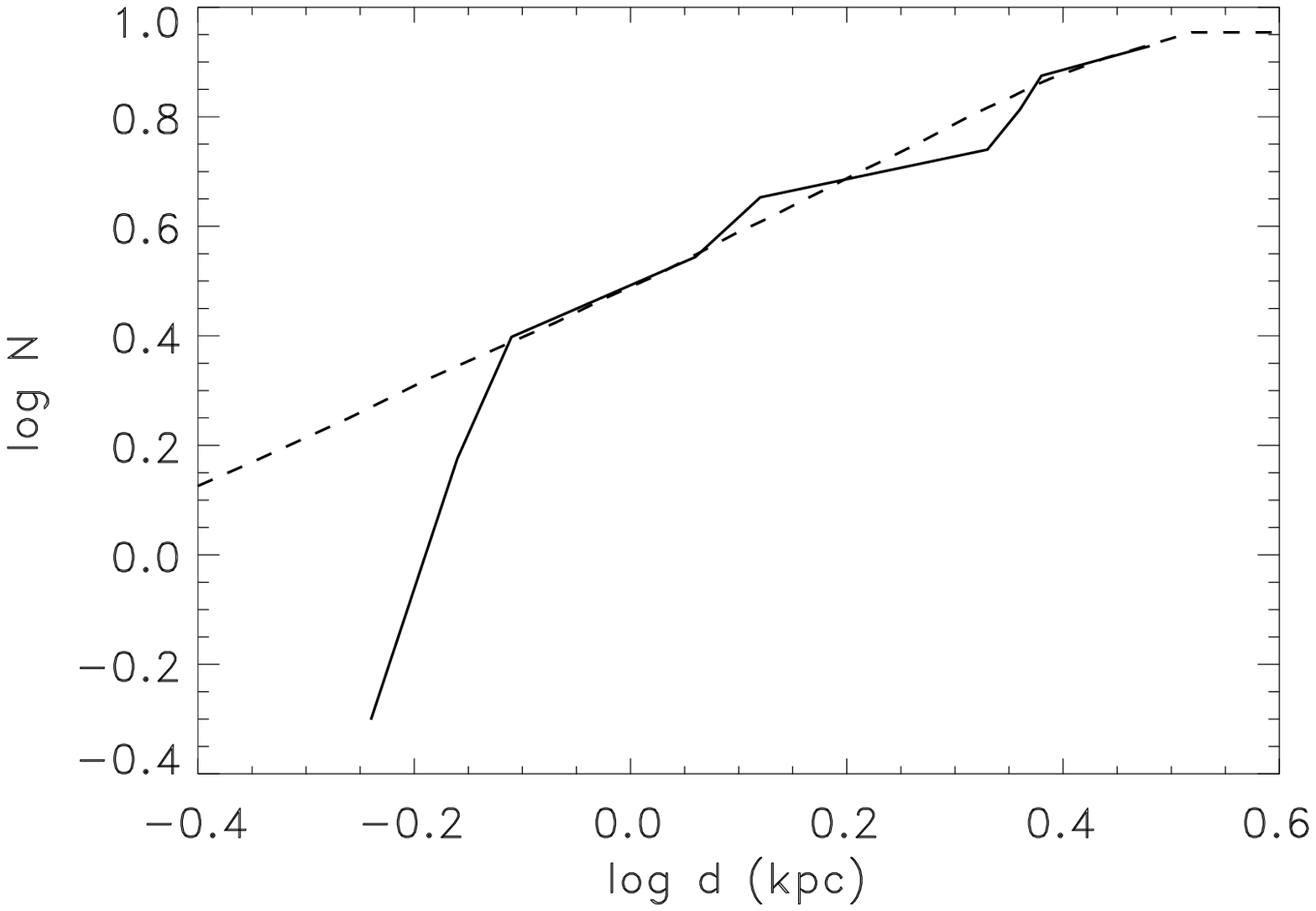}
\caption{{\bf Figure 4.} Cumulative distribution function (cdf) for distance with C99 
sample ({\it filled circles}) vs. cdf for synthetic sample with linear distribution 
({\it solid curve}).}
\endfigure

Table 7 shows how well $B$ is estimated by both estimators with idealized synthetic 
samples similar to those used in Section 2.1. The actual value of $B$ is $1.5$, 
$\alpha_S$ is $0.4$, and $d_{max}$ is 3000; the two are evaluated for a range of 
values of $\sigma_{{\rm log}\ d}$. The spatial distribution is the linear 
distribution appropriate to C99. In place of $\Gamma$ we have used 
$\Gamma^*\equiv$ dex($\Gamma$) to make the comparison more directly. 
\par
As was noted above the median $\sigma_{{\rm log}\ d}$ for the C99 sample is 
$0.086$ and the mean is $0.089$. For those values $\kappa$ has less bias than 
$\zeta$ or $\Gamma^*$ and is roughly comparable in precision. The bias in 
$\kappa$ is small for the relevant range in $\sigma_{{\rm log}\ d}$. As 
a matter of fact both $\zeta$ and $\Gamma^*$ seem to have bias that is independent 
of $\sigma_{{\rm log}\ d}$, with $\Gamma^*$ having slightly less than $\zeta$; for 
the given value of $\alpha$ it is at most of order 7 per cent and 8 per cent respectively. 
$\Gamma^*$ is somewhat insensitive to outliers, though not quite as much as $\zeta$. 
The bias in those two becomes smaller when $\alpha$ is reduced. In what follows 
we will use $\kappa$ and sometimes $\zeta$ in our comparisons of scales with C99 
and F08.

\begintable{7}
\caption{{\bf Table 7.} Distance ratio estimates (with standard deviations) using 
$\kappa$, $\zeta$,  and $\Gamma^*$ for different logarithmic uncertainties 
$\sigma_{{\rm log}\ d}$ with $B = 1.5$}
\vskip 0.1 in
\hrule
\vskip 0.1 in
{\settabs 4 \columns
\+$\sigma_{{\rm log}\ d}$&$\kappa$&$\zeta$&$\Gamma^*$\cr
\vskip 0.1 in
\hrule
\vskip 0.1 in
\+$0.05$&$1.494\pm 0.006$&$1.379\pm 0.003$&$1.392\pm 0.003$\cr
\+$0.075$&$1.488\pm 0.009$&$1.383\pm 0.010$&$1.396\pm 0.011$\cr
\+$0.1$&$1.473\pm 0.008$&$1.376\pm 0.004$&$1.391\pm 0.004$\cr
\+$0.125$&$1.459\pm 0.010$&$1.377\pm 0.012$&$1.394\pm0.012$\cr
\+$0.15$&$1.436\pm 0.006$&$1.371\pm 0.003$&$1.392\pm 0.002$\cr
\vskip 0.1 in
\hrule
}
\endtable
\par
For the C99 sample the CKS distances give $\kappa = 0.99\pm 0.07$, with the error estimated 
using the jackknife method as before. $\zeta$ is virtually identical, $0.97\pm 0.09$, which 
suggests no strong distance dependence is present. The median value of ${\cal R}_{CKS}$ for 
C99 is $0.98$, which agrees with $\kappa$ and $\zeta$.  There are three of the eight values 
that are substantially smaller than unity and two substantially larger. 
\par
It appears that the CKS scale is in close agreement with the C99 distances, and were it not 
for the possibility of $R$-dependence there would seem to be a contradiction with our result 
from H07.  Fig.~5 shows a plot of ${\cal R}_{CKS}$ from both the H07 sample and C99 as a 
function of log $R$. (The error bars for the C99 points are based, like those for H07, on 
$\alpha_{CKS} = 0.35$ but are combined with the estimated uncertainties for the spectroscopic 
parallaxes.)  Clearly ${\cal R}_{CKS}$ decreases with increasing $R$, at least for medium to 
large nebul{\ae}. However, caution is called for because the displacement caused by a 
parallax error for H07 or a distance error for C99 follows a track very similar to the 
radius dependence in the distance ratio, as illustrated by the dashed curve which has been 
displaced to the upper right. As we will demonstrate below, correlated errors like those 
we have here with ${\cal R}$ and $R$ connected through the standard distance $d_{std}$ 
can distort a relation. Nevertheless we will show in the next section that the dependence is 
real and is caused by the $S$-$R$ relation used.  

\beginfigure{5}
\includegraphics[height=6cm,width=7.5cm]{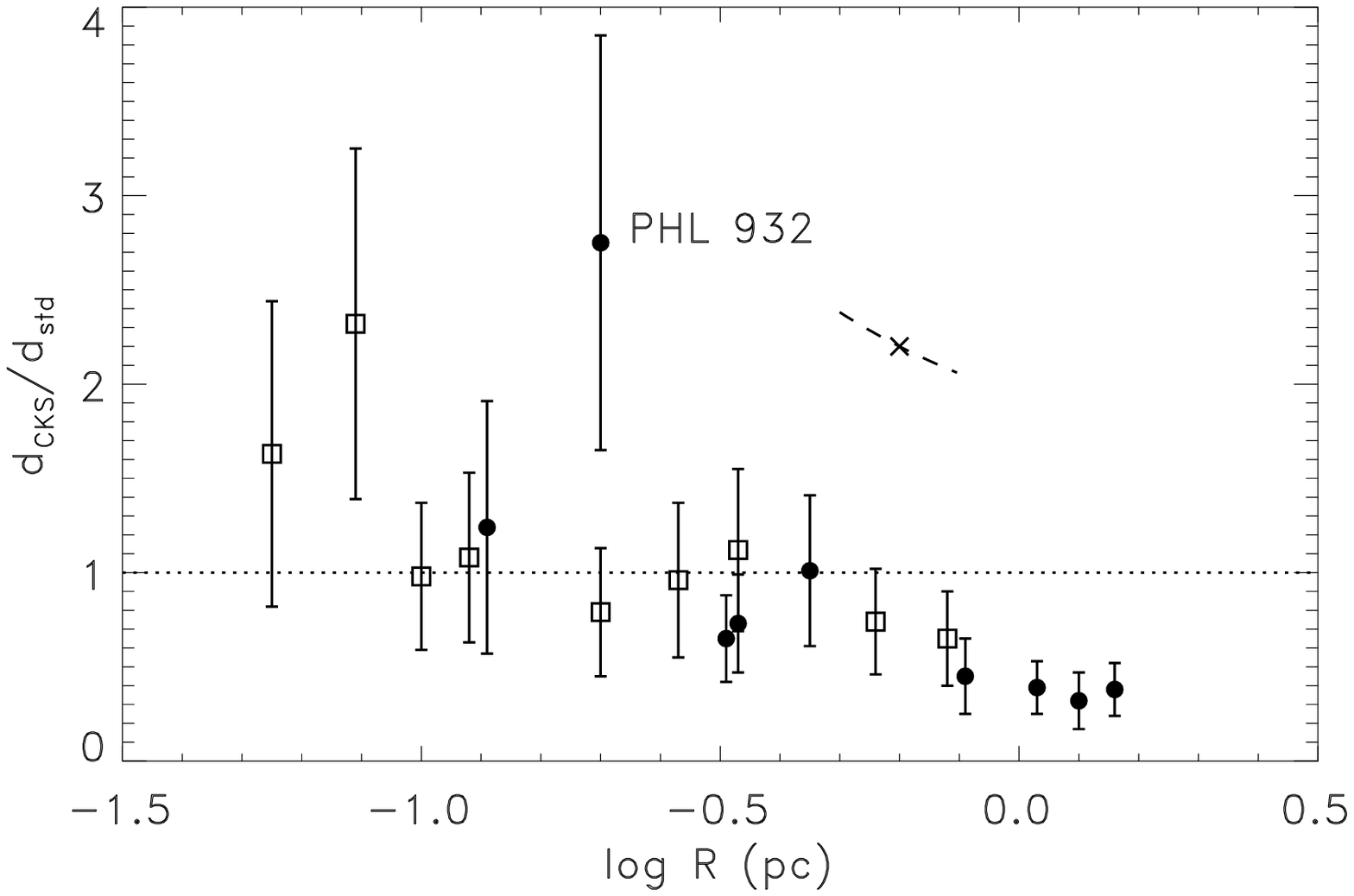}
\caption{{\bf Figure 5.} Distance ratio ${\cal R}_{CKS}$ {\it vs.} 
log $R$ with the latter calculated using the H07 parallaxes ({\it filled 
circles}) or C99 spectroscopic parallaxes ({\it open squares}). 
The {\it dashed curve} represents the trajectories of positive and negative distance 
errors.}
\endfigure

\par
Concentrating on the range in log $R$ over which the H07 and C99 samples overlap, namely 
-1 to 0, the distance ratios for the two -- unweighted mean ${\cal R}_{CKS} = 0.91\pm 0.14$ 
for H07 after correcting for sample selection bias, $\kappa = 0.93\pm 0.07$ for C99 -- 
agree very well, which is encouraging.  All but two of the C99 nebul{\ae}, K 1-27 and NGC 3132, 
are inside this range whereas only five H07 planetaries are. (K 1-27 does not have a CKS distance; 
also, it is a special case.). The mean log $R$ for the H07 sample over this interval is 
$-0.46\pm 0.13$ while that for C99 is $-0.59\pm 0.11$. 
\par
The correlation coefficient $r$ for the H07 distances compared to CKS is $0.55$; that for C99 
vs.~CKS is $r = 0.93$. A look at fig.~7 (upper left panel) of C99 confirms the correlation. 
Perhaps the use of statistical distance as a confirmation criterion for central star companions 
accounts for some of the relatively high correlation with C99. 
\par
There are at least three reasons not to use the P96 spectroscopic parallaxes: (1) that data set 
was compiled from a variety of sources and is therefore not homogeneous; (2) the calibrations for 
the different sources may well depart from that for C99 and not necessarily be consistent with 
the latter or with H07; and (3) at this point we do not have error estimates, even approximate 
ones. This data set will be considered briefly in Section 3.3 and limited use made of it in 
Section 6. 
\par
Stanghellini, Shaw, \& Villaver (2008, hereafter SSV) revised the CKS scale using data on 
planetary nebul{\ae} in the Magellanic Clouds as well as Galactic calibration objects. 
Their scale is essentially identical to the CKS scale for optically thin nebul{\ae}; for 
optically thick planetaries the slope of the relation between the optical thickness 
parameter $\tau$ (inversely proportional to $T_b$) and ionized mass parameter $\mu$ 
(proportional to $M_i$) is a little steeper, and the transition between the two is moved to 
a lower $R$ value than with CKS. (A referee has pointed out that the $\varphi$ values 
and fluxes from CKS were carried over by SSV, but see the exception noted below.) 
Our results for the CKS scale apply to the SSV scale as well because the distances from the two 
are virtually identical for the H07 and C99 samples, to within a factor of about $1.01$. The lone 
exception is PuWe 1 from the H07 sample, whose SSV distance of 416 pc in their table 3 is 
incorrect; it should instead be 142 pc based on their calibration. (The error is traceable to 
truncation of the angular radius by the output format they used, which caused 1200 arcsec to be 
misread as 200 arcsec.)  In particular the same $R$-dependence as that of CKS is present. 

\subsection{3.3 \hdskip The \lq long' Z95 scale.}
\par
Table 8 shows the distance values from Z95 for eight planetary nebul{\ae} in common with 
the H07 sample (and overlapping the CKS sample in Table 5) and eight in common with C99 
together with the imposter PHL 932 as well as the individual distance ratios ${\cal R}_Z$ 
along with their estimated uncertainties $\sigma_{\cal R}$. As with CKS the calculations 
do not include the latter object. The value chosen for $\alpha_Z$, $0.42$, is higher than for 
CKS as explained below. 
\par
We estimate $\alpha_S$ in much the same way as before, except that there is no bias from sample 
selection correlation as with CKS. For the synthetic T93 subsample we use the values for 
uncorrelated distances. Six of the eight belong to that group; combining the values for the 
synthetic subsamples quadratically in that ratio when $\alpha=0.35$ we get 
$\alpha_Z^\prime = 0.278$ and a ratio $0.79$, while using our entire H07 synthetic sample we 
have $0.280$ and ratio $0.8$.   The value of $\alpha_Z^\prime$ from Eq.~(5) is $0.342$, so the 
above values indicate $\alpha_Z$ is around $0.43$. 
\par
It is possible to estimate $\alpha_Z$ using the C99 sample as well, in a fashion analogous 
to that used with parallaxes. Approximating the errors in C99 as Gaussian with $\alpha_1 =0.20$, 
which is reasonably close, we can write
$$\alpha_S^\prime={\Sigma_{i=1}^N\ (d_Z^\prime - \kappa d_C^\prime)^2-\alpha_1^2\over\Sigma_{i=1}^N\ 
d_Z^{\prime\ 2}}\ \ \eqno{(8)}$$
recognizing that this estimate is biased not only because we are using observed values as before 
but also because the errors in $d_C^\prime$ are lognormal. Experiments with synthetic data based 
on our C99 model for the actual $\alpha_2$ ranging from $0.20$ to $0.40$ show that 
$\alpha_S^\prime$ consistently underestimates the true value by 20 per cent except at the lowest 
value, for which the underestimation is 25 per cent. There is the same problem as before with 
non-solutions, roughly 10 per cent of all samples ($N = 8$) for $\alpha_S = 0.40$ and 45 per cent 
at $\alpha_S = 0.20$.  For Z95 we obtain $\alpha_Z^\prime = 0.336$; correcting for bias, we find 
$\alpha_Z = 0.42$. Hence we conclude that this value is probably correct. 

\begintable{8}
\caption{{\bf Table 8.} Distances, distance ratios ${\cal R}_Z$ 
and uncertainties for the Z95 scale (H07 top, C99 bottom)}
{\settabs 3 \columns
\vskip 0.1 in
\hrule
\vskip 0.1 in
\+Name&$d_Z$ (pc)&${\cal R}_Z\pm\sigma_{\cal R}$\cr
\vskip 0.1 in
\hrule
\vskip 0.1 in
\+A 7&700&$1.04\pm 0.54$\cr
\+A 21&710&$1.31\pm 0.68$\cr
\+A 24&1900&$3.65\pm 1.68$\cr
\+A 31&1010&$1.63\pm 0.72$\cr
\+NGC 6720&1130&$1.60\pm 0.95$\cr
\+NGC 6853&480&$1.19\pm 0.50$\cr
\+NGC 7293&420&$1.96\pm 0.83$\cr
\+PHL 932&3330&$11.19\pm 5.21$\cr
\+PuWe 1&900&$2.47\pm 1.08$\cr
\vskip 0.1 in
\hrule
\vskip 0.1 in
\+A 33&2920&$2.52\pm 1.14$\cr
\+K 1-22&3430&$2.58\pm 1.18$\cr
\+Mz 2&2700&$1.25\pm 0.62$\cr
\+NGC 246&990&$1.71\pm 0.86$\cr
\+NGC 1535&2140&$0.93\pm 0.43$\cr
\+NGC 3132&1500&$1.95\pm 1.16$\cr
\+NGC 7008&1310&$1.90\pm 0.95$\cr
\+Sp 3&2620&$1.10\pm 0.55$\cr
\vskip 0.1 in
\hrule
}
\endtable
\par
The mean ${\cal R}_Z$ for the H07 planetaries is $1.85\pm 0.30$ and the median $1.62$. The 
weighted mean is $1.47\pm 0.22$.  Once again the weighted mean is the smallest, presumably 
because of weighting bias.  The expected weighting bias with $\alpha_S = 0.42$ is $-0.60$, 
some 60 per cent greater than the observed $-0.38$. The value of $\kappa$ with C99 is 
$1.56\pm 0.26$, while $\zeta$ is $1.62\pm 0.22$; the uncertainties are estimated using the 
jackknife. The difference between H07 and C99 suggests there may be a radius dependence in 
the opposite direction from that for CKS, but in Fig.~6 there is no obvious trend of 
distance ratio with log $R$. In Section 5 we will confirm this dependence and explain why 
it exists. 

\beginfigure{6}
\includegraphics[height=6cm,width=7.5cm]{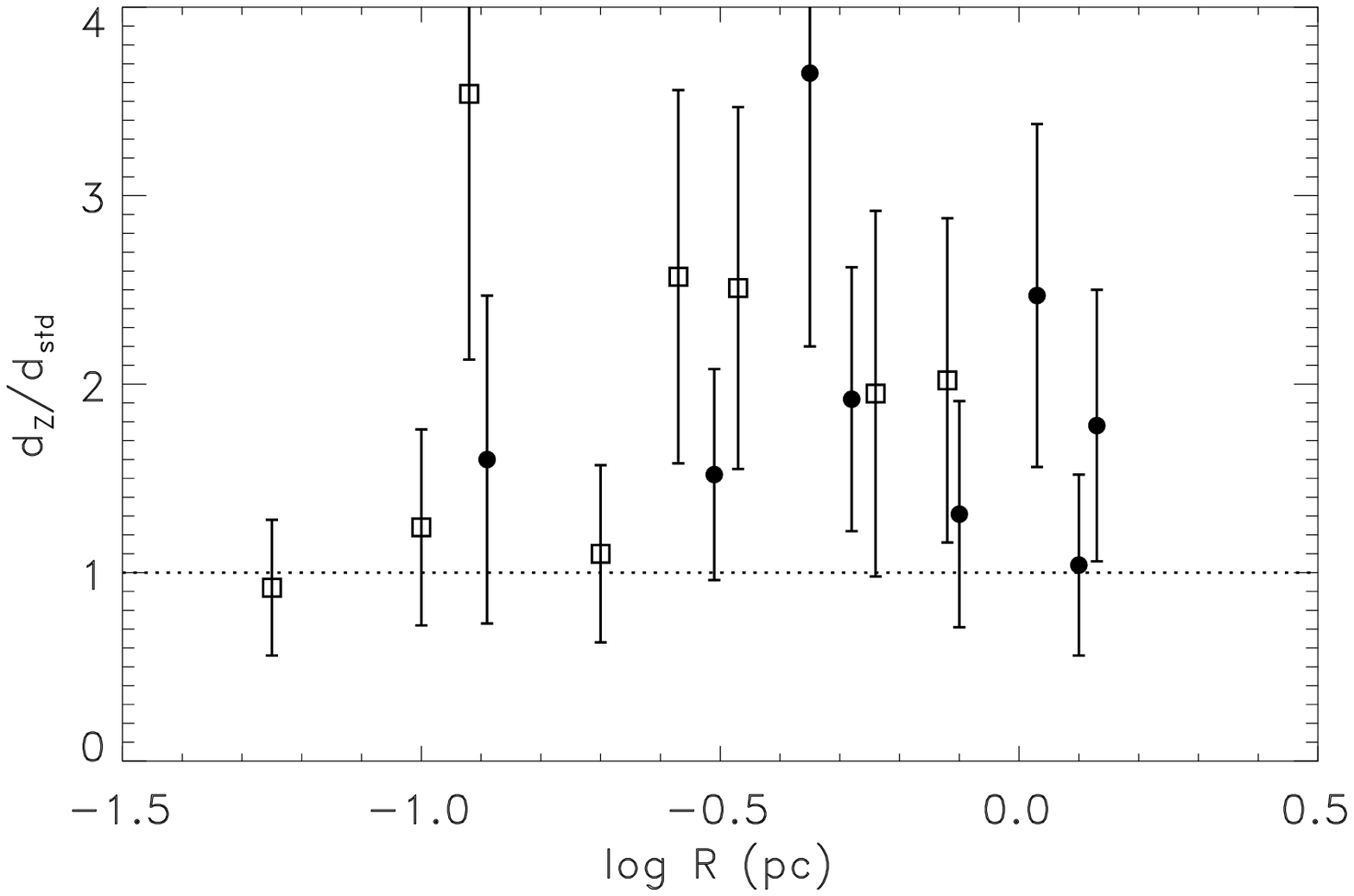}
\caption{{\bf Figure 6.} Distance ratio ${\cal R}_Z$ vs.~log $R$; symbols 
as in Fig. 5.}
\endfigure

\par
Clearly ${\cal R}_Z$ for the H07 sample differs significantly from unity on average, 
and as a matter of fact not one individual value is less than or equal to unity (as H07 
already found). The same is very nearly true of the C99 sample in the lower part of 
the table; only one value is less than unity, and that only slightly. Hence we conclude 
that the Z95 scale is substantially overestimated. If we ignore for now the radius 
dependence we can combine the H07 and C99 results to obtain a grand overall distance 
ratio for this scale. The uncertainties of the two estimated ratios are similar, the 
sample sizes are virtually the same, and the median relative errors are very nearly 
identical, so we simply weight them equally to arrive at a grand mean value, which is 
$1.71\pm 0.21$. The ratio of the latter number to the CKS value $0.89\pm 0.08$ (weighted 
mean of the ratios $0.72$ and $0.99$ we found for CKS in Sections 3.1 and 3.2) is quite 
close to the factor of 2 from Ph02 mentioned in the Introduction. 
\par
Overall the H07 results are more or less consistent with the C99 ones in the region of 
overlap, as before taken to be $-1<{\rm log}\ R\ {\rm (pc)}\leq 0$.  The mean distance 
ratio over this range with the H07 parallaxes is $1.94\pm 0.45$, while $\kappa$ for the C99 
sample is $1.46\pm 0.29$. The difference between the two values, while substantial, is 
smaller than the combined uncertainties. 
\par
The value of $r$ for Z95 vs.~H07 is $0.39$, even lower than for CKS. That for Z95 
vs.~C99 is somewhat higher, $0.56$, which is similar to what we found with CKS vs. 
H07 but considerably lower than for CKS vs.~C99. 
\par
Recall that the Z95 scale we have used is the mean of two scales, one based on $M_i$-$R$ 
and the other on $T_b$-$R$. In Section 4 we consider the latter by itself. 

\subsection{3.5 \hdskip Frew's (F08) mean $S_\alpha$ scale.}
\par
Frew (F08) obtained a relation between H$\alpha$ surface brightness $S_\alpha$ and 
$R$ using a variety of calibrating objects. (He also obtained relations for several 
different types of nebul{\ae}; in this paper we consider only the former because of the 
paucity of classifications.) Table 9 contains distances derived from the 
extinction-corrected H$\alpha$ data in tables 7.1 and 9.5 and the relation itself 
eq.~(7.1) in F08. Distances from the latter table have {\bf not} been used for our H07 
and C99 samples because often those are either ones from one of the specialized relations 
or the ones from the calibration sources rather than from the mean relation. Hereafter 
unless we are referring to some other set of distances taken from F08 the \lq F08 scale' is 
the scale based on that mean $S_\alpha$-$R$ relation; other scales from F08 will have an 
identifier appended. Using our Eq.~(5) we found $\alpha_F^\prime = 0.14$. We have chosen to 
use instead a more conservative value of $0.20$ for $\alpha_F$ in keeping with the 
corrections we have found necessary before, especially with very small values.  

\begintable{9}
\caption{{\bf Table 9.} Distances, distance ratios ${\cal R}_F$ and uncertainties 
for F08 statistical scale (H07 top, C99 bottom)}
{\settabs  3 \columns
\vskip 0.1 in
\hrule
\vskip 0.1 in
\+Name&$d$ (pc)&${\cal R}_F\pm \sigma_{\cal R}$\cr
\vskip 0.1 in
\hrule
\vskip 0.1 in 
\+A 7&590&$0.87\pm 0.31$\cr
\+A 21&440&$0.81\pm 0.28$\cr
\+A 24&850&$1.63\pm 0.44$\cr
\+A 31&410&$0.66\pm 0.16$\cr
\+A 74&640&$0.85\pm 0.45$\cr
\+NGC 6720&900&$1.28\pm 0.57$\cr
\+NGC 6853&300&$0.74\pm 0.16$\cr
\+NGC 7293&210&$0.98\pm 0.20$\cr
\+PuWe 1&400&$1.10\pm 0.25$\cr
\+Sh 2-216&80&$0.62\pm 0.13$\cr
\+Ton 320&470&$0.88\pm 0.24$\cr
\vskip 0.1 in
\hrule
\vskip 0.1 in
\+A 33&1300&$1.12\pm 0.28$\cr
\+K 1-14&4890&$1.63\pm 0.43$\cr
\+K 1-22&1360&$1.02\pm 0.27$\cr
\+Mz 2&1970&$0.91\pm 0.29$\cr
\+NGC 246&790&$1.36\pm 0.45$\cr
\+NGC 1535&1740&$0.75\pm 0.21$\cr
\+NGC 3132&1050&$1.36\pm 0.60$\cr
\+NGC 7008&990&$1.43\pm 0.47$\cr
\+Sp 3&1890&$0.79\pm 0.27$\cr
\vskip 0.1 in
\hrule
\vskip 0.1 in
}
\endtable

\par
The unweighted mean ${\cal R}_F$ for the H07 sample is $0.95\pm 0.09$, the median is 
$0.87$, and the weighted mean is $0.79\pm 0.06$. The difference of $-0.16$ between 
the weighted and unweighted means, interpreted as weighting bias, is smaller than we 
found for CKS and Z95 but qualitatively consistent with our lower value for $\alpha_F$. 
The value of $\kappa$ for the C99 sample is $1.11\pm 0.16$ and $\zeta$ is $1.12\pm 0.10$; 
the uncertainty was estimated using the jackknife. The ${\cal R}_F$ values are plotted 
against log $R$ in Fig.~7; there is no obvious trend but perhaps a drop in ${\cal R}_F$ 
at the largest $R$.  In fact there is a radius dependence similar to that with CKS but 
milder and over a smaller range, evidence for which is presented in Section 4.1. In 
Fig.~8 no distance-dependence is evident.  
\par
Combining the two values we have for the overall F08 distance ratio $0.99\pm 0.08$, to be 
compared with $0.89\pm 0.08$ for CKS and $1.71\pm 0.21$ for Z95.  The respective 
correlations with F08 are $r = 0.79$ for H07 and $0.82$ for C99.  Of course H07 and C99 
were given high weight in the F08 calibration because of their (in most cases) small 
relative errors; on the other hand, F08 had a large calibration set (over 120 objects) and 
included many distances that we have rejected.  Hence we would not expect the overall 
distance ratio to be nearly unity or the high correlation. 
\par
Very recently Frew et al. (2014, hereafter F14) have slightly revised the $S_\alpha$-$R$ 
relation. Results based on it are essentially the same:  unweighted mean for H07 
$0.96\pm 0.09$ and weighted mean $0.81\pm0.07$; for C99 $\kappa = 1.15\pm 0.06$ and 
$\zeta = 1.15\pm 0.10$, and for the two combined we have $1.01\pm 0.08$. Respective 
correlations are $0.78$ and $0.83$.

\beginfigure{7}
\includegraphics[height=6cm,width=7.5cm]{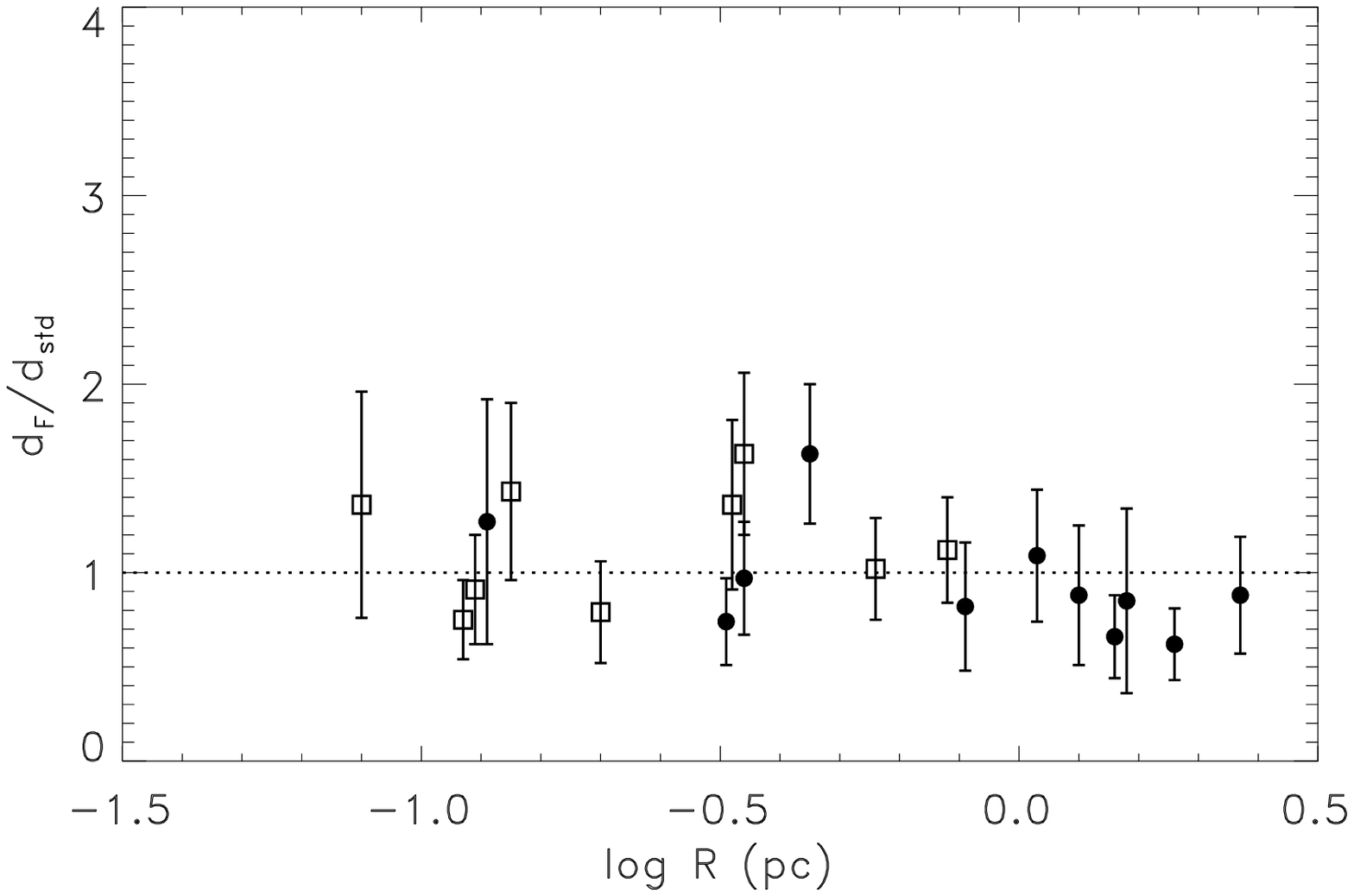}
\caption{{\bf Figure 7.} Distance ratio ${\cal R}_F$ vs.~log $R$; symbols as in Fig. 5.}
\endfigure

\beginfigure{8}
\includegraphics[height=6cm,width=7.5cm]{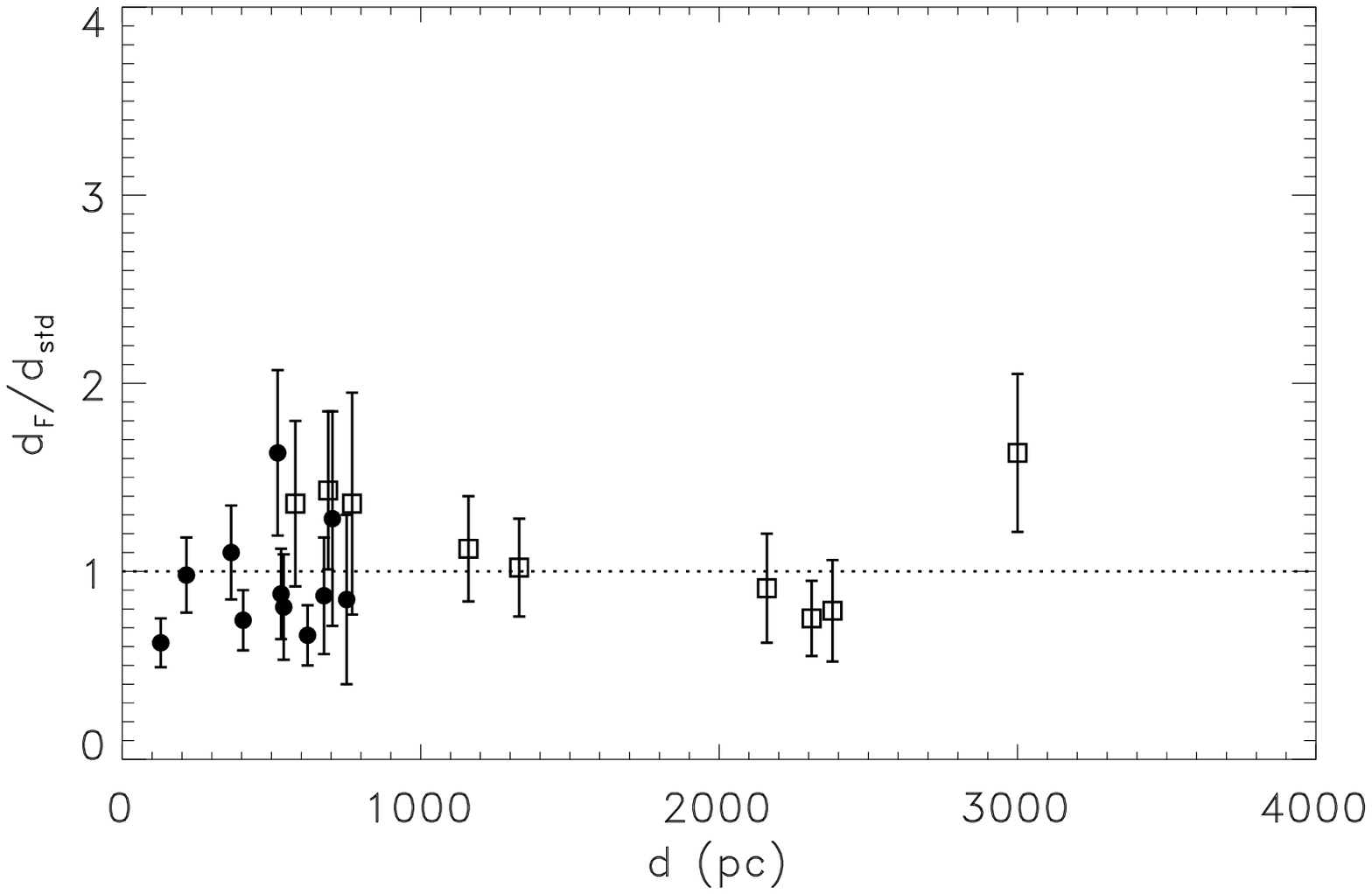}
\caption{{\bf Figure 8.} Distance ratio ${\cal R}_F$ vs.~$d$; symbols as in Fig.~5.}
\endfigure

\par
Using the F08 scale to indirectly compare H07 and C99 in the region of overlap, 
we find a mean ${\cal R}_F$ for the former of $1.09\pm 0.16$ and $\kappa$ for the latter of 
$0.91\pm 0.16$, confirming that the H07 and C99 distance scales are consistent with each 
other.  Accordingly C99 is treated as a primary distance standard on the same footing as 
H07. 
\par
There are no objects in common between the P96 spectroscopic parallaxes, termed here P96-s, 
and H07, but there are two with C99, namely NGC 246 and NGC 3132. For the former the C99 
distance is $1.2$ times larger than the P96-s value, 580 pc {\it vs.} 470 pc.  With the 
latter the C99 distance of 770 pc is more than $1.5$ times larger than the P96-s value 
510 pc. 

\begintable{10}
\caption{{\bf Table 10.} Comparison of spectroscopic distances from P96 with F08 
statistical distances}
{\settabs 4 \columns
\vskip 0.1 in
\hrule
\vskip 0.1 in
\+Name&$d_{P,s}$&$d_F$&${\cal R}_{F/P}$\cr
\vskip 0.1 in
\hrule
\vskip 0.1 in
\+He 2-36&780&2000&$2.56$\cr
\+LoTr 5&420&870&$2.07$\cr
\+NGC 246&470&790&$1.68$\cr
\+NGC 1514&400&800&$2.00$\cr
\+NGC 2346&690&1460&$2.12$\cr
\+NGC 3132&510&1050&$2.06$\cr
\vskip 0.1 in
\hrule
}
\endtable

\par
The P96-s distances are compared with the F08 scale using $\kappa$ and $\zeta$ as with C99. 
The distances for the six objects in common are shown in Table 10 together with the distance 
ratios. The overall F08/P96 distance ratio is $\kappa = 2.13\pm 0.14$; the same with $\zeta$.  
Remarkably, in each individual case the distance ratio is equal to or not far from $2.00$, 
indicating a calibration problem. The correlation between P96-g and F08 is very high, namely 
$0.97$. (Once again there is the issue of inclusion of these nebul{\ae} in the F08 
calibration sample, but that hardly explains such a high value.)  The mean value is larger 
than that for F08/C99 by roughly a factor of two, and the difference is clearly 
statistically significant. It is also larger than the F08/H07 value, by a factor of more 
than two. These findings imply that the P96-s distances can be used provided the distances 
are doubled and we choose a reasonable value for $\alpha_P$. Unfortunately we are unable to 
estimate $\alpha_P$ directly from this comparison, but the high correlation with F08 together 
with the relatively small $\alpha_F$ suggests that it is not large.   

\vskip 0.2 in
\section{4 \hdskip The brightness temperature $-$ radius relation}

\subsection{4.1 \hdskip The $T_b$-$R$ relation.}
\par
We have $T_b$ for a number of the H07 and C99 nebul{\ae} based on published 5 GHz flux 
densities from A92 and Z95. For several objects the $T_b$ values given in table 3 of Z95 are 
inadequate because they are only given to two decimal places; this is of course particularly 
a problem for the larger nebul{\ae}. We have recalculated those, and the logarithms of the 
new values are included in Table 11. The $\varphi$ values used in the calculations are those 
from Tables 1 and 6 for the sake of consistency.

\begintable{11}
\caption{{\bf Table 11.} $S_5$ and log $T_b$ for H07 and C99 sample nebul{\ae}}
\vskip 0.1 in
\hrule
\vskip 0.1 in
{\settabs  4 \columns
\+Name&$S_5$ (mJy)&Ref.&log $T_b$ (K)\cr
\vskip 0.1 in
\hrule
\vskip 0.1 in
\+A 7&$305.0$&Z95&$-1.43$\cr
\+A 21&$327.0$&A92&$-1.21$\cr
\+A 24&$36.0$&Z95&$-1.69$\cr
\+A 31&$101.9$&Z95&$-2.12$\cr
\+A 33&$14.0$&Z95&$-1.87$\cr
\+K 1-22&$11.5$&Z95&$-1.60$\cr
\+Mz 2&$75.0$&A92&$1.00$\cr
\+NGC 246&$248.0$&Z95&$-0.46$\cr
\+NGC 1535&$160.0$&Z95&$1.52$\cr
\+NGC 3132&$230.0$&Z95&$1.26$\cr
\+NGC 6720&$360.0$&Z95&$0.72$\cr
\+NGC 6853&$1325.0$&Z95&$-0.06$\cr
\+NGC 7008&$217.0$&A92&$0.32$\cr
\+NGC 7293&1292.0&A92&$-0.68$\cr
\+PHL 932&$10.0$&Z95&$-2.03$\cr
\+PuWe 1&$84.7$&Z95&$-2.38$\cr
\+Sp 3&$61.0$&A92&$0.53$\cr
\vskip 0.1 in
\hrule
\vskip 0.1 in
}
\endtable

\par
Our log $R$ values based on H07 and C99 are graphed against the log $T_b$ values in Fig.~9. 
$R$ for NGC 6720 has been corrected slightly for the Lutz-Kelker effect discussed in Section 
2.2; in the other cases the correction was deemed insignificant and has been omitted. PHL 
932, HDW 4, and DeHt 5 are included for illustrative purposes along with K 1-27 from C99; 
they were not used in any of the least squares solutions presented in Table 12.  H07 and C99 
define fairly similar $T_b$-$R$ relations. As expected the imposters' statistical distances 
would be overestimated. K 1-27 may be an outlier because of an incorrect spectroscopic 
parallax, but we consider it more likely because it is not a planetary nebula. 
\par
We write the $T_b$-$R$ relation in the customary form 
$${\rm log}\ R=e+f\cdot{\rm log}\ T_b\ ;\ \eqno{(9)}$$
the values found for the coefficients by various authors or in the present paper using least 
squares fitting to Eq.~(9) are listed in Table 12 together with the correlation coefficients 
$r$ for log $R$ with log $T_b$. The relation we find here for H07 and C99 combined has almost 
identical slope to that of Z95 eq.~(7) but a slightly different zero point. 

\begintable{12}
\caption{{\bf Table 12.} Values for intercept $e$ and slope $f$ for 
the $T_b$-$R$ relation obtained by various authors and correlation 
coefficient $r$} 
\vskip 0.1 in
{\settabs 4 \columns
\vskip 0.1 in
\hrule
\vskip 0.1 in
\+$e$&$f$&$r$&Ref.\cr
\vskip 0.1 in
\hrule
\vskip 0.1 in
\+$-0.56\pm 0.05$&$-0.32\pm 0.02$&$-0.80$&Z95-$T_b$\cr
\+$-0.33\pm 0.08$&$-0.42\pm 0.05$&$-0.85$&Z95-$M_i$\cr
\+$-0.51\pm 0.05$&$-0.35\pm 0.02$&$-0.84$&VdSZ\cr
\+$-0.35$&$-0.39$&$-0.90$&BL01\cr
\+$-0.86\pm 0.03$&$-0.36\pm 0.02$&&Ph02, $d_p^*\leq 1$ kpc\cr
\+$-0.63\pm 0.04$&$-0.27\pm 0.03$&&Ph02, $d_p > $1 kpc\cr
\+$-0.66\pm 0.10$&$-0.33\pm 0.07$&$-0.82$&H07, this paper\cr
\+$-0.71\pm 0.06$&$-0.29\pm 0.05$&$-0.92$&C99, this paper\cr
\+$-0.68\pm 0.05$&$-0.32\pm 0.04$&$-0.93$&H07+C99\cr
\+$-0.64\pm 0.09$&$-0.24\pm 0.04$&$-0.89$&LMC\cr
\+$-0.53\pm 0.25$&$-0.29\pm 0.11$&$-0.92$&SMC\cr
\vskip 0.1 in
\hrule
\vskip 0.1 in
\tabletext{ $^*$The projected distance $d_p$ is given by $d\cdot$ cos $b$, 
with $b$ the galactic latitude.}
}
\endtable

\par
Using eqs.~(1), (3), (4) and (5) in Z95 and assuming a constant filling factor 
$\epsilon = 0.6$, the $M_i$-$R$ relation yields a $T_b$-$R$ relation with $e=-0.33$ and 
$f=-0.42$ as given in Table 12. As a result of the difference in their $T_b$-$R$ relations 
the two distance scales that are combined to give the Z95 mean distances have a 
$T_b$-dependent difference which can be discerned in table 3 of Z95. Specifically, the 
steeper slope for the Z95-$M_i$ relation introduces a growing {\it overestimation} of 
distance with increasing $R$, the opposite of CKS and SSV.  The expected distance ratio 
[mean/$T_b$] increases from roughly unity at small $R$ to a factor of $2.2$ for 
$R = 2$ pc. 

\beginfigure{9}
\includegraphics[height=6cm,width=7.5cm]{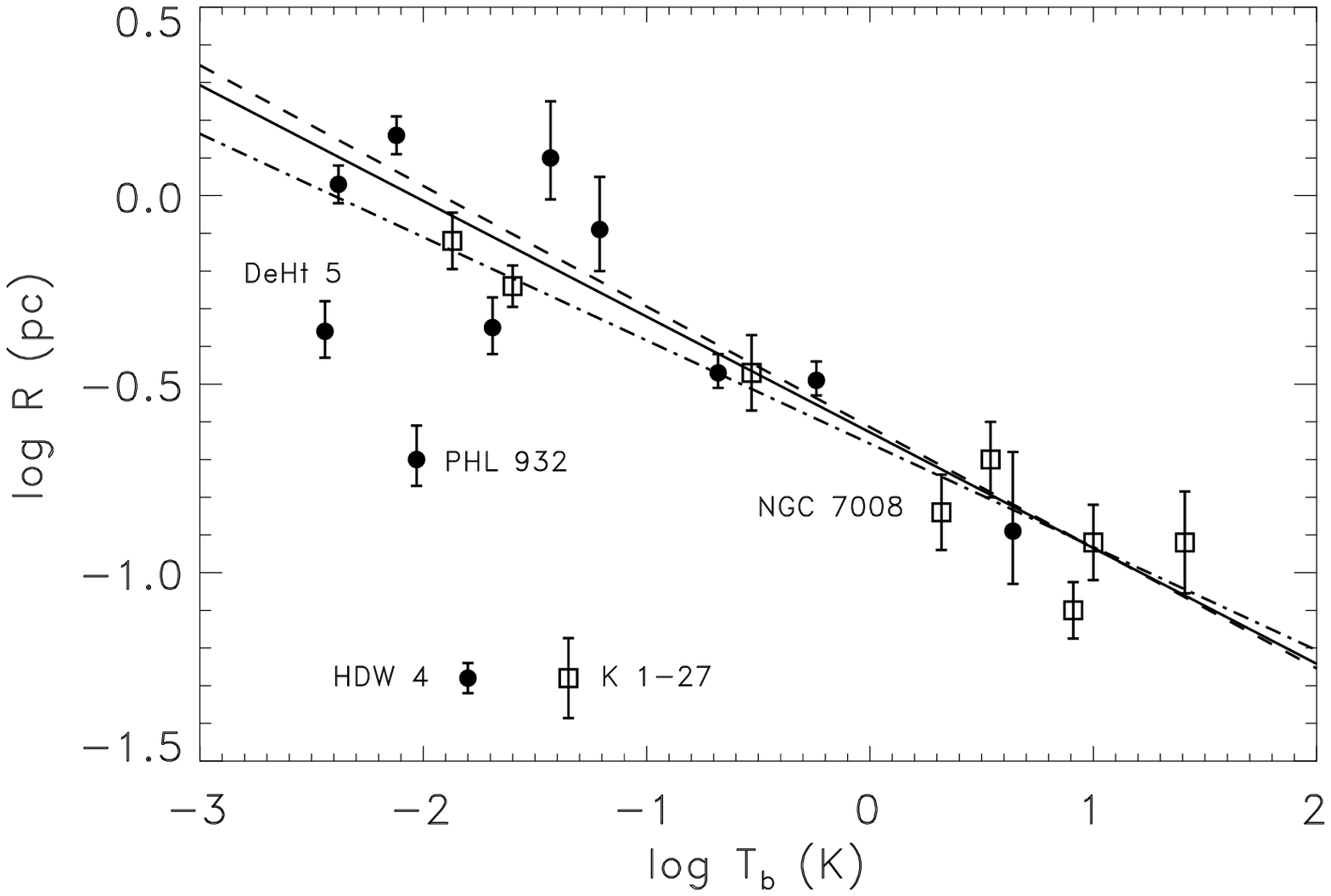}
\caption{{\bf Figure 9.} Radius $R$ vs.~5 GHz brightness temperature $T_b$ with 
symbols as in Fig.~5. The $T_b$-$R$ relation from least squares fitting to the H07 
data is shown by the {\it dashed line}; the relation from the C99 data is shown by the 
{\it dot-dashed line}; and the fit to both sets is the {\it solid line}.}
\endfigure

\par
The slope of the Z95 $T_b$-$R$ relation is very close to that for H07 and C99 together, 
suggesting that the scale based on it, termed Z95-$T_b$, might be nearly free of radius 
dependence.  The value of $\alpha_{Z,T}^\prime$ found for the Z95-$T_b$ distances with the 
H07 sample is $0.281$, very close to the value quoted in Z95 but lower than the value we 
found earlier for the mean Z95 scale. Correcting based on the H07 synthetic sample as for 
the mean Z95 case, we estimate $\alpha_{Z,T}$ to be $0.35$.  Comparison with C99 gives a 
much lower $\alpha^\prime_{Z,T}$, namely $0.12$, but such a low value is highly implausible. 
We thus adopt the value $0.35$. The unweighted mean ${\cal R}_{Z,T}$ for H07 is $1.12\pm 0.16$, 
the weighted mean is $0.91\pm 0.14$, the median is $1.00$, and the correlation is $0.51$. 
With C99 we find $\kappa = 1.13\pm 0.11$ and $\zeta = 1.21\pm 0.11$; the median is somewhat 
larger, $1.30$.  The correlation is $0.88$. Distance ratios are presented in Table 13. Fig.~10 
indicates no radius dependence with Z95-$T_b$, and Fig.~11 shows no distance dependence. 

\begintable{13}
\caption{{\bf Table 13.} Distances, distance ratios ${\cal R}_{Z,T}$ 
and uncertainties for the Z95-$T_b$  scale (H07 top, C99 bottom)}
{\settabs 3 \columns
\vskip 0.1 in
\hrule
\vskip 0.1 in
\+Name&$d_{Z,T}$ (pc)&${\cal R}_{Z,T}\pm\sigma_{\cal R}$\cr
\vskip 0.1 in
\hrule
\vskip 0.1 in
\+A 7&410&$0.61\pm 0.28$\cr
\+A 21&440&$0.81\pm 0.37$\cr
\+A 24&1080&$2.07\pm 0.82$\cr
\+A 31&540&$0.87\pm 0.33$\cr
\+NGC 6720&930&$1.32\pm 0.71$\cr
\+NGC 6853&350&$0.86\pm 0.31$\cr
\+NGC 7293&280&$1.30\pm 0.46$\cr
\+PHL 932&1790&$6.01\pm 2.80$\cr
\+PuWe 1&410&$1.12\pm 0.42$\cr
\vskip 0.1 in
\hrule
\vskip 0.1 in
\+A 33&1620&$1.40\pm 0.54$\cr
\+K 1-22&1980&$1.49\pm 0.58$\cr
\+Mz 2&2290&$1.06\pm 0.46$\cr
\+NGC 246&690&$1.19\pm 0.53$\cr
\+NGC 1535&1940&$0.84\pm 0.34$\cr
\+NGC 3132&1250&$1.62\pm 0.87$\cr
\+NGC 7008&1010&$1.46\pm 0.65$\cr
\+Sp 3&2090&$0.88\pm 0.38$\cr
\vskip 0.1 in
\hrule
}
\endtable

\beginfigure{10}
\includegraphics[height=6cm,width=7.5cm]{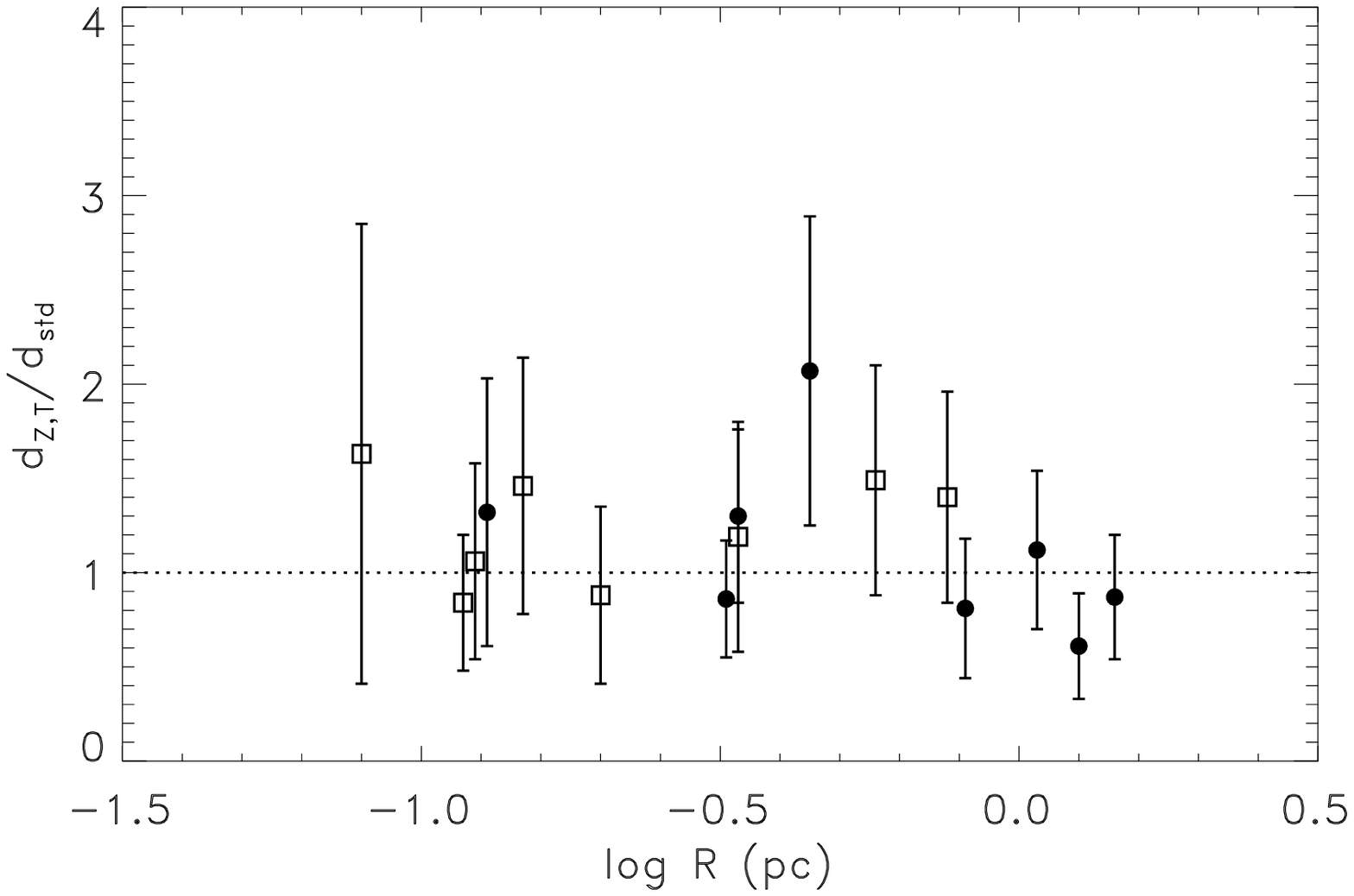}
\caption{{\bf Figure 10.} Distance ratio $d_{Z,T}/d_{std}$ vs.~log $R$ for H07 
and C99 samples; symbols as in Fig.~5.} 
\endfigure

\beginfigure{11}
\includegraphics[height=6cm,width=7.5cm]{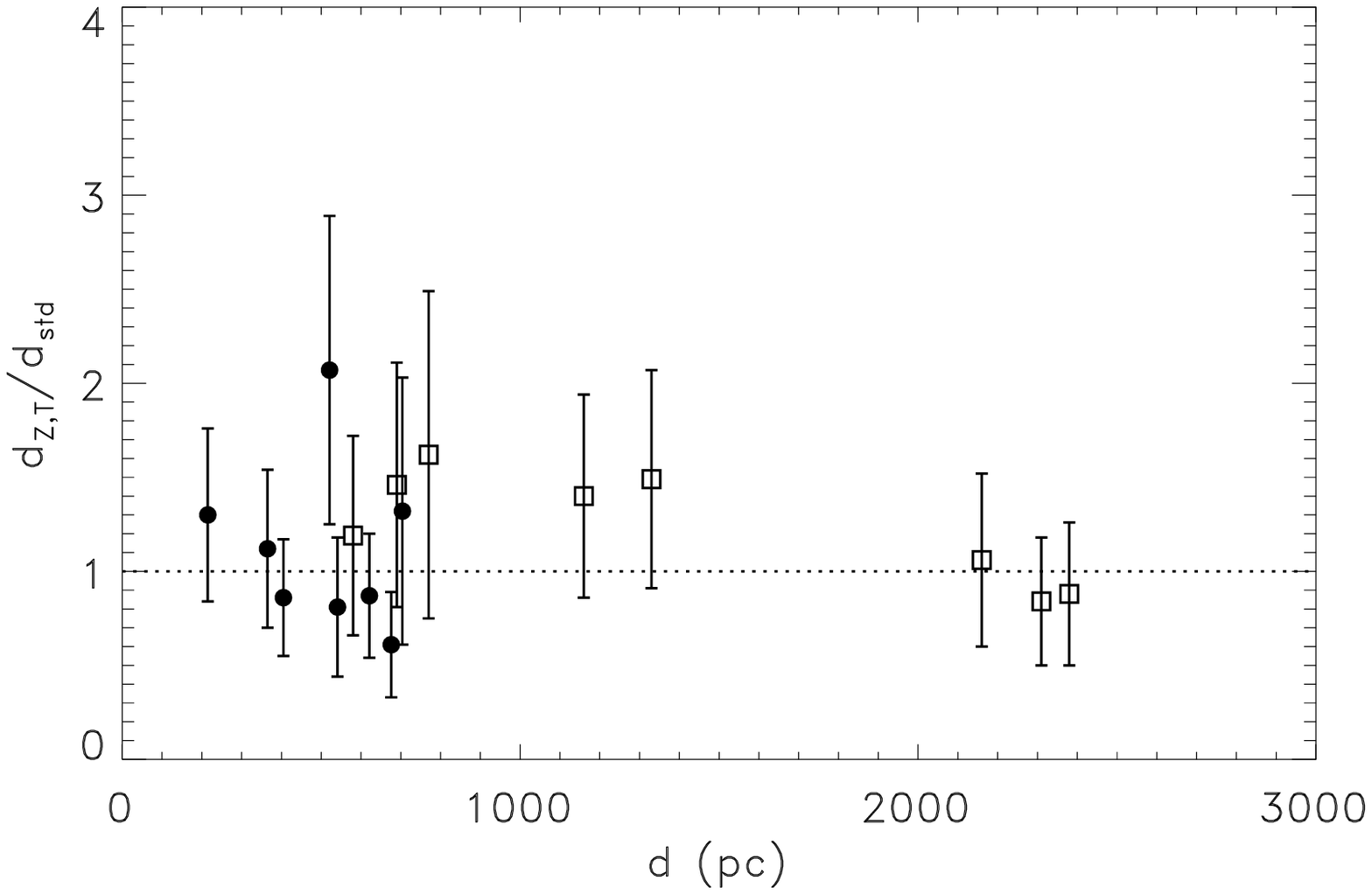}
\caption{{\bf Figure 11.} Distance ratio $d_{Z,T}/d_{std}$ vs.~$d$; symbols as in Fig.~5.}
\endfigure

The F08 sample is truncated at $2.04$ kpc according to the final distance values in table 
9.5, which are a mixture of ones from $S_\alpha$-$R$ relations and ones from other methods; 
hence there is truncation bias with F08. Applying similar truncation to the Z95-$T_b$ values 
should tend to counter the bias in distance ratio arising from the F08 truncation.  Indeed, 
for the 77 nebul{\ae} having $d_{Z,T} < 2.1$ kpc we get $\kappa = 0.99\pm 0.03$ and 
$\zeta = 1.01\pm 0.03$. There is a weak but statistically significant radius dependence. 
Least squares fitting gives a relation of the form 
$$ {\rm log}\ (d_{Z,T}/d_F) = 0.06\pm 0.02 + 0.08\pm 0.02 \ {\rm log}\ R\  .\eqno{(14)}$$
We attribute this dependence to underestimation at large $R$ for F08, with evidence given 
below. The correlation $r$ of the truncated samples is $0.82$. Fig.~12 shows the distance 
ratio for Z95-$T_b$/F08 as a function of $R_F$ for those. In our judgment the Z95-$T_b$ 
scale is good enough to serve as a standard. 

\beginfigure{12}
\includegraphics[height=6cm,width=7.5cm]{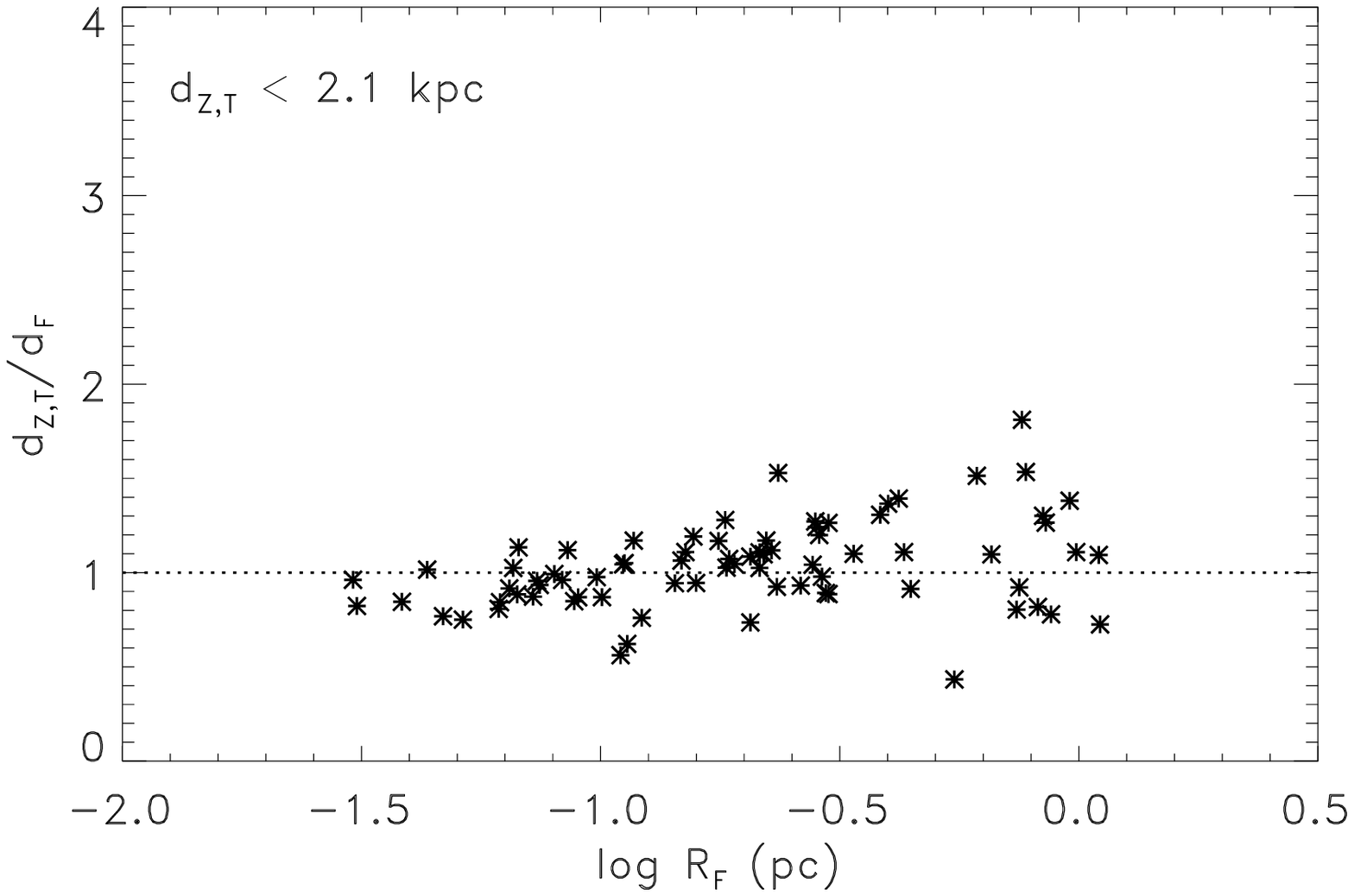}
\caption{{\bf Figure 12.} Distance ratio $d_{Z,T}/d_F$ vs.~log $R_F$ for truncation 
of Z95 sample at $d_{Z,T}< 2.1$ kpc.}
\endfigure

\par

\par
The $M_i$-$R$ relation implies such a different slope because of how it was calibrated. In 
Fig.~13 we present synthetic data for a large ($N=8000$) sample of planetaries all of which 
are assumed to obey the mass-radius relation $M_i = M_0 (R/R_0)^{0.9}$ with $M_0 = 0.2 M_
\odot$ and $R_0 = 0.122$ pc. The apparently arbitrary choice for the exponent comes from a 
least squares fit of revised $M_i$ values to $R_i$ values, both corrected using F08 
distances. Each value of $M_i$ from the Z95 calibration was multiplied by the $2.5$ power of 
the ratio of F08 distance to Z95 calibration distance, based on eq.~(4) of Z95, and the 
corresponding $R$ was multiplied by the distance ratio. The logarithmic slope found by least 
squares was $0.91\pm 0.08$ instead of the value $1.31\pm 0.07$ from eq.~(5) in Z95; although 
our value is probably closer to the true slope because of the more accurate F08 distances it 
may well still be slightly too high. The \lq true' $R$ values for the synthetic sample were 
selected at random over the interval $0.05$-$0.5$ pc; random (Gaussian) relative errors in 
$R$ (corresponding to errors in distance) with standard deviation $\alpha = 0.35$ were then 
added to give \lq measured' values $R^\prime$. The \lq calculated' value of $M_i$, which we 
call $M_i^\prime$, was obtained using the same equation, multiplying $M_i$ by 
$(R^\prime/R)^{2.5}$.  Least squares fitting of log $M_i^\prime$ to log $R^\prime$ yields a 
value $1.31\pm 0.01$ for the logarithmic slope of the $M_i$-$R$ relation, and the 
correlation is $r=0.89$. Clearly correlation of the errors in $M_i$ and $R$ tends to steepen 
the relation. 

\beginfigure{13}
\includegraphics[height=6cm,width=7.5cm]{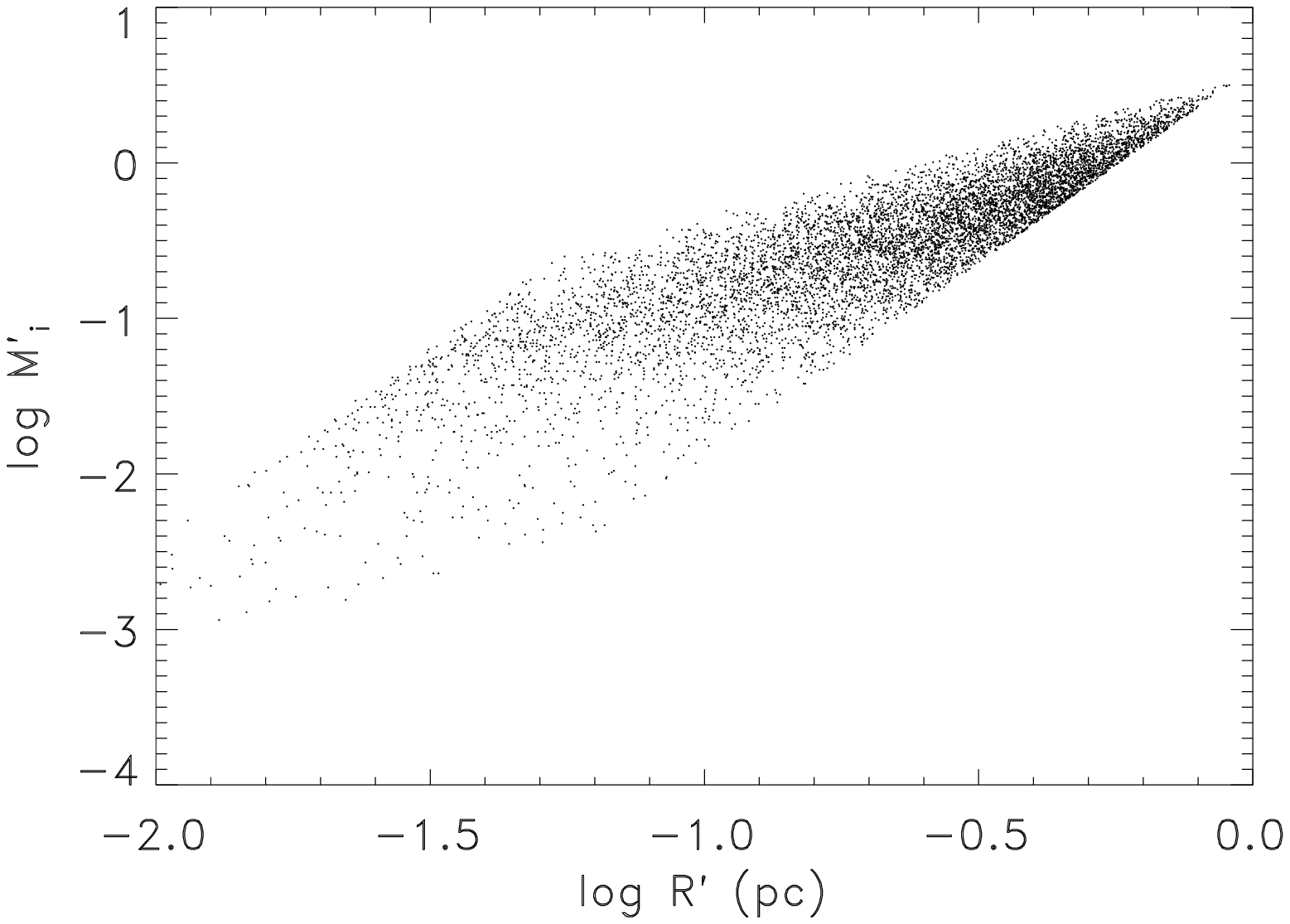}
\caption{{\bf Figure 13.} Log of simulated ionized mass $M_i^\prime$ vs.~log $R^\prime$ for 
synthetic data ($N=8000$) (see text for recipe).}
\endfigure
 
\par
Our finding for the Z95-$M_i$ scale coincidentally sheds new light on the first \lq 
long' statistical scale known to us. Seaton's original calibration in S66 for optically 
thin nebul{\ae} was based on estimation of $M_i$, which (as was customary at the time) 
he took to be constant.  He used measures of the [O II] $\lambda$3727 \AA\ line pair to 
estimate the quantity $x\equiv N_e\ T_e^{-1/2}$ for each of 14 calibration nebul{\ae}; 
as usual $N_e$ is the electron temperature and $T_e$ the electron temperature.  Using 
the $x$ values and the $\epsilon$ values estimated from photographs together with the 
H$\beta$ surface brightness he could then obtain a value for the radius, $R(N_e)$ (see 
eq.~(6) of S66). The nebular mass $M_i$ can then be shown to be proportional to the 
quantity $R^3(N_e)\epsilon x$. Eight objects were identified as optically thin on 
spectroscopic grounds, but two including NGC 2392 were ignored because of small $R(N_e)$ 
and hence small $R^3(N_e)\epsilon x$ (cf. fig.~1 of S66). 
\par
A problem with the S66 approach is that, like $M_i$ and $R$ in Z95, the quantities 
$R(N_e)^3\epsilon x$ and $R(N_e)$ have correlated errors because they depend on (mostly) the 
same quantities. Errors in distance cause errors in $R$ which tend to move a data point 
along a line of slope 3; $R^3(N_e)\epsilon x$ is highly sensitive to those. Similar 
correlations arise from errors in $\epsilon$ and $x$ because they are elements of the 
estimation of $R$ and hence distance in S66.  As a result, modest errors cause large 
deviations which could remove an object from inclusion in the mean. A striking 
illustration of this is NGC 2392. The S66 value for log $x$ was $-0.11$; Kingsburgh \& 
Barlow (1992) found it to be $-0.89$. In consequence the value of log $R(N_e)$ changed from 
$-1.86$ to $-0.77$ and $R^3(N_e)\epsilon x$ changed from $-6.99$ to $-3.72$, placing NGC 
2392 in the midst of the calibration nebul{\ae}. Correcting the S66 values using F08 or, 
for NGC 6210 and HD 138403, Z95-$T_b$, we find the S66 distance scale is reduced by 31 per 
cent. 
\par
The radius dependence of ${\cal R}_{CKS}$ obviously results from an incorrect value for 
the logarithmic slope of the $S$-$R$ and $T_b$-$R$ relations for the larger nebul{\ae}. 
The CKS and SSV value of $f$ for optically thin nebul{\ae} is the original Shklovsky 
value $-0.2$, which causes distances to be increasingly underestimated as $S$ and $T_b$ 
decrease and $R$ increases, the radius dependence seen in Fig.~5. This problem was already 
noted in F08. 
\par
Thus far our $T_b$-$R$ relation includes only nebul{\ae} having log $R\geq -1.2$. There is a 
considerable body of data on Magellanic Cloud planetaries (Shaw et al.~2001, 2006; 
Stanghellini et al.~2002, 2003). Table 12 contains the values of $e$ and $f$ from 
least squares fits to these data also.  The two samples are at reasonably well-known 
distances and can used to calibrate statistical scales, as SSV stated. Although those 
authors recommend using the photometric radius, for consistency with our results we chose 
the geometric mean of the two optical dimensions to obtain the results in Table 12. We have 
calculated $T_b$ using eq.~(6) from CKS to convert $F_{{\rm H}\beta}$ to $S_5$ (in mJy). 
Only those nebul{\ae} having extinction values $c$ were included. 
\par
From the table we see that $f$ for the LMC sample is closer to the Shklovsky value 
than either our standard samples or the SMC one. In Fig.~14 we have added MC data to our 
Fig.~9. Those data appear to match up with our H07-C99 relation where they overlap. Also 
shown for reference are lines schematically representing three evolutionary stages: a dotted 
line with $f=-0.33$, a dashed line with $f = -0.2$, and a solid line having $f=-0.5$. In F08 
the $S_\alpha$-$R$ relation in eq.~(7.1) implies a slope of $-0.277$, slightly less steep 
than we find for C99. (The slope in F14 is $-0.275$.) That value is roughly midway between 
$-0.2$ and $-0.33$; a single line with that slope would fit well too. In Table 12 there 
is a trend of $f$ decreasing with decreasing $T_b$, but the differences are not statistically 
significant. 
\par
Schneider \& Buckley (1996) proposed a quadratic form for the (logarithmic) $T_b$-$R$ 
relation rather than a single power law. However, their formula explicitly approached the 
Shklovsky slope at large $R$, whereas we find a steeper slope there. 
\par
We have estimated the rms relative error in $R(T_b)$, essentially $\alpha_{MC}$, from the 
least squares fits for the two MC samples. Nebul{\ae} having log $T_b\ > 2.8$ seem to lie on 
a tightly constrained line and are presumably optically thick; leaving those out we find $0.31$ 
and $0.27$. These values are only upper limits to the intrinsic spread for a mixture of types. 

\beginfigure{14}
\includegraphics[height=6cm,width=7.5cm]{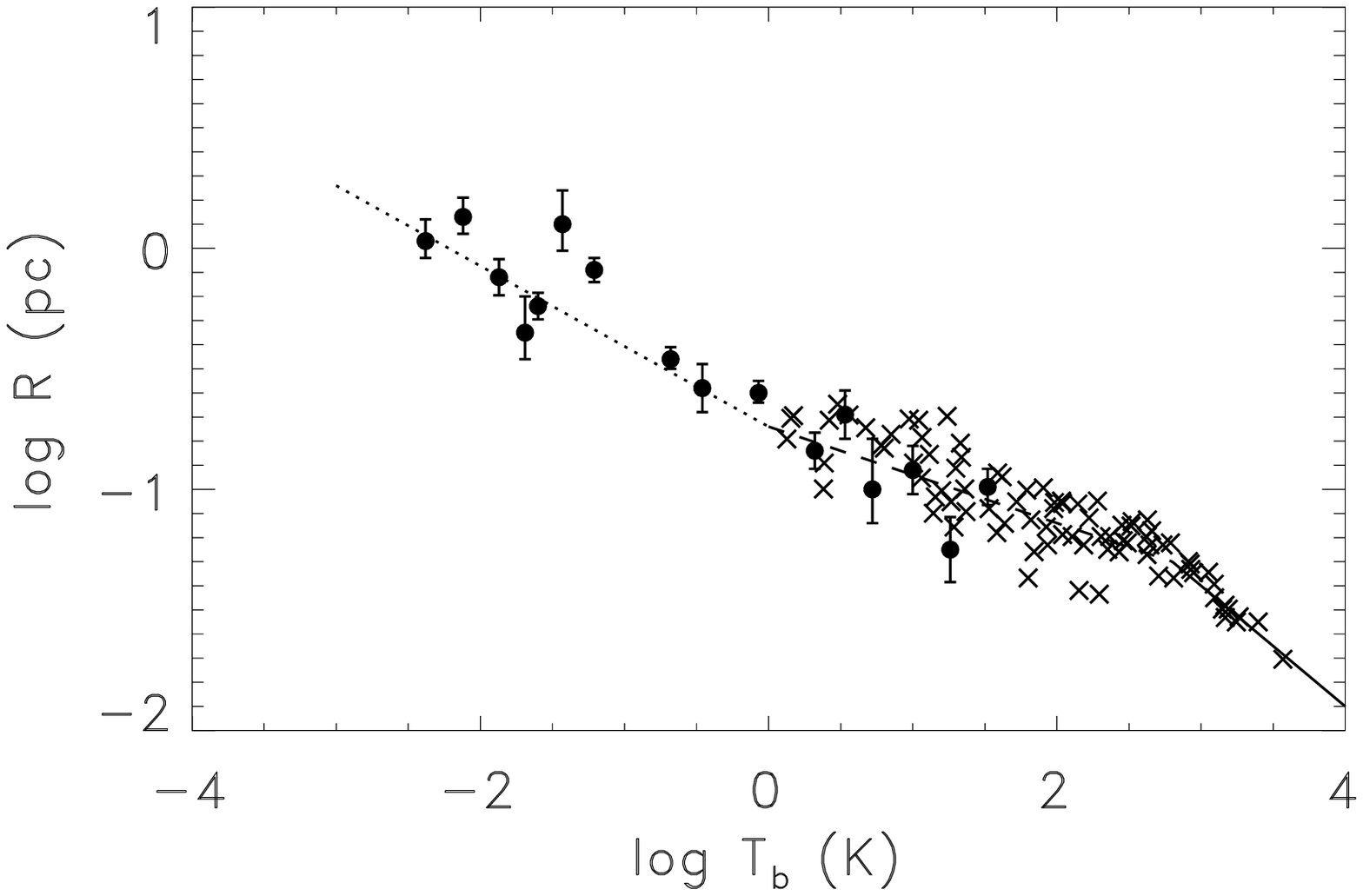}
\caption{{\bf Figure 14.} Plot of log $R$ vs.~log $T_b$ with data on Magellanic Cloud 
nebul{\ae} added to H07 and C99. Here both H07 and C99 objects are plotted with {\it filled 
circles}; MC objects are shown by {\it crosses}. The lines are described in the text.} 
\endfigure

\beginfigure{15}
\includegraphics[height=6cm,width=7.5cm]{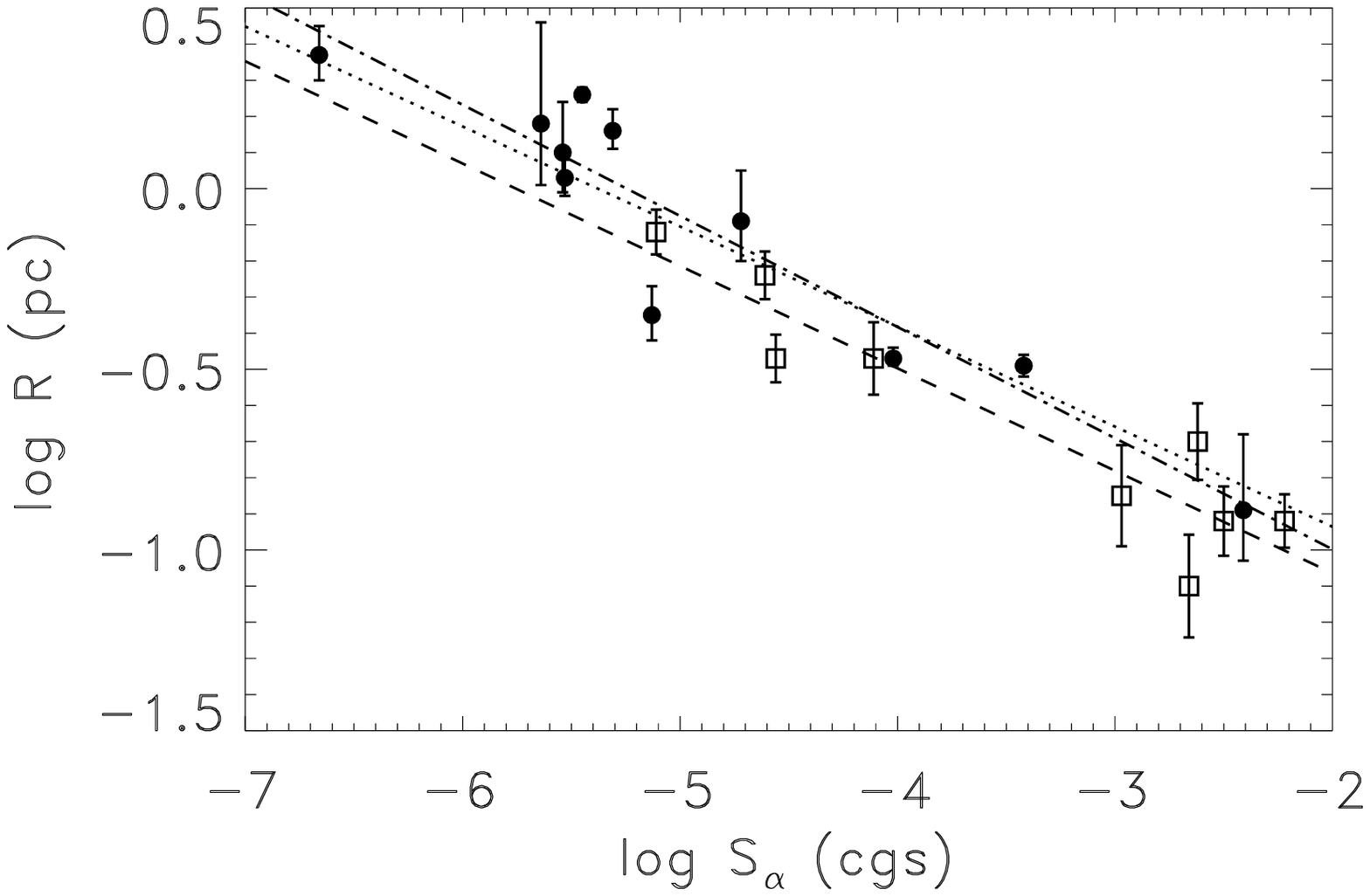}
\caption{{\bf Figure 15.} Same as Fig.~9 but with log $S_\alpha$ instead of log $T_b$ 
with symbols as in Fig.~5. {\it Dashed line:} least squares fit for C99; 
{\it dot-dashed line:} fit for H07; {\it dotted line:} mean relation according to F08 
eq.~(7.1). }
\endfigure

\par
To examine possible underestimation at large $R$ with the F08 scale we first look at $S_\alpha$ 
instead of $T_b$ as shown in Fig.~15. We find least-squares slopes with H07 and C99 of 
$-0.31\pm 0.03$ and $-0.28\pm 0.03$ respectively, more precise than those in Table 12 but 
basically the same (and consistent with the F08 and F14 slopes). All of the largest objects lie 
on or above the dotted line corresponding to the mean relation from F08.  Second, we compare 
the calibration distances from table 7.1 of F08 with those predicted by the F08 mean 
relation for the nebul{\ae} with log $R_F>-0.4$ based on that table; there are 37 of those, 
shown in Fig.~16. For $+0.05<$ log $R_F<+0.45$ (the largest) all 9 distances based on the 
mean relation are underestimates, by as much as a factor of slightly more than 2. Fewer 
than half of these, including A 31 which is not labelled, are H07 objects. The mean 
distance ratio for all nine is $1.52\pm 0.11$ compared to $0.95\pm 0.07$ for the remainder. 
We would say, then, that F08 distances are underestimated by a factor of roughly 1.5 for 
log $R_F>+0.05$. As mentioned previously the F08 angular diameters are larger than those we 
have used by about $0.05$ in the log, so it seems that nebul{\ae} larger than 1 pc in 
radius in our system have distances significantly underestimated. This factor is to be 
compared to the corresponding CKS value of 2.5 for the three nebul{\ae} in Fig.~5 with log 
$R> 0$. The same effect is visible in figs.~2 and 4 of F14. For the 11 calibration objects 
therein we have $1.38\pm 0.09$. 

\beginfigure{16}
\includegraphics[height=6cm,width=7.5cm]{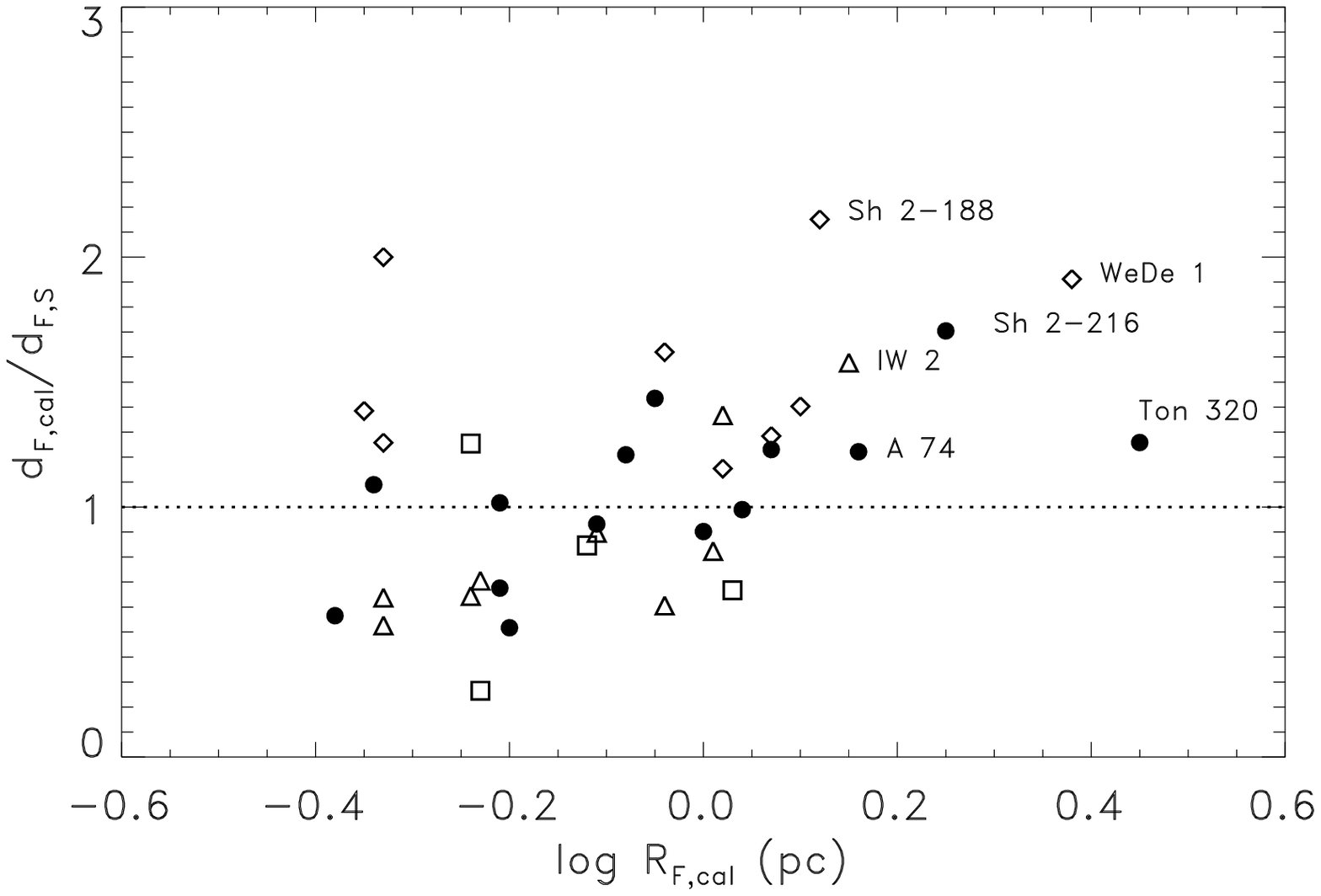}
\caption{{\bf Figure 16.} Distance ratio $d_{F,cal}/d_{F,S}$ {\it vs.} log $R_{F,cal}$. 
{\it Filled circles}: trigonometric and spectroscopic parallax calibrators from F08; 
{\it open diamonds}: cluster membership calibrators; {\it open triangles}: expansion calibrators; 
{\it open squares}: extinction calibrators.}
\endfigure

\subsection{4.2  \hdskip  The $T_b$-$R$ relation and nebular evolution.}
Traditionally the $S$-$R$ relation and the $T_b$-$R$ relation have been thought to reflect 
in a gross sense the evolution of the nebula. Recently this evolution has been described in 
some detail by Jacob et al. (2013), based on hydrodynamical modelling coupled to the 
evolution of the central star. The authors found that their models agree well with the F08 
$S_\alpha$-$R$ relation and with the Magellanic Cloud relation presented in F08 (cf.~their 
fig.~4). The same may be said of the F14 relation; see fig.~4 of that paper. 
\par
Our results seem mostly consistent with these. The Jacob et al. tracks stop at $R=1$ pc, so 
they do not show the steepening slope at large $R$ or the F08/F14 underestimation. (The 
figure uses F08's definition of radius, which differs a little from ours.) In our Fig.~14 
there is also steepening at small $R$ for the Magellanic sample where the nebul{\ae} are 
optically thick.  The model tracks show a flattening, qualitatively the same as our 
steepening because the axes are reversed. The points for log $T_b> 2.8$ in Fig.~14 are few 
but fit tightly around a line of slope $f\simeq -0.5$. In F14 a theoretical value 
corresponding to $f = -0.61$ is given for the optically thick case. 
\par
We have looked at the ISM interaction classes of Wareing, Zijlstra, \& O'Brien (2007) for 
the H07 and C99 objects in Fig.~9. The range is from WZO 1 to 2; none of these are class 
2/3 or 3. Class 2 is unsurprisingly to be found almost entirely among some of the larger 
objects, but some of those are 1/2 or even 1. For instance, A 7, one of the largest, is 
class 1. In Fig.~15 the very largest objects, Ton 320 and Sh 2-216, are the only ones 
classified as having strong interaction with the ISM, as might have been expected; their 
WZO classes are 3 and 2/3, respectively. The interaction is expected to cause a brightening, 
but we do not see that for Ton 320, apparently the one most strongly affected. However, its 
angular diameter is not from A92. Possibly Sh 2-216 has been affected, but it is situated 
near a clump of nebul{\ae} with lower classes.  

\section{5 \hdskip Gravity, expansion, and extinction distances}

\subsection{5.1 \hdskip Gravity distances.}
\par
The basic idea behind gravity distances was stated in Section 1.1. The formula used is of 
the form (M\'endez et al.~1988) 
$$d^2= 3.82\ \times 10^{-11}{M_*\ F_{\lambda 5480}\over g}10^{0.4V_0}\ \eqno{(10)}$$
where $M_*$ is the star's mass (from the evolutionary track in the log $g$-$T_{eff}$ 
plane), $g$ is surface gravity, $T_{eff}$ is effective temperature, $F_{\lambda5480}$ 
is the flux at 5480 \AA, and $V_0$ is the apparent visual magnitude corrected 
for extinction. $F_{\lambda5480}$ and $g$ are taken from the model atmosphere fitted to the 
spectrum. 
\par
The largest truly homogeneous set of gravity distances known to us is that of N01 for 
hydrogen-rich central stars of old planetaries. Because they are old and hence fairly 
large these nebul{\ae} more nearly resemble those of the H07 sample than the ones in 
the C99 sample; indeed, no C99 object is in the set. Data for the eleven H07 central 
stars having N01 distances are in Table 14.  Also included are spectral types from 
Napiwotzki \& Sch\"onberner (1995, hereafter NS95). The sample includes PHL 932, DeHt 5, 
and HDW 4, which are imposters; however, that fact is irrelevant here because no assumed 
value of $M_i$ plays a r\^ole in gravity method distances. PHL 932 does not have an 
anomalously large distance ratio, instead the smallest. DeHt 5 and HDW 4 have ratios 
comfortably inside the range of the others. 
\par
The median relative error for the entire N01 sample calculated from the stated 
uncertainties is $0.30$, slightly less than for the CKS scale ($0.35$). We consider this a 
lower bound to $\alpha_N$ and will instead adopt the value $0.35$. In our analysis we treat 
the uncertainties as symmetric even though they are not, as though the pdf were normal. 
While not rigorous, this treatment probably makes little difference in our results. 
\par
The mean ${\cal R}_N$ with all objects included is $1.43\pm 0.12$. The median is slightly 
larger, $1.47$. The weighted mean including all objects is $1.23\pm 0.13$.  Again, it is 
likely that the difference is attributable to weighting bias; our H07 model predicts 
$-0.29$ for $\alpha_S = 0.3$.  There must be a fairly large overestimation relative to the 
H07 distances: Of the eleven objects only PHL 932 has ${\cal R}_N<1$. 
\par
The degree of overestimation seems unrelated to spectral type. The cause of such a distance 
overestimation is unlikely to lie with either $M_*$ or $F_{\lambda 5480}$; in either case 
that is improbable. (For the latter it would mean overestimating $T_{eff}$ by a factor two.) 
If overestimation of both were to blame at least one would have to be at least 40 per cent 
too high. One can easily imagine log $g$ being systematically too small by something like 
$0.20$. Indeed in both P96 and B09 it was suggested that log $g$ is responsible. 

\begintable{14}
\caption{{\bf Table 14.} Spectral types from NS95 and gravity distances from N01 along 
with the corresponding distance ratios ${\cal R}_N$ and uncertainties for the N01 scale}
{\settabs 4 \columns
\vskip 0.1 in
\hrule
\vskip 0.1 in
\+Name&Spectral type&$d_N$ (pc)&${\cal R}_N\pm\sigma_{\cal R}$\cr
\vskip 0.1 in
\hrule
\vskip 0.1 in
\+A 7&DAO&700&$1.04\pm 0.66$\cr
\+A 31&--&1000&$1.61\pm 0.76$\cr
\+A 74&DAO&1700&$2.26\pm 1.38$\cr
\+DeHt 5&DA&510&$1.48\pm 0.45$\cr
\+HDW 4&DA&250&$1.20\pm 0.36$\cr
\+NGC 6720&DA&1100&$1.56\pm 0.80$\cr
\+NGC 6853&sdO/DAO&440&$1.09\pm 0.34$\cr
\+NGC 7293&DAO&290&$1.35\pm 0.37$\cr
\+PHL 932&hgO(H)&240&$0.81\pm 0.21$\cr
\+PuWe 1&DAO&700&$1.92\pm 0.65$\cr
\+Sh 2-216&DAO&190&$1.47\pm 0.36$\cr
\vskip 0.1 in
\hrule
}
\endtable
\par
Fig.~17 hints at $R$-dependence, with the highest ${\cal R}_N$ at $R > 1$ pc. However, the 
sample is small and the errors are large. 

\beginfigure{17}
\includegraphics[height=6cm,width=7.5cm]{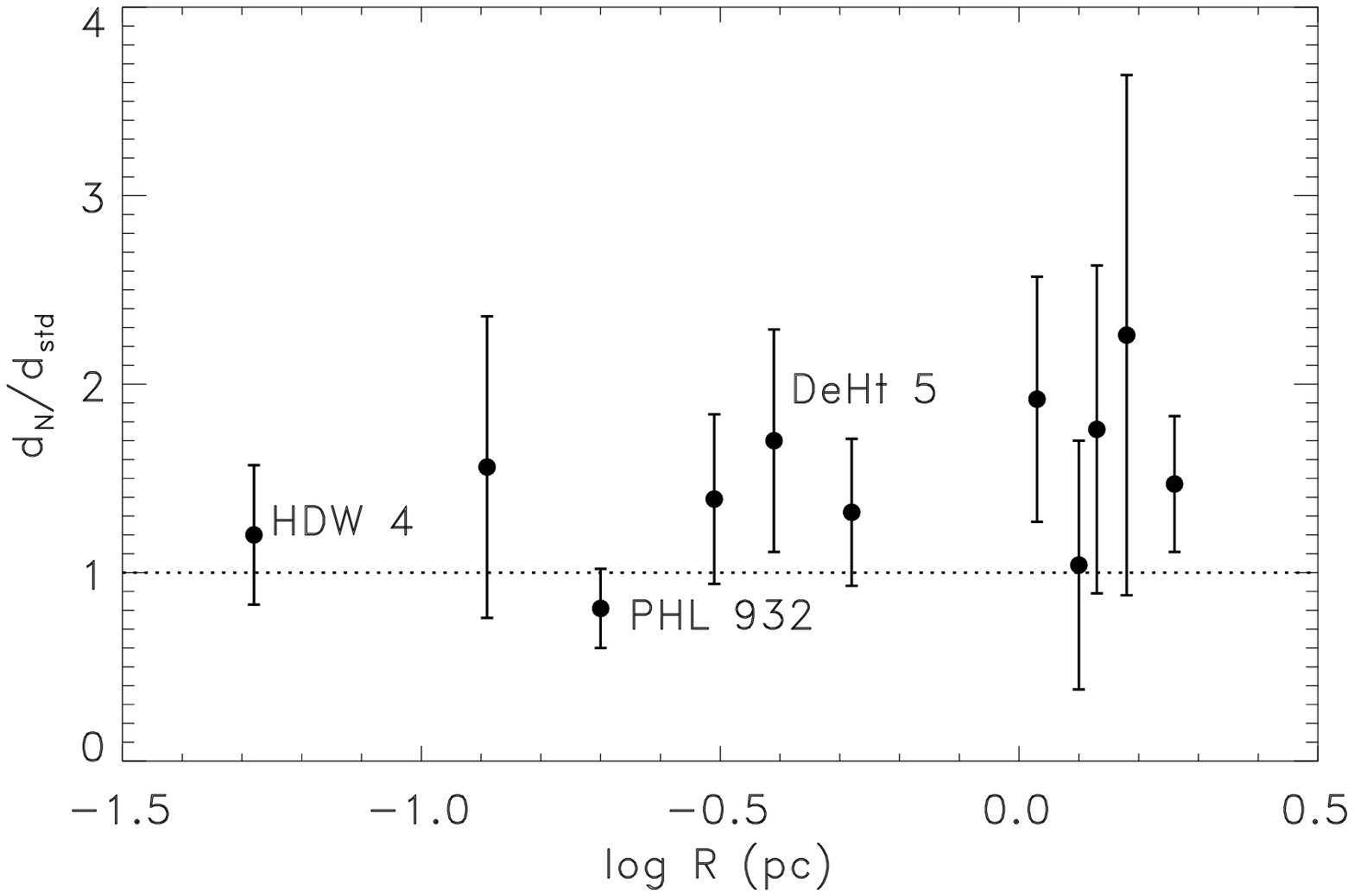}
\caption{{\bf Figure 17.} Distance ratio ${\cal R}_N$ vs.~log $R$; 
symbols are as in Fig.~5.}
\endfigure

\par
The N01 scale is also tested using the F08 scale as intermediary; leaving out objects that 
are not planetaries there are 20 nebul{\ae} in common. Unlike H07 the F08 sample includes 
smaller nebul{\ae} and some that are distant. We find $\kappa = 1.04\pm 0.16$ and 
$\zeta = 1.27\pm 0.15$. The differences between ${\cal R}_N$ and $\zeta$ and between 
$\kappa$ and $\zeta$ indicate a distance dependence while the former difference also 
suggests a possible radius dependence.   
\par
Fig.~18 is a plot of the N01/F08 distance ratio as a function of the F08 distance $d_F$. 
The overestimation is entirely for planetaries with $d_F < 1$ kpc and hence consistent with 
our finding for H07. At larger $d_F$ there seems to be underestimation, and $\kappa$ weights 
larger distances more heavily, which we believe to be the reason for the difference between 
$\kappa$ and $\zeta$. With $\kappa$ the low distance ratios at 2-3 kpc almost cancel the 
high ones inside 1 kpc; if there were a few more nebul{\ae} out there it would have. 
With $\zeta$ the ratios are weighted more evenly, but the value is lower than for H07 
most obviously because it includes the underestimates at large distances. The low distance 
ratios at large $d_F$ with this sample may partly be due to positive errors in the F08 
distances.  However, considering the fairly high precision of F08 we do not expect errors 
to contribute so much in the way of overestimation, and the difference between ${\cal R}_N$ 
and $\zeta$ argues against that explanation. The central stars of K 2-22 and HDW 11 which 
are not post-AGB objects are marked as such in this figure as well as Fig.~20. 

\beginfigure{18}
\includegraphics[height=6cm,width=7.5cm]{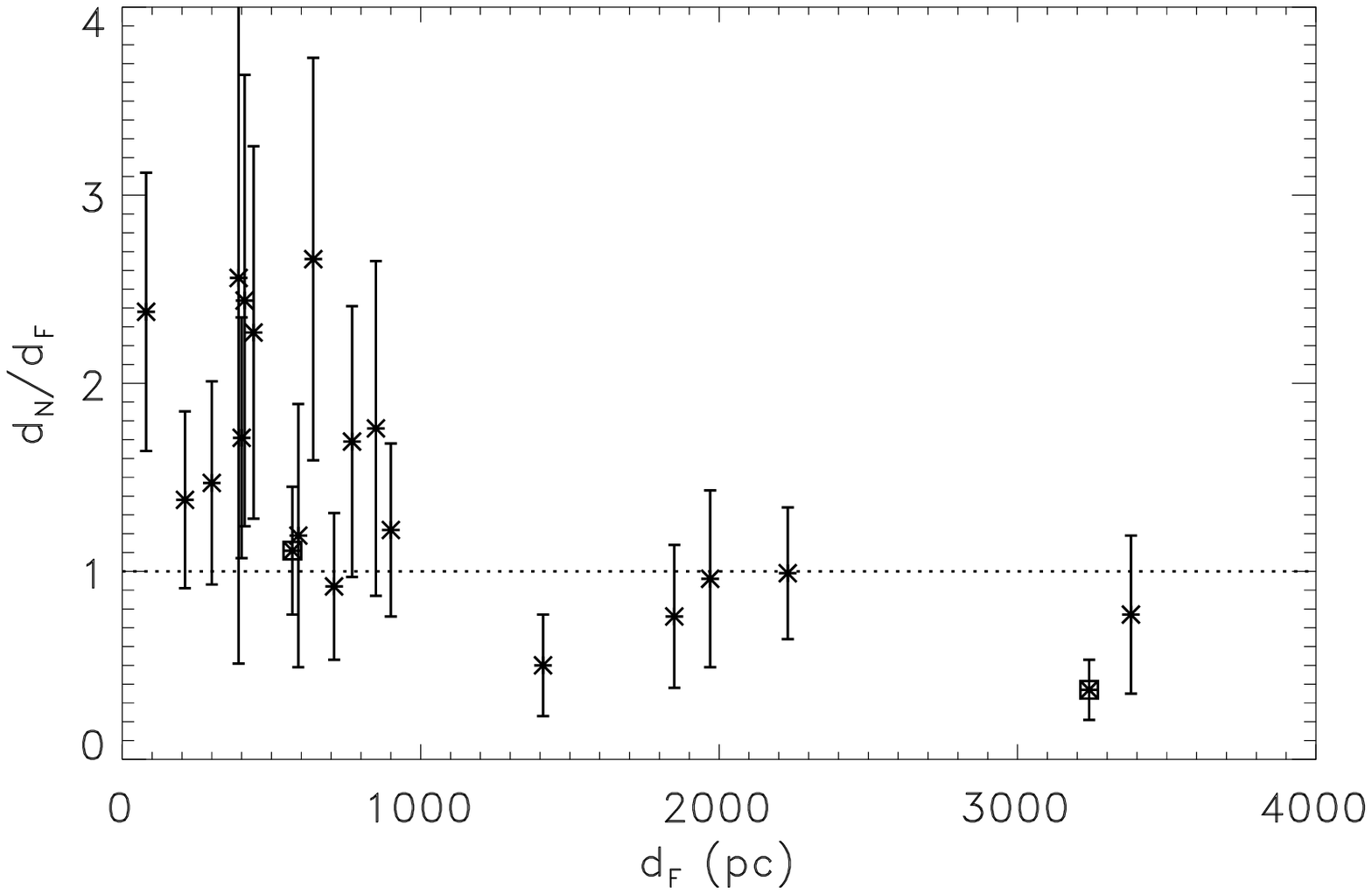}
\caption{{\bf Figure 18.} Plot of distance ratio $d_N/d_F$ vs. distance $d_F$; 
{\it asterisks} denote F08 objects; additionally those whose central stars are not post-AGB 
are marked with a {\it square}.}
\endfigure

\par
The effect of scatter in the standard distances, in this case F08, is shown in Fig.~19 
using synthetic data ($N = 8000$). This example was based on an exponential space 
distribution like that of N01 with scale height $z_0 = 250$ pc, distance scale factor 
$B=1.5$ (roughly the value for N01 vs.~H07), $\alpha_1=0.20$ (for F08) and 
$\alpha_2 = 0.3$, and a distance cutoff of 2 kpc. (Remember that the F08 sample 
in table 9.4 is basically truncated at that value.)  The filled circles are means in bins 
of 0.2 kpc. Note the elevation of $d_2/d_1$ at small $d_1$ and the depression at large 
$d_1$. The former is associated with negative errors in $d_1$; because of the geometry 
more objects will be scattered in by those than are scattered out by positive errors. 
Comparison with Fig.~18 suggests that the effect at large $d_1$ is insufficient to explain 
the behaviour of the real sample. There are only a few data points, but it seems to us 
that there is a real distance dependence with N01. If so we cannot explain it. 

\beginfigure{19}
\includegraphics[height=6cm,width=7.5cm]{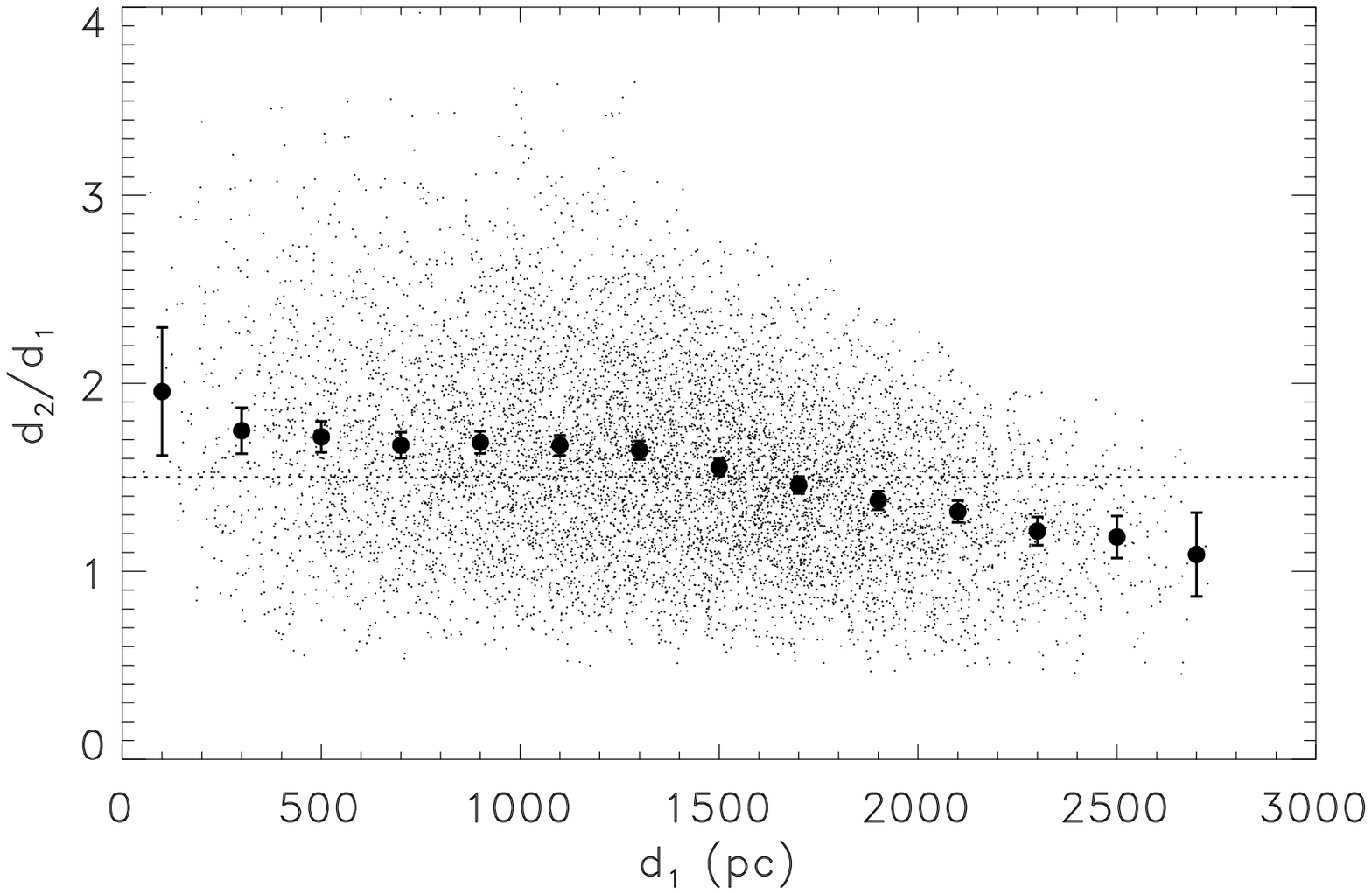}
\caption{{\bf Figure 19.} Distance ratio $d_2/d_1$ vs. $d_1$ for synthetic data with 
true ratio $B=1.5$; see text.}
\endfigure

\par
Fig.~20 shows the ratio $d_N/d_F$ as a function of log $R_F$ with $R_F$ based on 
the F08 $\varphi$ values and distances. (Recall that the former tend to be a little 
larger than those used with H07 and C99, about $0.05$ in the log, so the $R_F$ values 
will likewise tend to be that much larger.)  The most severe overestimation appears to be 
mainly at large $R$; the mean distance ratio is $2.00$ for $R_F > 1$ pc vs.~$1.22$ for the 
smaller objects. However, many of the latter have $d_F > 1$ kpc while none of the former do, 
so distance dependence accounts for some of the difference. As we showed in Section 4.1 
there is underestimation of distance with F08 for $R > 1$ pc that can qualitatively explain 
the larger apparent overestimation seen in the figure than in Fig.~17. There is thus no 
persuasive evidence of $R$-dependence with N01. 

\beginfigure{20}
\includegraphics[height=6cm,width=7.5cm]{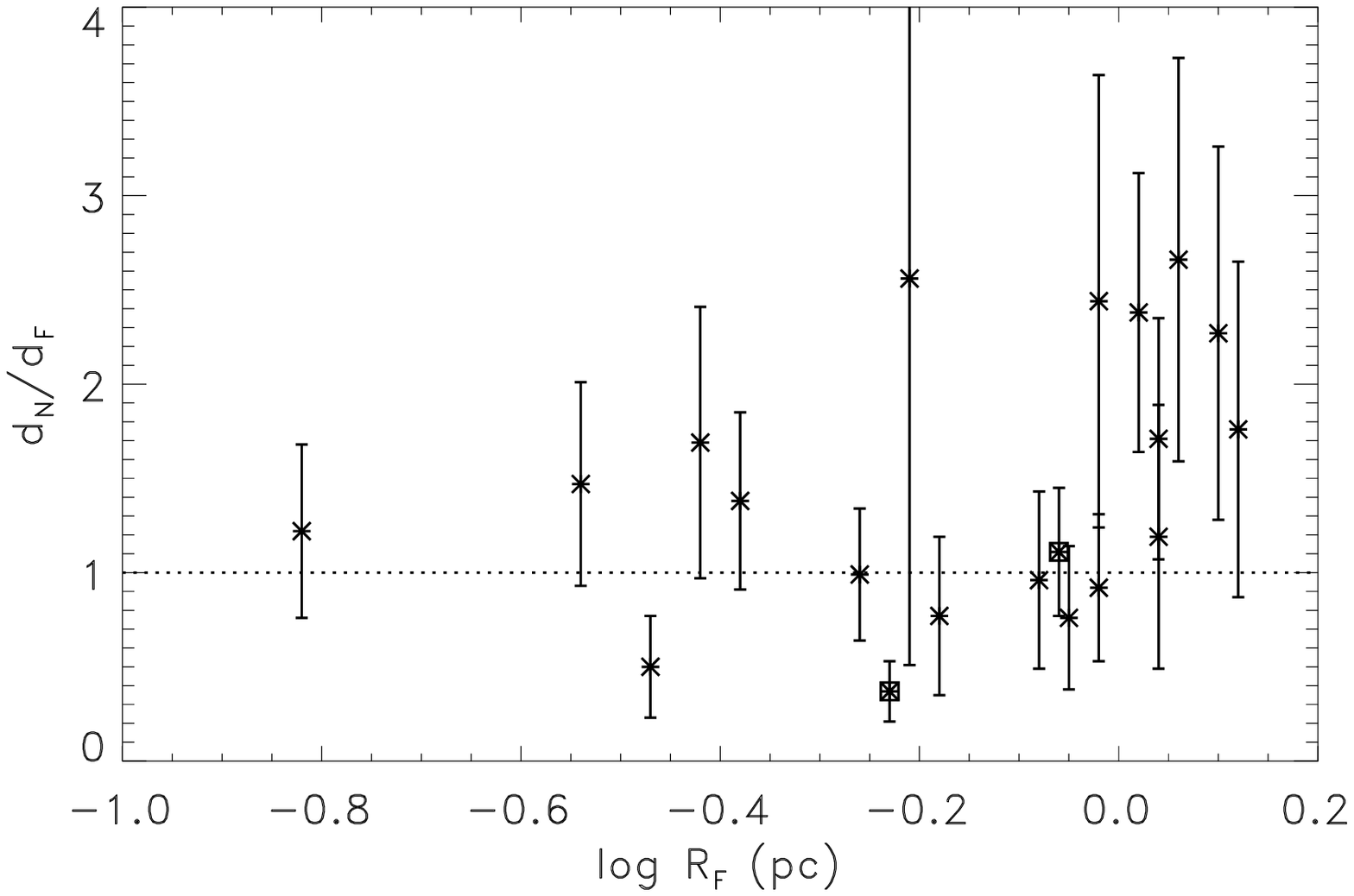}
\caption{{\bf Figure 20.} Distance ratio $d_N/d_F$ vs. log $R_F$; symbols as in Fig.~13.}
\endfigure

\par
Considering $d_N/d_{std}$ as a function of $V$ for just the objects with standard distances 
less than 1 kpc, we see signs of an increase going to fainter stars. For the four brightest 
stars, with $V<15$, the weighted mean distance ratio is $1.06\pm 0.15$; for the nine stars 
with $15<V<17$ it is $1.26\pm 0.16$; while for the four faintest stars ($17\leq V< 18$) it 
is $2.08\pm 0.57$. Breaking the ratios down according to $M_V$ instead, we have a ratio for 
the three brightest stars of $0.94\pm 0.16$, for the eleven stars with $M_V$ between 6 and 
8 $1.40\pm 0.16$, and for the three faintest stars $1.36\pm 0.33$. Thus the effect seems to 
depend primarily on $V$ rather than $M_V$, raising suspicion of some observational effect. 
\par
The correlation coefficient for the N01 and H07 distances is $0.89$; for the N01 and 
F08 distances it is $0.73$. Since we believe the precision of F08 is comparable to that 
of the original H07 parallaxes, we conjecture that the larger typical distances for the 
F08 nebul{\ae} along with the possible N01 distance dependence may weaken the correlation 
in the latter case. 
\par
A compilation of gravity distances was presented in table 6.6 of F08.  There seem to be two 
main groups of references, one associated with M\'endez and the other with Rauch and 
Werner. However, if we eliminate the distances that predate N01, the vast majority of those 
remaining are from the Rauch-Werner group. These approximate a homogeneous sample, so we 
will use them, designating the set F08-g.  
\par
Ten of the objects in the table have two or more distance estimates, allowing us to 
estimate a typical relative distance error.  The value we get is $\alpha_{F,g}^\prime =0.36$, 
slightly greater than the internal estimate for N01. 
\par
Using Eq.~(8) with the F08 distances for the F08-g set we get $\alpha_{F,g}^\prime = 0.30$, which 
may underestimate $\alpha_{F,g}$ a bit. However, we once again choose the value $0.35$. 
\par
The unweighted mean distance ratio for the F08-g distances compared to H07 ($N = 11$) is 
$1.39\pm 0.19$ and the weighted mean $1.09\pm 0.12$; the difference is consistent with 
weighting bias. The distance ratio for NGC 246, the one C99 object, is $1.50$. The median 
distance ratio for all twelve is $1.41$, and only two have ratios less than unity. 
Obviously the F08-g distances for this sample are overestimated, very much like the N01 ones.  
Yet comparing F08-g with F08 distances ($N = 18$) we find $\kappa = 0.79\pm 0.09$ and 
$\zeta = 1.01\pm 0.09$. Looking at the distance ratio as a function of distance (Fig.~21) we 
see a pattern similar to that for N01, namely overestimation for $d<$ 1 kpc and 
underestimation beyond. Note that this figure and the next include F08 data.  There is no 
evidence that the overestimation is mainly at large $V$ as with N01. There might be 
overestimation at large $R$, but it is hard to tell because there are so few points and 
those have large error bars. 

\beginfigure{21}
\includegraphics[height=6cm,width=7.5cm]{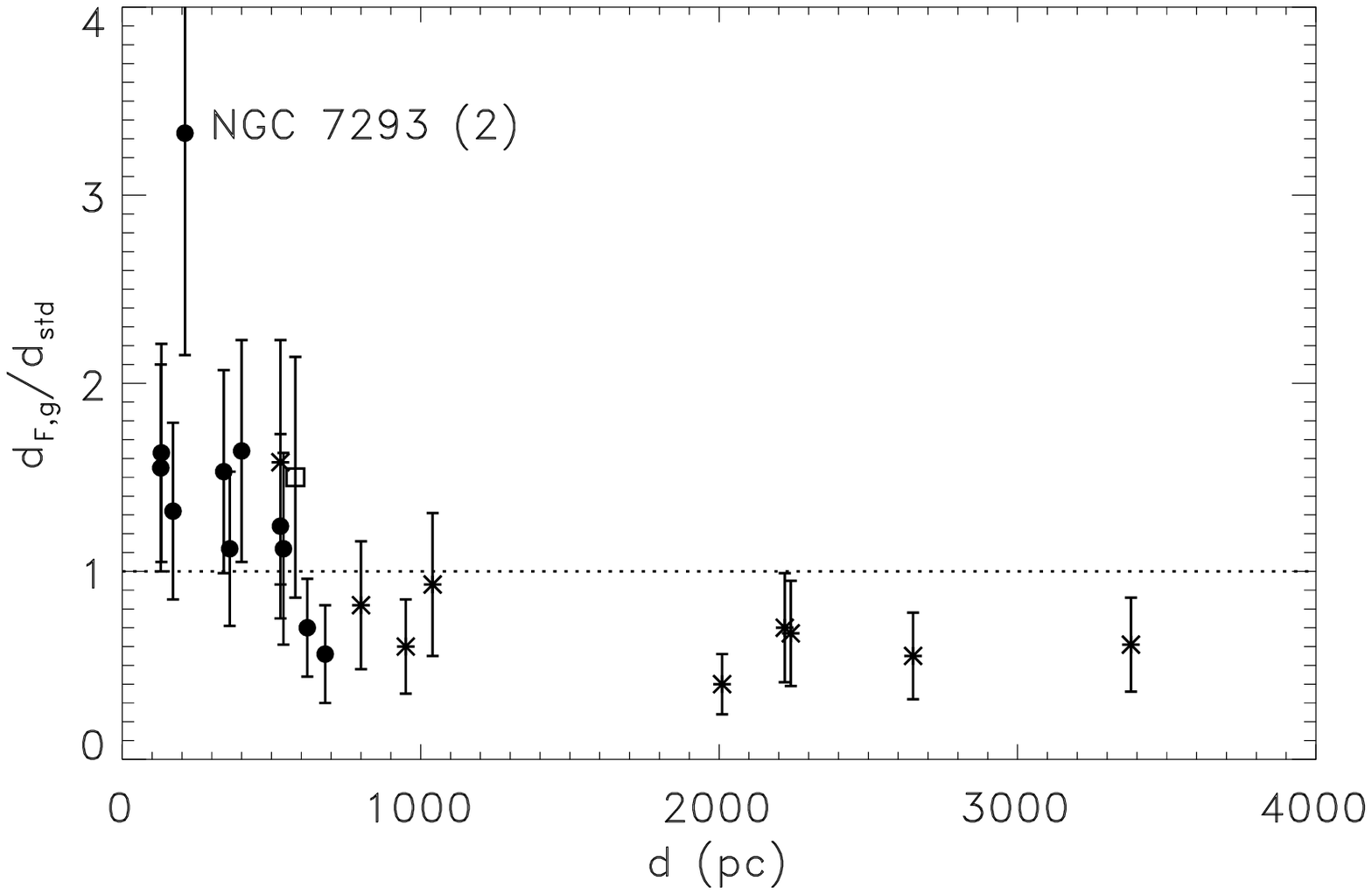}
\caption{{\bf Figure 21.} Distance ratio $d_{F,g}/d_{std}$ vs.~$d$; symbols as in Figs.~5 
and 10. Value for NGC 7293 is second one in table.}
\endfigure

\beginfigure{22}
\includegraphics[height=6cm,width=7.5cm]{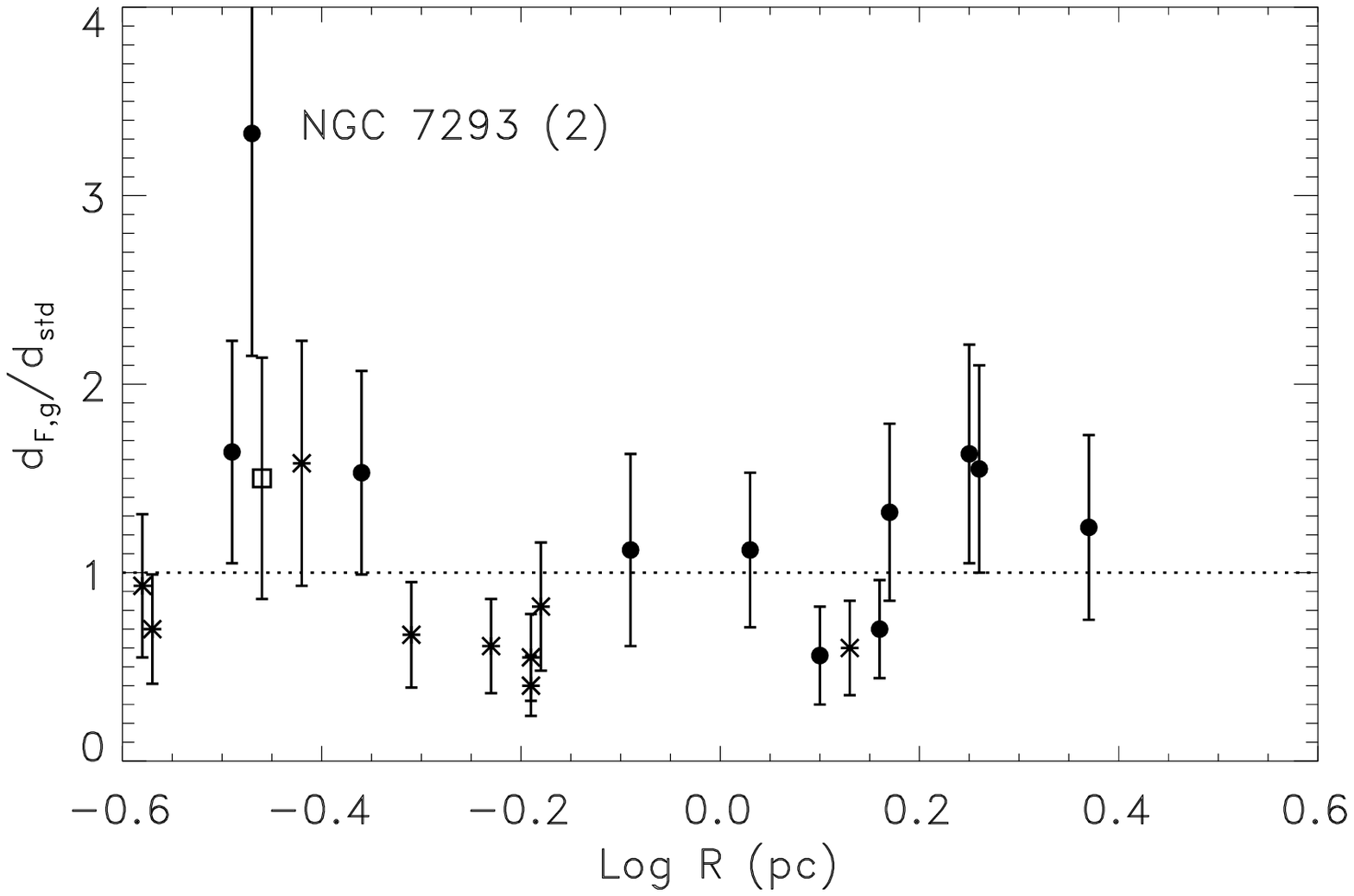}
\caption{{\bf Figure 22.} Distance ratio $d_{F,g}/d_{std}$ vs.~log R; symbols as in Figs.~5 
and 10.}
\endfigure

\par
Table 6 of P96 was a compilation of mean gravity distances from various sources, so it is 
inhomogeneous; however, the vast majority are from the Rauch-Werner group. We denote these 
by P96-g. It includes 13 of the H07 objects and two from C99. Some involve earlier distances 
from Napiwotzki (1993) and thus may not be independent of N01. Errors have been estimated 
using $\alpha_{P,g}=0.35$; with H07 Eq.~(5) gives $\alpha_{P,g}^\prime = 0.14$ (far too low 
in our opinion) while Eq.~(8) with F08 yields $\alpha_{P,g}^\prime = 0.27$. The mean distance 
ratio for the H07 objects is $1.15\pm 0.09$ and the weighted mean $0.99\pm 0.10$, a small 
weighting bias consistent with $\alpha^\prime$ for that set. For the two C99 objects 
$\kappa = 1.02$. Nine of the fifteen have ratios greater than unity, indicating little if 
any overestimation. For the F08 distances ($N=29$) we find $\kappa = 0.84\pm 0.06$ and 
$\zeta = 0.92\pm 0.06$. The median $d_{P,g}/d_F$ is $0.86$, with 18 of 29 ratios less 
than unity. Hence there may be a slight underestimation overall.  We find no indication of 
$R$-dependence in Fig.~22. Looking at the ratio as a function of $d_F$ (Fig.~23) there is a 
pattern like that of N01 but weaker. The correlation $r$ with H07 is $0.84$ and with C99 $0.83$. 
\par

\beginfigure{23}
\includegraphics[height=6cm,width=7.5cm]{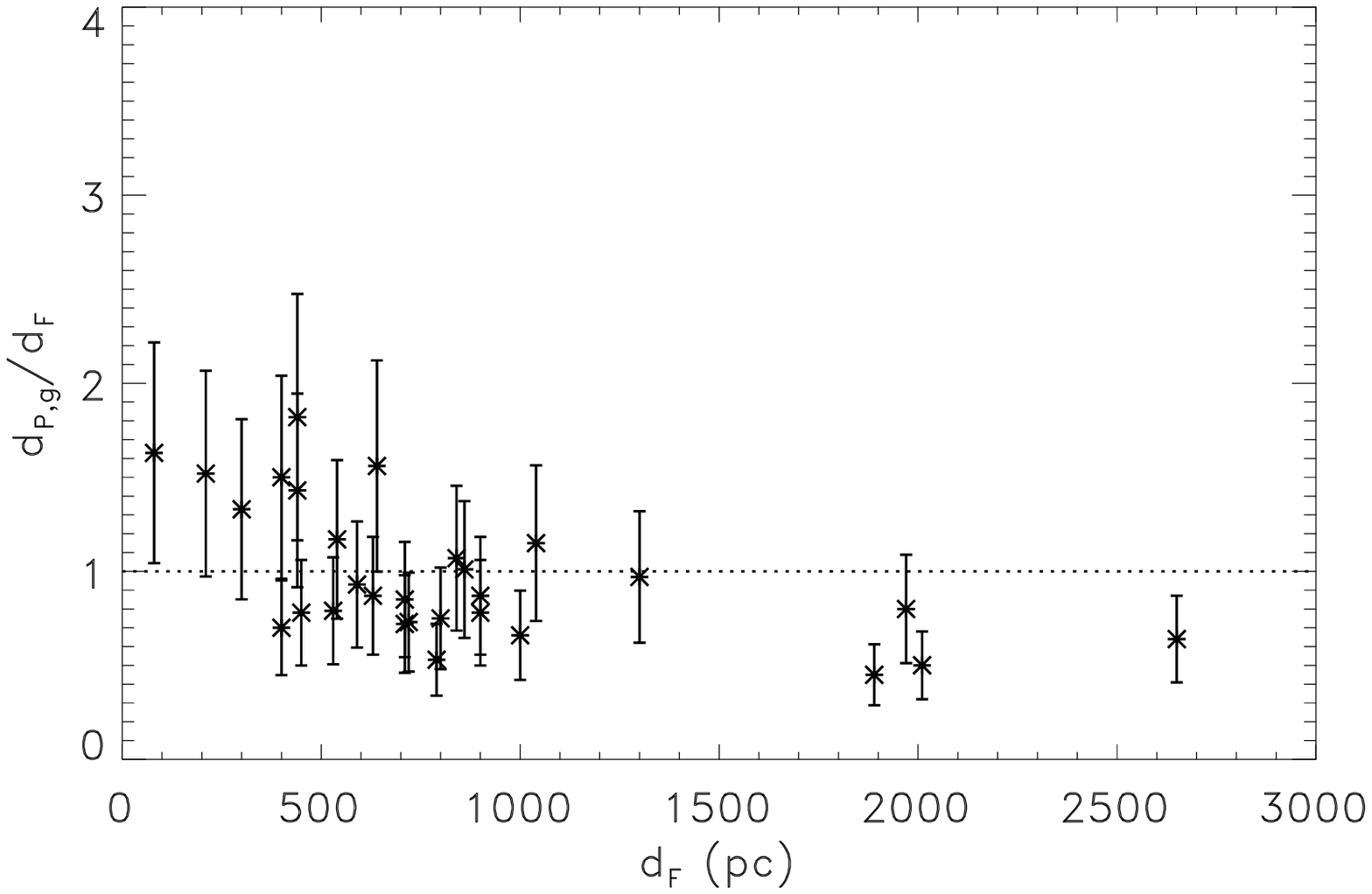}
\caption{{\bf Figure 23.} Distance ratio $d_{P,g}/d_F$ vs.~$d_F$; symbols as in Figs.~5 and 
10.}
\endfigure

There is independent evidence that error in log $g$ can cause overestimation. Some central 
stars of planetaries are PG 1159 type, for example that of NGC 246. For the GW Vir stars, a 
subset of that class that are non-radial pulsators, $g$ can be estimated using 
asteroseismology. Table 6 of C\'orsico \& Althaus (2006) compares spectroscopic estimates of 
$T_{eff}$, $M_*$, and log $g$ with those from models that fit the pulsations for four \lq 
naked' GW Vir stars (those without nebul{\ae}). The spectroscopic $g$'s are all smaller, by 
average $\Delta {\rm log}\ g$ of $-0.25\pm 0.06$. The differences in log $T_{eff}$ are 
negligible (mean $\Delta {\rm log}\ T_{eff} = -0.04\pm 0.05$); the mean $\Delta {\rm log}\ 
M_*$ is $-0.04\pm 0.01$. Applying Eq.~(10) we find that these differences imply 
$\Delta {\rm log}\ d = 0.11$, which corresponds to a distance overestimation of a factor 
$1.29$. However, asteroseismology involves modelling; in addition the average spectroscopic 
log $g$ for these high-gravity GW Vir stars is $7.38$, higher than 49 of the 60 values in 
table 6.6 of F08 and 24 of 27 in table 2 of N01. One does not know whether the same effect 
might occur at lower $g$. 

\subsection{5.2 \hdskip Expansion distances.}
Expansion distances are obtained by comparing an apparent angular expansion rate derived 
from measured changes in images from different epochs with a Doppler expansion velocity. The 
basic formula relating the angular expansion rate $\dot\theta$ and expansion velocity $V_R$ 
to the distance assuming spherical symmetry (not necessarily realistic) is, with $d$ in 
parsecs, $\dot\theta$ in mas yr$^{-1}$, and $V_R$ in km s$^{-1}$, 
$$d_{exp}={211\ V_R\over \dot\theta}\  \eqno{(11)}$$
\par
Potential problems with this approach are well known. In particular there are questions of 
interpretation such as the relation between ionization-front velocity and the gas flow 
velocity in the case of optically thick nebul{\ae} or between the latter and the 
motions of knots or the apparent \lq edge' of the nebula as well as departures from 
spherical symmetry. In recent years progress has been made with the astrophysical analysis 
in terms of ionisation fronts and shocks by Mellema (2004, hereafter M04) and studies using 
integrated star-nebula models (with coupling of stellar evolution and hydrodynamics codes) 
such as that of Sch\"onberner et al.~(2005, hereafter S05). There is now a recognized need 
for a correction factor $F$ generally greater than unity applied to the na\"{\i}ve 
expansion distance from Eq.~(11) to find the true distance. From that equation we have 
$$ F = \dot\theta^\prime\cdot {d\over 211\ V_R^\prime}  \  \eqno{(12a)}$$
where $\dot\theta^\prime$ is the measured angular expansion rate, $V_R^\prime$ is the 
measured expansion velocity, and $d$ is the true distance. The factor multiplying 
$\dot\theta^\prime$ is the inverse of the predicted expansion rate.  However, the relation 
can also be written 
$$ F = {\dot\theta^\prime\over 211\ V_R^\prime}\cdot\ d \  .\eqno{(12b)}$$
The first factor on the rhs is actually a kind of parallax. To compare distance scales 
we now obtain $F$ in the same way we did ${\cal R}$ previously, namely by multiplying 
the \lq parallax' by the distance. The difference is that whereas ${\cal R}$ is the 
ratio of the estimated distance to the actual distance $F$ is the inverse, the ratio 
of the actual distance to the estimated one, hence the term \lq correction factor.'  
This approach is, we believe, the proper one to use when evaluating expansion 
distances rather than dividing the expansion distance by the corresponding standard 
distance, especially if (as is often the case) the relative accidental error in 
$V_R^\prime$ is significantly smaller than that in $\dot\theta^\prime$. As before, it 
has the virtue of allowing us to avoid truncation bias by including zero and negative 
values of $\dot\theta^\prime$. (When looking at individual values we sometimes still 
may wish to look at the distance ratio.)  
\par
Two homogeneous samples of expansion distances obtained using {\it HST} are considered. The 
first is tiny, consisting of but three objects imaged and measured by Palen et al.~(2002). 
The second is a sample of fifteen nebul{\ae} measured by Hajian and collaborators and 
analyzed by Frew; the results for these after the latter's application of his own values of 
$F$ are taken from his table 6.5. We refer to this as the Hajian-Frew sample. 
\par
(Purely for reference we give here the formula to estimate $F$ for expansion distances 
compared with trigonometric parallaxes without inversion.  It is analogous to $\kappa$ but 
for parallaxes instead of distances:
$$F = \Sigma_{i=1}^N{\dot\theta_i^\prime\over 211\ V_{R,i}}\cdot\bigl(\Sigma_{i=1}^N\pi_i^\prime\bigr)^{-1}\ .\eqno{(13)}$$  
There is also a form analogous to our $\zeta$ that could be used if a single large parallax 
dominates. The uncertainty can be estimated using the jackknife.)

\begintable{15}
\caption{{\bf Table 15.} Correction factors $F$ for Palen et al.~(2002) {\it HST} 
expansion distances from F08 distances; $F_g$ from gradient method, $F_m$ from 
magnification method}
\vskip 0.1 in
\hrule
\vskip 0.1 in
{\settabs  4 \columns 
\+Name&$d_F$ (pc)&$F_g\pm \sigma_F$&$F_m\pm \sigma_F$\cr
\vskip 0.1 in
\hrule
\vskip 0.1 in
\+IC 2448 &2470&$1.75\pm 0.89$&$1.80\pm 0.61$\cr
\+NGC 6578&2290&$1.40\pm 0.79$&$1.15\pm 0.37$\cr
\+NGC 6884&2420&$1.55\pm 1.03$&$1.10\pm 0.30$&\cr
\vskip 0.1 in
\hrule
\vskip 0.1 in
}
\endtable

\par
The Palen et al.~expansion distances are compared with the F08 distances in Table 
15. They were derived using velocities $V_m$ given in that paper instead of catalog values. 
Rather than combine the results from the gradient and magnification methods (described in 
the reference) we have kept them separate.  The unweighted mean $F_g$ for the gradient 
method is $1.57\pm 0.10$, which is higher than one might expect; that for the magnification 
method is a more conventional $F_m = 1.35\pm 0.23$. The F08 distances for these three are 
practically identical, which means it is unfeasible to evaluate the correlation $r$ for 
distances from this sample and F08. 
\par
We do not estimate $F$ for the Hajian-Frew HST expansion distances because it has already 
been taken into account; also, we do not have the angular expansion rates. To evaluate these 
distances overall we use $\kappa$ and $\zeta$. Since there are F08 distances for only eight 
of them we use distances from F14 and Z95-$T_b$. Table 16 shows the individual distance 
ratios. The uncertainties given are combinations of the respective $\alpha$'s with the 
individual values of relative error according to F08. The median of those is $0.22$, 
whereas using Eq.~(8) with the F14 distances yielded $\alpha_{H-F}^\prime = 0.26$ and with 
the Z95-$T_b$ distances $0.35$. It thus appears that these distances are not as precise as the 
H07, C99, F08, and F14 ones but perhaps a little more so than Z95-$T_b$. 

\begintable{16} 
\caption{{\bf Table 16.} Distance ratios for Hajian-Frew sample vs. F14 and Z95-$T_b$ scales}
\vskip 0.1 in
\hrule
\vskip 0.1 in
{\settabs  3\columns
\+Name&${\cal R}_{H-F}$ (F14)&${\cal R}_{H-F}$ (Z95-$T_b$)\cr
\vskip 0.1 in
\hrule
\vskip 0.1 in
\+BD+30$^\circ$ 3639&$0.60\pm 0.15$&$0.64\pm0.24$\cr
\+\ \ \ \ \ =He 2-438&&&&\cr
\+IC 418&$0.85\pm 0.23$&$0.93\pm 0.36$\cr
\+IC 2448&$0.85\pm 0.26$&$0.69\pm 0.29$\cr
\+J 900&$1.24\pm 0.36$&$1.42\pm 0.58$\cr
\+NGC 3132&$1.09\pm 0.42$&$0.96\pm 0.46$\cr
\+NGC 3918&$0.96\pm 0.28$&$1.18\pm 0.48$\cr
\+NGC 5882&$0.89\pm 0.27$&$1.01\pm 0.42$\cr
\+NGC 5979&$0.72\pm 0.23$&$0.69\pm 0.30$\cr
\+NGC 6326&$1.80\pm 0.57$&$1.78\pm 0.77$\cr
\+NGC 6543&$1.36\pm 0.45$&$1.35\pm 0.59$\cr
\+NGC 6565&$0.81\pm 0.23$&$0.64\pm 0.26$\cr
\+NGC 6826&$1.69\pm 0.53$&$3.00\pm 1.26$\cr
\+NGC 6886&$1.37\pm 0.38$&$1.73\pm 0.69$\cr
\+NGC 6891&$1.10\pm 0.32$&$1.14\pm 0.47$\cr
\+NGC 7026&$2.30\pm 0.68$&$1.86\pm 0.76$\cr
\vskip 0.1 in
\hrule
\vskip 0.1 in
}
\endtable

\par
Ratios are with one exception (NGC 6826) fairly close for F14 and Z95-$T_b$. For F14 we have 
$\kappa = 1.16\pm 0.11$, $\zeta = 1.09\pm 0.10$, and the correlation is $0.71$. With 
Z95-$T_b$ we have $\kappa = 1.18\pm 0.15$, $\zeta = 1.14\pm 0.13$, and $r=0.61$.
The respective medians are $1.10$ and $1.14$. For the eight having F08 distance values 
$\kappa = 0.95\pm 0.12$. There may be a slight overestimation, but that is unclear. 

\subsection{5.3 \hdskip Interstellar extinction distances.}
\par
We have tested two more or less homogeneous samples of extinction distances against the H07 
and C99 standards and the F08 and Z95-$T_b$ statistical scales: Gathier, Pottasch, \& Pel 
(1986, hereafter G86) and the large sample of Giammanco et al.~(2011, hereafter G11). In 
addition we considered distances based on interstellar absorption in the Na D line (NS95). 
\par
There are 11 G86 nebul{\ae} with F08 distances, while all 12 have Z95-$T_b$ distances. 
The values of $\kappa$ and $\zeta$ are $0.84\pm 0.17$ and and $0.84\pm 0.18$ respectively 
for the F08 values and $0.92\pm 0.20$ and $0.91\pm 0.18$ for the Z95-$T_b$ ones, hinting 
that the scale is slightly underestimated overall. Correlation coefficients are $0.09$ and 
$-0.11$, which indicate highly unreliable individual distances. Values of $\alpha_{G86}^\prime$ 
are $0.53$ and $0.50$. 
\par
Seven H07 objects including DeHt 5 have Na D absorption distances. The mean of the distance 
ratios ${\cal R}_{Na}$ is $1.09\pm 0.12$, the median is $1.11$, and the weighted mean is 
$0.96\pm 0.12$.  The value of $r$ is $0.77$. No solution is obtained for $\alpha_{Na}^\prime$, 
suggesting that it is fairly small. The difference between the mean and the weighted mean, 
$0.13$, supports that inference. For fourteen objects having F08 distances 
$\kappa = 1.20\pm 0.15$, $\zeta = 1.17\pm 0.13$, and the median is $1.14$. The 
$\alpha_{Na}^\prime$ value for these is $0.34$, perhaps a bit high, and $r=0.59$, a marginal 
value. For five objects the Z95-$T_b$ distances give $\kappa = 1.20\pm 0.10$, 
$\zeta = 1.16\pm 0.08$, and median $1.15$.  Taking all these results together we regard the 
Na D scale as slightly overestimated overall but fairly accurate otherwise, between the old 
statistical scales and F08. 
\par
The sample in table 3 in G11 is mostly homogeneous with respect to data (using extinctions 
largely taken from A92) and homogeneous in its treatment. Only estimates were tested, not 
upper or lower limits to distance; the sample has 39 of these. However, four have 
$\varphi_{\rm Z95}\leq 1$ arcsec, which we felt makes the statistical distance uncertain, 
and one object was identified as a symbiotic star, so the sample ended up with only 34 
objects.  We do not have F08 distances for them, but we do have F14 and Z95-$T_b$ ones.  
\par 
Comparing G11 to F14 gives $\kappa=0.52\pm 0.08$ and $\zeta=0.59\pm 0.07$ whereas with 
Z95-$T_b$ $\kappa=0.56\pm 0.07$ and $\zeta=0.60\pm 0.08$, implying underestimation of nearly 
a factor of two. The median for F14 is $0.54$ and that for Z95-$T_b$ is $0.43$. Respective 
values of $r$ are $0.00$ and $-0.05$, so individual distances are unreliable.  The smaller 
\lq test' sample of extinction distances for six NGC objects in table 1 of G11 can be 
compared with the F08 distances; the results are $\kappa = 0.81\pm 0.24$, $\zeta=0.79\pm 0.19$, 
median $0.85$, and $r = -0.14$, basically the same as for G86. In figs.~4 and 5 of G11 we see 
that the chosen extinction value catches the \lq toe' of the extinction curve for four of 
them and runs along a gently sloping extinction curve in a fifth case, that of NGC 6842. The 
one well-defined solution is for NGC 6894, and its distance is underestimated (ratio $0.68$). 
There is however one strongly discrepant distance, that for NGC 6803 (950 pc as against 3930 
pc from F08). With that object omitted the values are $\kappa = 0.95\pm 0.16$, 
$\zeta = 0.91\pm 0.14$, $\alpha^\prime = 0.25$, and $r=0.63$.  G11's table 1 quotes two 
extinction values for NGC 6803 which according to their fig.~4 imply a distance more than 
twice as large as the one chosen, though still too low. The authors stated that as a general 
rule they selected the smallest extinction value when more than one was available. The 
extinction curves for the smaller sample seem to be mostly well-defined and suitable; perhaps 
the choices of extinctions and/or use of the nonlinear portions of the curves account for the 
underestimation problem with the larger sample. (We understand that the G11 results were 
somewhat preliminary and that improved values are forthcoming.) 
\par
Our judgment is that in general extinction distances have at present the following three 
problems: (1) the systematic errors can be considerable; (2) the correlations with standard 
scales are low enough to imply that individual distances cannot be trusted; and (3) there 
seem to be occasional outliers such as K 1-16 or NGC 6803.  One data set that seems to be 
an exception is Na D absorption, as noted above.  Any set for which $r$ is approximately 
zero can at best only be used to constrain the overall scale factor and not the slope.

\section{6 \hdskip Comparison of the {\it Hipparcos} parallaxes with the USNO system}
\par 
Frew (F08) made use of three {\it Hipparcos} parallaxes in his calibration sample; we, on 
the other hand, have eschewed the use of any for two reasons. First, the $\lambda$ values 
of most of the positive parallaxes are quite large. Correcting for bias effects with such 
large errors is virtually impossible. On the other side, selecting according to $\lambda$ 
is in our view undesirable even when $\lambda$ is fairly small (Smith 2006). Second, we 
believe that those parallaxes are systematically too large, truncation error aside.  
\par
In the preceding sections a consistent system of distances, the USNO system, for planetary 
nebul{\ae} has been developed, anchored using the H07 trigonometric parallaxes reinforced 
by those of B09 and expanded using other distance information: spectroscopic parallaxes, 
gravity distances, expansion distances, and statistical distances. We have emphasised the 
importance of checking for systematic differences between methods and between data sets for 
a given method. To incorporate the {\it Hipparcos} parallaxes into that system requires 
that they first be tested for systematic differences from it.  
\par
Only one object, PHL 932, is common to the {\it Hipparcos} and H07 parallax samples.  H07 
found the {\it Hipparcos} parallax from A98 to be $2.7$ times larger than theirs, slightly 
more than 2$\sigma$. This preliminary result was suggestive but not conclusive.  With van 
Leeuwen's (2007a, hereafter VL07) recent re-reduction of the {\it Hipparcos} data the error 
for PHL 932 was larger than the A98 one and rendered the difference no longer significant. 
However, the parallax ratio changed only very slightly, to $2.6$.  Only one object is 
common to the VL07 and C99 samples, NGC 246. The parallax ratio for it is $1.04\pm 1.53$. 
\par
Because the parallaxes from VL07 are improved over those presented in A98 we have used the 
former in our test. For the {\it Hipparcos} planetary nebula sample nine parallaxes improved 
in precision, three declined, and seven were essentially unchanged. 
\par
To further evaluate those parallaxes we combined distance estimates in the USNO system from 
all sources we consider to be reasonably accurate; we have included only those secondary 
distances with substantial correlation with our standards, typically $r\geq 0.6$. We have 
corrected for systematic error when deemed necessary, based on our estimates of same, 
and have assigned approximate relative errors and corresponding weights based on our 
intercomparisons of the various distance scales. Our primary distance estimates are those 
from H07 (+B09), C99, and F08, for which the typical $\alpha$'s are of order $0.2$, 
together with Z95-$T_b$, with $\alpha = 0.35$. Our supplemental or secondary sources are 
listed in Table 17; we consider all to have relative errors of roughly $0.35$, the same as 
Z95-$T_b$. Unit weight is assigned to that value, so the primary standards other than 
Z95-$T_b$ have weight 3. The assumed $\alpha$'s are mostly in reasonable agreement with 
those in table $7.4$ of F08.  The distances from supplementary sources are given in Table 
18, and the mean values from those are given together with primary estimates (when 
applicable) in Table 19. Also listed in this table are the parallax ratios ${\cal P}_H$ 
which are the mean distances multiplied by the respective {\it Hipparcos} parallaxes 
along with the respective uncertainties $\sigma_{\cal P}$ calculated using Eq.~(1a). The 
parallax ratios were then tested using the formalism employed previously with the H07 
parallaxes for testing other distance scales. 

\begintable{17}
\caption{{\bf Table 17.} Supplementary distance sources with adopted systematic 
corrections}
\vskip 0.1 in 
\hrule 
\vskip 0.1 in 
{\settabs 4 \columns 
\+Method&Ref.&$d$ range (kpc)&Multiplier\cr 
\+&&&for $d$&\cr
\vskip 0.1 in 
\hrule 
\vskip 0.1 in 
\+gravity&N01&$< 1$&$0.71$\cr
\+       &N01&$\geq 1$&$1.00$\cr
\+       &P96&all&$1.00$\cr
\+       &F08&$< 1$&$0.71$\cr
\+       &F08&$\geq 1$&$1.00$\cr 
\+expansion&F08 (H-F)&$< 2$&$1.00$\cr
\+spectroscopic&P96&all&$2.00$\cr
\vskip 0.1 in
\hrule
}
\endtable
\vskip 0.1 in

\begintable{18}
\caption{{\bf Table 18.} Distances $d$ in pc from supplementary sources}
\vskip 0.1 in
\hrule
\vskip 0.1 in
{\settabs  7 \columns
\+&&Source&&&\cr
\+Name&&N01&P96-g&F08-g&F08-p&P96-s\cr 
\+&&&&&(H-F)&\cr
\vskip 0.1 in
\hrule
\vskip 0.1 in
\+A 35&&--&--&150&--&400\cr
\+A 36&--&600&470&--&--\cr
\+BD+30$^\circ$ 3639&&--&--&--&1300&--\cr
\+He 2-36&&--&--&--&--&1560\cr
\+LoTr 5&&--&--&1900&--&840\cr
\+NGC 246&&--&420&620&--&940\cr
\+NGC 1360&&--&420&600&--&--\cr
\+NGC 1514&&--&--&--&--&800\cr
\+PHL 932&&170&520&--&--&--\cr
\vskip 0.1 in
\hrule
\vskip 0.1 in
}
\endtable

\par
The unweighted mean of ${\cal P}_H$ for all 19 is $2.14\pm 5.95$; the weighted mean 
(weighted as before by $(\sigma_{\cal R})^{-2}$, not by $w$) is $2.40\pm 0.47$.  The median 
${\cal P}_H$ for the entire sample is $2.65$; 14 of 19 ${\cal P}_H$ values are greater than 
unity, and 13 are larger than 2. For just those large ($\varphi\geq 160$ arcsec) objects for 
which parallaxes are more accurately measurable -- namely A 35, A 36, LoTr 5, NGC 246, NGC 
1360, NGC 1514, and PHL 932 -- the unweighted mean ${\cal P}_H = 2.38\pm 0.19$ and the 
weighted mean $2.40\pm 0.49$. For the six of those whose \lq best estimate' distances are 
less than 1 kpc the unweighted mean is $2.36\pm 0.23$ and the weighted mean $2.39\pm 0.50$. 
The median ${\cal P}_H$ for the six is $2.50$. Two objects, A 35 and NGC 1514, have a large 
fraction of the weight; A 35 has almost half, but even removing it yields a weighted mean of 
$2.37\pm 0.66$. Its \lq best estimate' distance is the mean of the F08-g distance of 150 pc 
and the corrected (by a factor of two) P96 spectroscopic distance of 400 pc. (There is no 
F08 or F14 distance for it because it is an imposter.)  Without the P96-s correction the 
weighted mean becomes $1.82\pm 0.39$. With both A 35 and NGC 1514 removed the weighted mean 
becomes $2.17\pm 0.75$. It appears the {\it Hipparcos} parallaxes for large nebul{\ae} are 
overestimated by at least a factor 2 but more likely $2.5$, consistent with the H07 result 
for PHL 932. 

\begintable*{19}
\caption{{\bf Table 19.} Distances in pc from primary sources and weighted mean supplementary 
distances with total weights together with resulting parallax ratios}
\vskip 0.1 in 
\hrule 
\vskip 0.1 in 
{\settabs 9 \columns 
\+Name&&Source&&&&&Mean&$w$&${\cal P}_H\pm \sigma_{\cal P}$\cr
\+&&H07&C99&F14 &Z95-$T_b$&Other&$d$&\cr
\vskip 0.1 in 
\hrule 
\vskip 0.1 in 
\+A 35&&--&--&--&--&280&280&2&$2.35\pm 0.74$\cr
\+A 36&&--&--&770&640&540&670&6&$2.88\pm 1.88$\cr
\+BD+30$^\circ$ 3639&&--&--&2180&2040&1300&1980&5&$0.40\pm 4.65$\cr
\+He 1-5&&--&--&3720&4570&--&3930&4&$21.14\pm 7.97$\cr
\+He 2-36&&--&--&2640&2910&1560&2480&5&$-2.80\pm 5.74$\cr
\+He 2-138&&--&--&3770&3490&--&3490&1&$3.48\pm 8.66$\cr
\+He 3-1333&&--&--&9560&--&--&9650&3&$-94.74\pm 142.53$\cr
\+Hu 2-1&&--&--&4070&5060&--&4320&4&$16.16\pm 32.84$\cr
\+LoTr 5&&--&--&910&6930&1370&2070&6&$2.50\pm 2.37$\cr
\+M 2-54&&--&--&7710&14460&--&9400&4&$6.77\pm 33.33$\cr
\+NGC 40&&--&--&1070&1080&--&1070&4&$-1.68\pm 3.89$\cr 
\+NGC 246&&--&580&770&690&660&670&10&$1.42\pm 2.04$\cr
\+NGC 1360&&--&--&560&590&510&550&6&$2.02\pm 1.36$\cr
\+NGC 1514&&--&--&650&740&800&700&5&$2.65\pm 1.21$\cr
\+NGC 2346&&--&--&1340&1600&--&1500&4&$-0.56\pm 4.28$\cr
\+NGC 2392&&--&--&1390&1210&--&1260&4&$7.56\pm 5.31$\cr
\+PHL 932&&300&--&--&--&350&320&5&$2.83\pm 1.50$\cr
\+SaSt 2-12&&--&--&3580&--&--&3580&3&$30.97\pm 12.87$\cr
\+SwSt 1&&--&--&2810&4460&--&4460&1&$37.38\pm 29.15$\cr
\vskip 0.1 in
\hrule
}
\endtable

\par
There is no indication of truncation according to $\pi^\prime$ with the sample, given the 
negative values and ones with large $\lambda$; hence truncation bias from selection can be 
ruled out.  The T93 list was used in selecting the {\it Hipparcos} sample. However, as we 
have shown those distances are not strongly correlated with the Z95 distances and probably 
are not correlated with the Z95-$T_b$ or F08 scales either. Even if they were, however, 
${\cal P}_H$ would be systematically underestimated, not overestimated. The 1 kpc sample 
may well have truncation bias, but here again it would cause underestimation, not 
overestimation.   
\par
It has been pointed out by van Leeuwen (private communication) that our analysis to this 
point is based on the normality assumption and hence does not precisely model the actual 
errors. Strictly speaking he is correct; however, the deviations from normality are quite 
small in most cases with the 1 kpc sample, because the relative errors of our \lq best 
estimates' are mostly small, as inferred from the weights $w$ in Table 19. To check his 
point we have sampled the distribution of distance ratios obtained from the errors in $d$ 
and $\pi^\prime$ using synthetic data obtained with a psuedo-random number generator to 
produce large ($N=2\times 10^7$) samples. As can be seen in Fig.~24 the distance ratio 
distribution from these synthetic data is actually below the Gaussian for distance ratios 
less than unity in the case of A 35, which means that the Gaussian {\it overestimates} the 
probability that the ratio is unity or less.  For the other five objects the curves are 
fairly similar to the corresponding Gaussians though with the peak shifted to lower ratios 
as with A 35. 

\beginfigure{24}
\includegraphics[height=6cm,width=7.5cm]{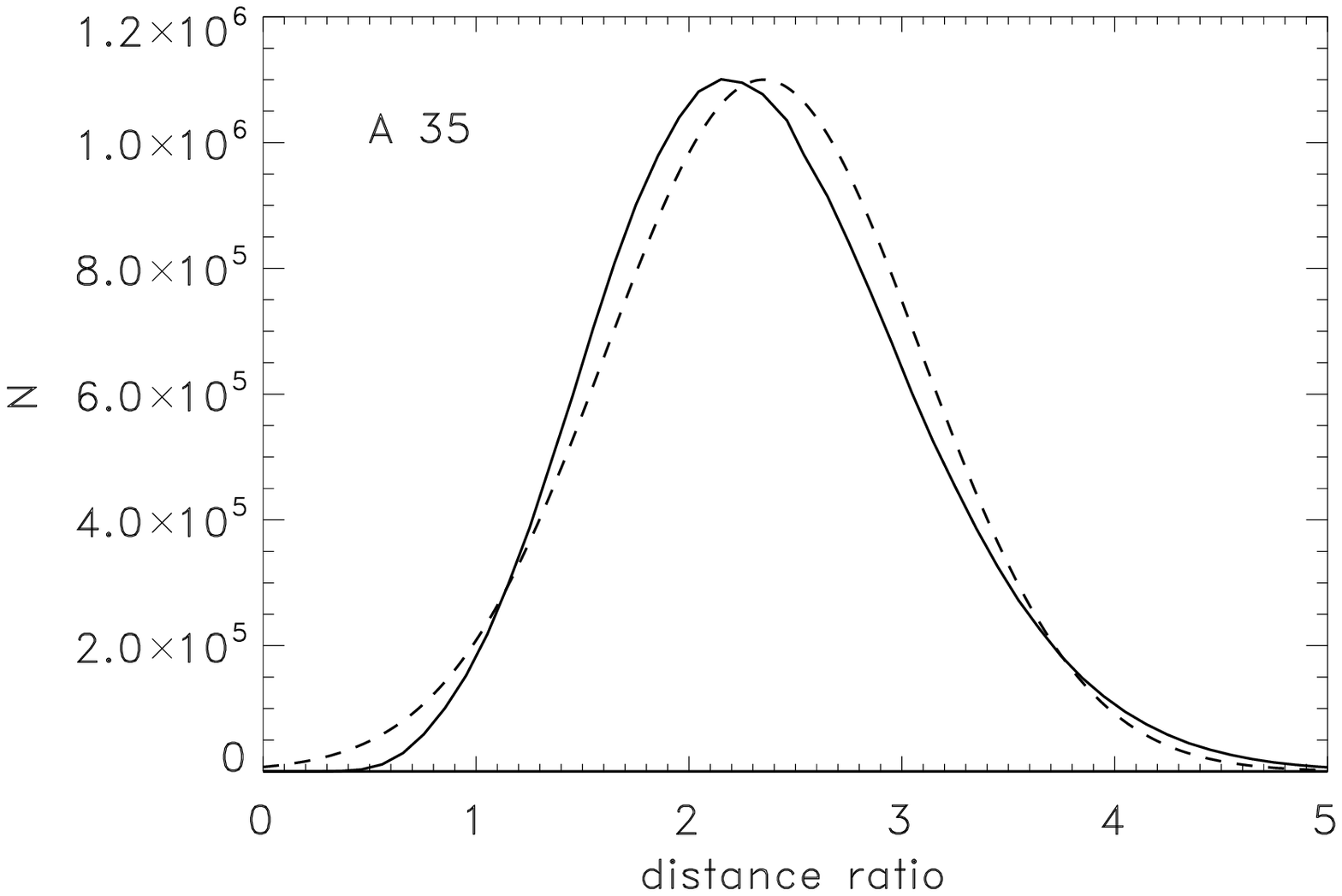}
\caption{{\bf Figure 24.} Distribution of distance ratios for A 35. The {\it solid 
curve} is the number of values of the given ratio found using synthetic data (see text) 
with the values for the {\it Hipparcos} parallax and the \lq best estimate' distance and the 
respective errors; the {\it dashed curve} is a Gaussian with parameters given in Table 20 
and maximum matched to the distribution for the synthetic data.}
\endfigure  

\par
What is the probability of getting the observed values of ${\cal P}_H$ if the true parallax 
is the inverse of the \lq best estimate' distance given the {\it Hipparcos} 
$\sigma_\pi^\prime$?  Synthetic data (same $N$) give the values presented in Table 20. Also 
presented therein for comparison are the probabilities from the Gaussian approximation.  As 
we have said the Gaussian values are if anything conservative. 

\begintable{20}
\caption{{\bf Table 20.} Probability of given ${\cal P}_H$ being equalled or exceeded by 
chance given true distance equal to \lq best estimate' and {\it Hipparcos} $\sigma_\pi^\prime$}
\vskip 0.1 in
\hrule
\vskip 0.1 in
{\settabs 3 \columns
\+&$p(\geq {\cal P}_H)$&\cr
\vskip 0.1 in
\+Name&Gaussian&synthetic\cr
\vskip 0.1 in
\hrule
\vskip 0.1 in
\+A 35&$0.034$&$0.014$\cr
\+A 36&$0.155$&$0.146$\cr
\+NGC 246&$0.417$&$0.419$\cr
\+NGC 1360&$0.221$&$0.214$\cr
\+NGC 1514&$0.072$&$0.063$\cr
\+PHL 932&$0.111$&$0.105$\cr
\vskip 0.1 in
\hrule
\vskip 0.1 in
}
\endtable

\par
It seems highly unlikely the H07 parallaxes could be underestimated by as much as a factor of 
two.  The B09 results are not consistent with such a change. Furthermore, it would greatly 
exacerbate the problem with the gravity distances and be inconsistent with the Hajian-Frew 
corrected expansion distances.  
\par
To summarize, our \lq best estimate' distances indicate that the {\it Hipparcos} parallaxes 
for the larger planetary nebul{\ae} are overestimated by a factor of $2.5$. This result 
seems to us to be fairly robust. We have chosen not to apply a uniform correction to 
these parallaxes for this error in order to use them. 

\section{7 \hdskip Calibration strategies and the \lq short' and \lq long' statistical scales}

In Section 1.1 we commented on the inclusive and eclectic strategies employed in selecting 
calibration samples for these scales.  In our opinion CKS used the latter: Calibration was 
based on two sets of extinction distances, Kaler \& Lutz (1985) and G86; two examples of 
cluster membership, Ps 1 and NGC 2818; and two spectroscopic parallaxes, those of NGC 246 
and NGC 1514. Expanding shell and gravity distances were explicitly excluded. None of the 
calibration objects they used except NGC 246 belong to H07 or C99; their distance for it was 
430 pc, 36 per cent less than our \lq best estimate.' Their distance for NGC 1514 was 6 per 
cent lower than our \lq best estimate.' The Kaler \& Lutz distances seem to be more or less 
correct in the mean (results omitted here) and the G86 ones perhaps a bit underestimated. The 
large disparity in cluster distances prevents using $\kappa$. For those two based on the Z95-$T_b$ 
distances we have $\zeta = 1.17$, while F14 gives $0.93$. On the whole we expect a short 
scale, as is the case by 11 per cent. 
\par
Obviously different selections lead to substantially different calibrations. If the 
expansion distances had been used but without the astrophysical correction referred to in 
Section 5.2 they would also have been underestimated, by roughly 25 per cent. On the other 
side, if the seventeen gravity distances from M\'endez et al.~(1988) had not been excluded 
(because they seemed systematically too high and because they were model-dependent) they 
would have made CKS longer. As a matter of fact, for those compared to F08 we find 
$\kappa = 1.57$ ($N$ = 9), while with Z95-$T_b$ we have $\kappa = 1.62$ ($N$ = 15). Assuming 
simply weighting by number they would have changed the CKS distance ratio from $0.89$ to $1.18$.  
\par
At first glance the Z95 calibration appears to be an extreme example of the eclectic 
strategy: It used a single calibration sample taken from an earlier paper by Zhang (1993) 
and modelling of the evolution of nebula-central star systems to obtain evolutionary tracks of 
distance-independent observational quantities, e.g.~$T_b$ for the nebula and $T_*$ for the 
central star. Position on an evolutionary track in the $T_b$-$T_*$ diagram is related to 
properties of the underlying model systems which can yield distances: central star mass $M_*$, 
luminosity $L_*$, and surface gravity $g$. In practice it was not quite so simple.  For 
example, values for the observed quantity $T_*$ were obtained in four different ways. Also, 
two different methods were used to obtain distance estimates, one using $M_*$, $L_*$, and 
$g$ in a formula like Eq.~(9) with $g$ from the track rather than high-resolution 
spectroscopy and the other using $L_*$ from the track together with estimated stellar flux 
over all wavelengths, thus bypassing $g$. 
\par
In Section 4 we showed the two scales Z95-$M_i$ and Z95-$T_b$ to be systematically different 
even though they used the same calibration sample. The difference arose because of 
correlation of random errors with the former scale, a flaw in the methodology resulting in 
miscalibration. 
\par
The approach taken in F08 exemplifies the inclusive approach, with a large ($>120$) 
calibration sample incorporating a wide variety of distance estimates. As we have shown some 
of those, in particular from the gravity method, the extinction method, and a few {\it Hipparcos} 
parallaxes, have systematic errors. In contrast to our ideal approach the compilations of data 
in F08 are in many cases heterogeneous, with numerous sources for each method. Sparse data sets 
or even single values are included. Yet as we have shown the F08 scale is basically on the USNO 
system. It thus appears that systematic errors had little net effect. To be sure, Frew 
recognized some of the systematic errors and mitigated them with some success, for example  
with the expansion method. In our view the only defect of the F08 scale (and the F14 one as 
well) is the radius dependence (underestimation) we believe is present at large $R$, in the 
same sense as with CKS and SSV but substantially less severe. It arises from the assumption 
of a single power law. With the F14 statistical scale the short-long dichotomy is resolved. 
\par
The agreement of the mean of the \lq long' statistical distances for the bulge planetaries 
with the distance of the galactic centre was for some time regarded as a fairly strong 
argument in favour of those. In Z95 and VdSZ the distance distributions of nebul{\ae} 
satisfying certain criteria for bulge membership were presented having a peak around 8 kpc 
(the presumed distance of the galactic centre). However, as N01 noted the Z95 distance 
distribution of bulge planetaries looked somewhat skew, hinting that it is sculpted by 
observational selection. A plot in $d_p$-$z$ space, where $d_p$ is the projected distance in 
the galactic plane and $z$ is the absolute value of the galactic $z$ coordinate, confirms 
this conjecture and, we contend, reveals that this identification of the peak with the 
galactic centre is indeed dubious. Instead the distribution is what one would expect as a 
result of extinction near the galactic plane (see Smith 1976). Fig.~25 shows just such a 
plot for the Z95 \lq bulge' sample with $d_p$ based on the Z95-$T_b$ distances. Obviously 
the sample is strongly affected by some type of selection. A curve whose form represents the 
cutoff imposed by a limit on surface brightness dimmed by interstellar extinction having an 
exponential $z$-distribution with scale height $0.4$ kpc has been included, an idealized 
model. The MASH survey (Parker et al.~2006; Miszalski et al.~2008) contains a large number 
of new nebul{\ae} located in the bulge direction, confirming the Z95 sample's 
incompleteness. Thus a piece of evidence that seemed to favor the \lq long' scale in fact 
does not. F14 contains an analysis of data on the bulge planetaries. 

\beginfigure{25}
\includegraphics[height=6cm,width=7.5cm]{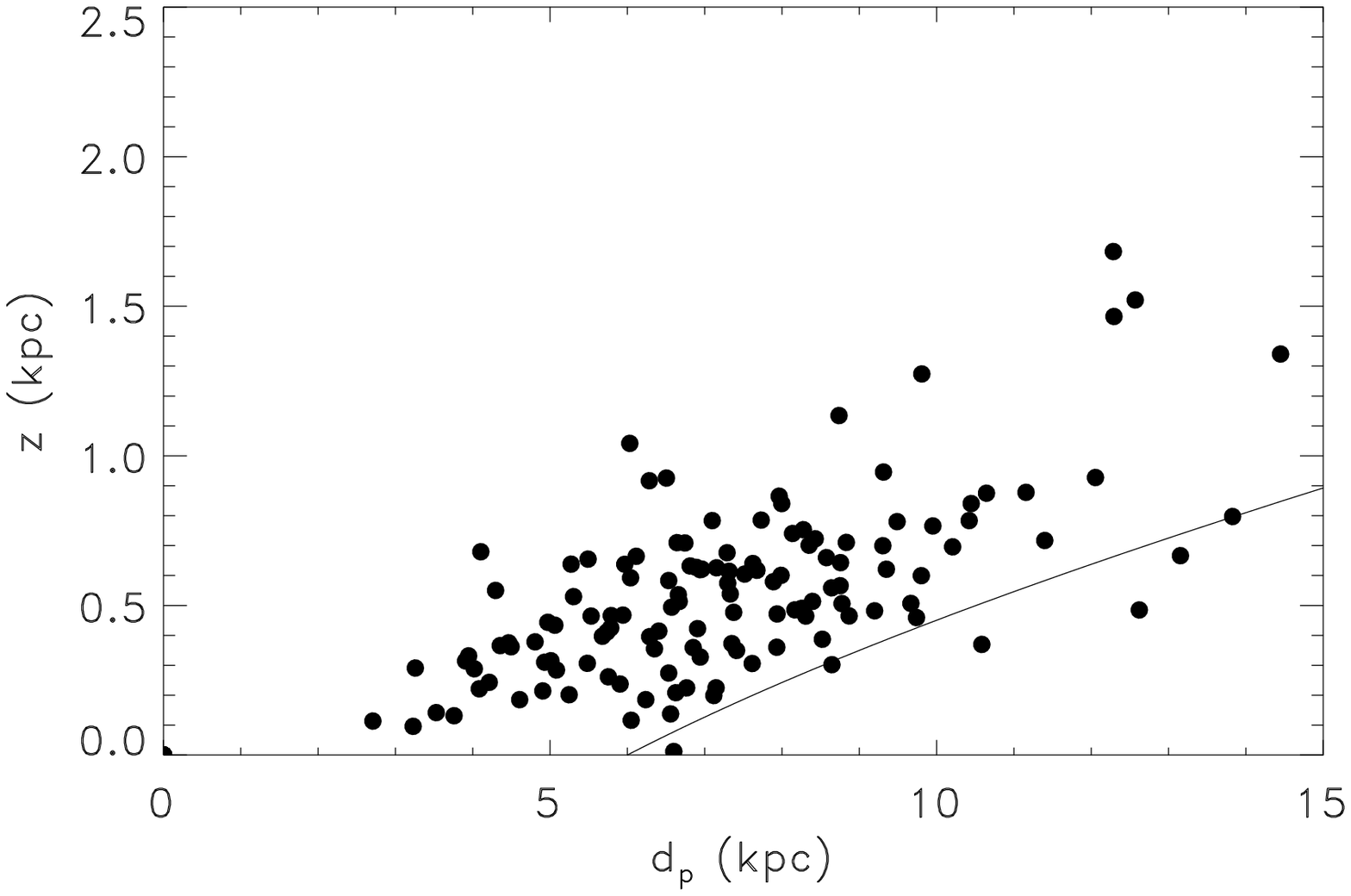}
\caption{{\bf Figure 25.}  Plot for galactic bulge objects from Z95 of absolute value of 
$z$-distance vs.~~projected distance $d_p$ (see footnote to Table 12) based on $d$ from 
Z95-$T_b$. The {\it solid curve} indicates how observational selection based on surface 
brightness in the presence of extinction might sculpt a lower envelope on the distribution.}
\endfigure

\section{8 \hdskip Summary and Discussion}
\subsection{8.1  \hdskip Comparing distance scales: some general conclusions}
\par
Although our primary subject in the present paper is construction of a system of distances for 
planetary nebul{\ae} we have considered the more general problem of comparing one distance 
scale to another, \lq standard' set of distances, both when the standard distances are based 
on trigonometric parallaxes but also when the distances have a normal or lognormal error pdf.  
Hence our findings may be of broader interest. 
\par
In the former case it is best to multiply the distances being tested by the respective 
parallaxes.  The alternative, converting the parallaxes to distances and then dividing those 
by the distances to be tested, yields a biased estimate. If possible one ought to avoid 
discarding negative parallaxes or imprecise ones; by not discarding one completely eliminates 
truncation bias.  Transformation bias is avoided by not converting parallax to distance. 
If weights are used based on the estimated errors $\sigma^\prime_{\cal R}$ in the distance ratio 
there is likely to be a weighting bias of the sort discussed by Smith (2006). Yet if weights 
which depend solely on the estimated parallax error $\sigma_\pi^\prime$ are employed the 
estimate may not be efficient, since the parallax error is just one contributing factor to 
the error in the distance ratio and will only be dominant if $\lambda$ is substantially larger 
than $\alpha$, the relative error in the test distances; see Eq.~(6). 
\par
In the latter case above (of a normal or lognormal error pdf for the standard distances) we 
find that Phillips's (Ph02) $\kappa$ and (Ph04) $\Gamma^*$ respectively can be useful for 
estimating the distance ratio, being superior to the mean of the individual distance ratios, 
which yields a biased and less precise estimate of the overall ratio. On the other hand, the 
Phillips estimators are potentially subject to domination by one or two extremely large 
values, as noted in Ph04. Also, in cases where the test distances have an error pdf which 
differs in form from that of the standard distances there can be bias; which of the two 
Phillips estimators is preferable depends on the parameters -- typical relative errors 
and/or typical logarithmic errors -- of the situation. Domination by a few large values, 
especially acute for $\kappa$, can be mitigated by using our proposed estimator $\zeta$ 
instead, at the cost of introducing a (usually) modest weighting bias. 
\par
When $\kappa$ and $\zeta$ differ substantially without domination by an outlier one can 
sometimes infer the presence of distance-dependent systematic error in one (or both) of the 
scales compared. The latter estimator is more nearly unaffected, although it can have 
weighting bias. 
\subsection{8.2  \hdskip  Calibration and testing of distance scales for planetary nebul{\ae}}
Because trigonometric parallaxes are independent of astrophysical models and (in principle, 
at least) free of systematic error they are a logical choice for serving as the basis for a 
system of distances. Other distance methods can then be tested using those parallaxes, 
either directly or indirectly, by using an intermediary distance scale. Some care is needed 
in using parallax data to avoid introducing bias. One potential complication with testing 
is a distance dependence in the intermediary scale or the one being tested. Another 
complication is the fact that distance methods for planetary nebul{\ae} can have 
radius-dependent systematic error. Here we have shown several ways it can arise in the 
calibration of statistical scales. By restricting distance scale comparisons to a narrow 
range in $R$ radius dependence can be mitigated as a cause of confusion. Indirect comparison 
ties the C99 spectroscopic parallaxes (but not the P96 ones) to the USNO trigonometric ones, 
and $T_b$-$R$ relations for the two are consistent. Similarly, overlap in $R$ ties the 
Magellanic Cloud $T_b$-$R$ relation to the one based on H07 and C99. The unified relation is 
mostly consistent with those of F08 and F14, which are nearly identical to each other; a 
possible exception is for $R> 1$ pc where the latter seem underestimated. The F14 results 
(and ours) resolve the issue of \lq short' vs.~\lq long' statistical scales more nearly in 
favor of the former. 
\par
As part of the construction of a system of distances we looked at several additional methods 
for which we found a few sizeable homogeneous data sets: gravity, expansion, and extinction. 
As others have noted (e.g.~B09) the N01 gravity scale seems to be too long for nearby 
objects (although we think it underestimates larger distances); gravity distances generally 
seem to have this problem, probably by underestimating $g$, with the possible exception 
of the heterogeneous sample in P96. Expansion distances need correction using astrophysical 
modelling (cf. F14 for an example) but can yield usable results. Extinction distances are 
highly problematic. Distances for one moderately large set were underestimated by almost a 
factor of two. Individual distance estimates can be close but mostly are not:~ correlation 
coefficients are generally near zero. The interstellar Na D absorption distances appear to 
be an exception, with overestimation of perhaps only 20 per cent and $r=0.77$. 
\par
For our anchor strategy to work the trigonometric parallaxes must be accurate; the 
{\it Hipparcos} ones are not, so we omitted those. Comparison of the VL07 parallaxes with 
our \lq best estimate' distances in the USNO system confirms the H07 preliminary finding of 
a systematic error in the parallaxes of approximately a factor of $2.5$ for the large 
nebul{\ae}. We have no explanation for this fact; at this time we can only point out several 
clues. There is no large overestimation in the {\it Hipparcos} parallaxes generally 
according to van Leeuwen's (2007b) validation study; therefore the cause must be something 
intrinsic to these particular objects. Two  distinguishing features are (1) the presence of 
nebulosity and (2) the relative faintness of the central stars -- as is well known planetary 
nebul{\ae} are often brighter than their central stars. One can argue that especially for 
the large-$\varphi$ objects the surface brightness is quite low, too low for them to show 
up. However, what was measured by {\it Hipparcos} was photon counts, not images. As far as 
we can tell the validation study considered only stars, not extended sources. While the 
contribution by that portion of a large nebula within the detector field at a given instant 
may be slight the effects on multiple stars in the vicinity of that nebula might add up to a 
significant effect on the nebula's parallax, through correlated errors. It is quite definite 
from A98 (section 3.1) that the nebulosity affected the parallax measurements for the small 
objects, not systematically but randomly, leading to noticeably larger uncertainties than 
for other stars of similar brightness; this is true of the VL07 parallaxes for those same 
ones.  For the five planetaries that are neither small nor large -- He 1-5, He 2-36, NGC 40, 
NGC 2346, and NGC 2392 -- the median is actually negative, $-0.55$.  In fact, three of the 
four negative VL07 parallaxes belong to this group. The angular diameters of the three 
according to A98 are 10, 36, and 55 arcsec, to be compared to the 38 arcsec {\it Hipparcos} 
field. (The fourth is one of the smallest nebul{\ae}.)  The factor of $2.5$ is decidedly 
odd. 
\par
Melis et al. (2014) very recently claimed that the {\it Hipparcos} parallaxes for the 
Pleiades are overestimated by about 10\% based on very-long-baseline radio interferometry. 
If so, that overestimation might be similar to what we have found with the nebul{\ae} but 
less severe, perhaps arising because of the reflection nebulosity in the cluster. The effect 
may be smaller because the relative brightness of the stars is greater. 

\subsection{8.3  \hdskip   Some suggestions for future work}
First and foremost, we crucially need more high-precision trigonometric parallaxes. It may 
seem as if this need will be more than satisfied by {\it Gaia}. However, if we are correct 
about the {\it Hipparcos} parallaxes having substantial systematic error there is then a 
concern that {\it Gaia}, which operates on the same principle, might also be afflicted.  
Hence there is a need for more {\it HST} and CCD parallaxes if for no other reason 
than to serve as a check on {\it Gaia} for these and similar objects.  Thus far all the 
accurate parallaxes we have are for objects with $\delta > -21^\circ$; the southern sky 
is practically untouched. As we already noted, almost all the H07 objects that are true 
planetaries have large $\varphi$; one might wish to consider especially such objects at 
galactic latitudes greater than 20$^\circ$ in absolute value as being more likely to be 
nearby. 
\par
Second, the gravity method needs to be refined; empirical corrections can be used, but it 
is preferable to understand why it overestimates distance, at least for the nearest nebul{\ae}, 
and remedy the problem or at least mitigate it.  Asteroseismology can be a useful testing 
tool as well as a means of obtaining accurate $g$'s for getting distances, but it is 
model-dependent and therefore must itself be checked. 
\par
Third, the expansion method might benefit from switching to a different wavelength 
region, using the Atacama Large Millimeter/Submillimeter Array (ALMA) to observe structure 
in the outer nearly-neutral regions of the nebul{\ae}. A digital counterpart of the 
Griffin \lq mask' technique (Griffin 1967) for radial velocity measurement using multiple 
molecular lines originating from a given structure in the nebula might facilitate precise 
velocity measurement. ALMA has extensive coverage of both northern and southern sky. Of 
course this might necessitate further detailed modelling along the lines of M04 and S05 to 
establish the expansion of the PDR on the outside. 
\par
Fourth, more companions of central stars are being found (De Marco et al.~2013) and hence 
more spectroscopic parallaxes may be forthcoming. This approach would benefit from a uniform 
reduction of all the available data using a single well-calibrated set of color-absolute 
magnitude relations or, failing that,  several cross-correlated ones. It is highly 
desirable to tie that calibration to the USNO system. 
\par
We believe the problem of calibrating a statistical distance scale will not be ripe 
for further efforts until significant progress has been made on the foregoing, especially 
the first. The F14 scale is probably the best statistical one we now have, though perhaps 
needing some modification at large $R$. 
\section*{ACKNOWLEDGMENTS}
The author has made liberal use of the SIMBAD database and VizieR catalogue access tool 
operated by CDS at Strasbourg, France, and the ADS query system provided by the Smithsonian 
Astrophysical Observatory and NASA in this research. The original description of the VizieR 
service was published in A\&AS 143, 23.  Thanks to Drs.~R. Ciardullo, H.~Harris, Q.~Parker, 
F.~van Leeuwen, R.~Napiwotzki, D.~Frew, J.~Phillips, and S.~Pottasch for helpful comments 
and criticisms and especially to Dr. Frew for providing a copy of his thesis and a preprint 
of his 2014 paper and also to Dr.  A.~Hajian for granting his permission to use the 
Hajian-Frew results. The author is also very grateful to two anonymous referees for a number 
of constructive criticisms and suggested improvements. Blame for any faults is of course the 
author's alone.

\section*{APPENDIX}
The quantities in this paper involved with distance estimation are sometimes represented by 
symbols that are not as familiar as the astrophysical ones.  Following is a list of those 
used in the present paper along with the page number of first appearance (for context) and 
equation number (if any).
\vskip 0.2 in
\item{$B$} : overall distance ratio for a scale [4]
\item{$d_S$} : true distance $d$ multiplied by scale factor $B$ for scale $S$  [4]
\item{$d_S^\prime$} : estimated distance in scale $S$ [4]
\item{$F$} : correction factor for expansion distance; ratio true to expansion distance 
[21; Eqs.~(11a), (11b), (12)]
\item{${\cal P}_H$} : ratio of {\it Hipparcos} parallax of an object to \lq best estimate' 
parallax [23]
\item{${\cal R}_S$} : distance ratio for an object in distance scale $S$ relative to a 
standard distance [4]
\item{$\alpha$} : relative error in a distance estimate in general [4] 
(Note: This departs from notation used in SE96, where $\alpha$ was the true relative 
parallax error $\sigma_\pi/\pi$.)
\item{$\alpha_S$} : relative error for distance scale $S$, assumed same for all objects [5]
\item{$\alpha_S^\prime$} : estimate of $\alpha_S$ from observations [8; Eqs.~(5), (8)]
\item{$\Gamma$} : Phillips estimator for logarithm of sample distance ratio [11; Eq.~(7)]
\item{$\Gamma^*$} : dex($\Gamma$) to give sample distance ratio [11]
\item{$\kappa$} : Phillips estimator for distance ratio [5; Eq.~(2)]
\item{$\lambda$} : relative parallax error $\sigma_\pi^\prime/\pi^\prime$ [2]
\item{$\pi$} : true trigonometric parallax of object [2]
\item{$\pi^\prime$} : measured trigonometric parallax of object [2]
\item{$\sigma_S$} : standard error of $d_S$  [4]
\item{$\sigma_S^\prime$} : estimate of $\sigma_S$ [4]
\item{$\sigma_\pi$} : true standard error of $\pi$  [2]
\item{$\sigma_\pi^\prime$} : estimate of $\sigma_\pi$ [2]
\item{$\sigma_{\cal R}$} : standard error of ${\cal R}_S$ [4; Eq.~(1)]
\item{$\sigma_{\cal R}^\prime$} : estimate of $\sigma_{\cal R}$ [5; Eq.~(1a)]
\item{$\zeta$} : modification of Phillips estimator $\kappa$ to mitigate excessive weight on 
large distances [6; Eq.~(3)]

\end